\documentclass[usenatbib,useAMS,referee]{biom_ours}
\usepackage{subcaption}
\usepackage{graphicx}
\usepackage{amsfonts, amssymb}
\usepackage{xr}
\makeatletter

\newcommand*{\addFileDependency}[1]{
\typeout{(#1)}
\@addtofilelist{#1}
\IfFileExists{#1}{}{\typeout{No file #1.}}
}\makeatother

\usepackage{ulem}
\usepackage{amsmath}
\usepackage{mathdots}
\usepackage{lscape}
\usepackage{lineno}
\usepackage{enumitem}
\usepackage[pdftex,dvipsnames]{xcolor} 
\usepackage{comment}
\usepackage{amssymb}
\usepackage{url} 
\usepackage{tikz}
\usetikzlibrary{matrix}
\usepackage{multirow}

\newcommand{\ADtext}[1]{{\color{purple}{#1}}} 

\title{Hidden Markov models with an unknown number of states and a repulsive prior on the state parameters}
\author{Ioannis Rotous, Alex Diana, Alessio Farcomeni, Eleni Material}
\date{January 2023}

{
\author
{Ioannis Rotous \vspace{-0.3cm}
\emailx{ir237@kent.ac.uk} \\
School of Mathematics, Statistics and Actuarial Science, University of Kent
\and
\vspace{-0.11cm}
Alex Diana\emailx{ad23269@essex.ac.uk} \\
Department of Mathematics, Statistics and Actuarial Science, University of Essex
\and
\vspace{-0.12cm}
Alessio Farcomeni\emailx{ alessio.farcomeni@uniroma2.it} \\
Department of
Economics and Finance, Tor Vergata University of Rome
\and
\vspace{-0.11cm}
Eleni Matechou\emailx{e.matechou@kent.ac.uk} \\
School of Mathematics, Statistics and Actuarial Science, University of Kent 
\and
\vspace{-0.11cm}
Andréa Thiebault \emailx{andrea.thiebault@cnrs.fr} \\
Institut des Neurosciences Paris-Saclay (NeuroPSI), CNRS UMR 9197, Université Paris-Saclay
}
}

\begin{document}

\clearpage
\date{{\it Received } . {\it Revised } .  {\it
Accepted } .}

\pagerange{\pageref{firstpage}--\pageref{lastpage}} 
\volume{64}
\pubyear{2008}
\artmonth{December}

\doi{10.1111/j.1541-0420.2005.00454.x}

\label{firstpage}


\begin{abstract}

Hidden Markov models (HMMs) offer a robust and efficient framework for analyzing time series data, modelling both the underlying latent state progression over time and the observation process, conditional on the latent state. However, a critical challenge lies in determining the appropriate number of underlying states, often unknown in practice. In this paper, we employ a Bayesian framework, treating the number of states as a random variable and employing reversible jump Markov chain Monte Carlo to sample from the posterior distributions of all parameters, including the number of states. Additionally, we introduce repulsive priors for the state parameters in HMMs, and hence avoid overfitting issues and promote parsimonious models with dissimilar state components. We perform an extensive simulation study comparing performance of models with independent and repulsive prior distributions on the state parameters, and demonstrate our proposed framework on two ecological case studies: GPS tracking data on muskox in Antarctica and acoustic data on Cape gannets in South Africa. Our results highlight how our framework effectively explores the model space, defined by models with different latent state dimensions, while leading to latent states that are distinguished better and hence are more interpretable, enabling better understanding of complex dynamic systems.
\end{abstract}


{
\maketitle}

\section{Introduction}
\label{sec:intr}

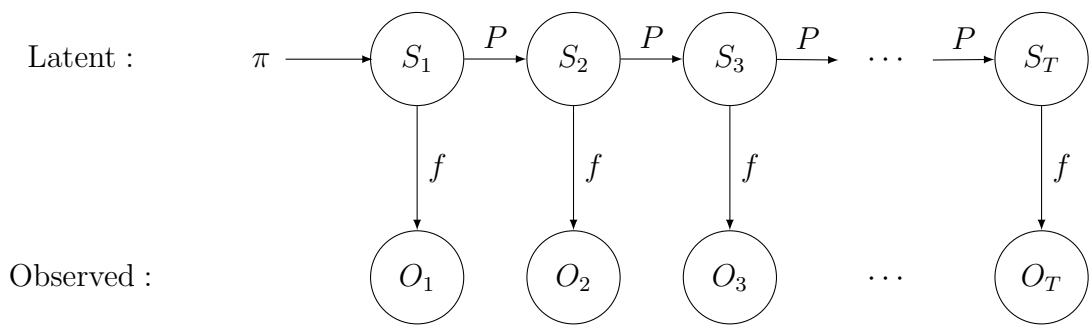
\begin{figure}
  \centering
  \begin{tikzpicture}
    \matrix[matrix of math nodes, column sep=2em, row sep=4em,
      cells={nodes={circle,draw,minimum width=3em,inner sep=0pt}},
      column 6/.style={nodes={rectangle,draw=none}},
      column 2/.style={nodes={rectangle,draw=none}},
      column 1/.style={nodes={rectangle,draw=none}}] (m) {
      \text{Latent}: & 
      \pi &  S_{1} & S_{2} & S_{3} & \cdots & S_{T}\\
      \text{Observed}: & &
      O_{1} & O_{2} & O_{3} & \cdots & O_{T}\\
    };
    \foreach \X in {3,4,5,6}
    {
      \draw[-latex] (m-1-\X) -- (m-1-\the\numexpr\X+1) node[midway,above]{$P$};
      \ifnum\X=6
        \draw[-latex] (m-1-7) -- (m-2-7) node[midway,right]{$f$};
      \else
        \draw[-latex] (m-1-\X) -- (m-2-\X) node[midway,right]{$f$};
      \fi
    }
    \draw[-latex] (-3.5,1.45) -- (-2.34,1.45);
  \end{tikzpicture}
  \caption{Illustration of a hidden Markov model evolution across $t = 1, 2, ..., T$ time points, with latent states $S_{t}$ and corresponding observations $O_{t}$, characterized by an initial latent distribution $\pi$, transition probabilities $P$, and emission distribution $f$.}
  \label{fig:markov-observations}
\end{figure}

Hidden Markov models (HMMs) are a powerful and well-established framework for analyzing time series data in cases where the studied system transitions between a set of hidden states over time. HMMs jointly model two processes: the underlying latent process of the hidden states, and the observation process, conditional on the states, as shown in Figure \ref{fig:markov-observations} \citep{cappe2009inference, zucchini2009hidden}. HMMs enable efficient modelling of the evolution of the latent states across time and, conditional on those latent states, explicit modelling of the data observation process, even in complex systems and processes with multiple latent states and complicated observation processes \citep{popov2017analysis}. 

{HMMs commonly employ a first-order Markov chain, where the evolution of the latent states depends only on the previous time point. Additionally, conditional on the latent state, they emit observables at the current time point, independent of the rest of the observables. {Further details can be found in Section \ref{subsec::hmm}, which describes the joint distribution of observables with latent states (Equation \ref{eq:hmm}). 
The efficacy of HMMs relies on the separation of the latent and observation processes and the use of algorithms that efficiently marginalize over the latent states, such as the forward/backward algorithm for computing the likelihood function and the Viterbi algorithm for finding the most likely sequence of hidden states \citep{zucchini2009hidden,bartolucci2012latent}. Hence, HMMs have proven to be powerful and easy-to-use tools, and are widely utilized in various fields, such as finance \citep{ryden1998stylized}, biology \citep{leroux1992maximum}, social science \citep{rabiner1989tutorial, zucchini2009hidden},  medicine \citep{farcomeni2017penalized}, and ecology \citep{schmidt2016ungulate, patterson2017statistical}, among others. }

One of the key challenges in employing HMMs for data analysis is the decision on the appropriate number of underlying states in the system. In practice, the true number of states is often unknown. It is standard 
practice to fix the number of latent states or fit models that consider different numbers
of latent states \citep{robert1993bayesian, chib1996calculating, robert1998resampling} in either a classical \citep{huang2017model}, or Bayesian framework \citep{berkhof2003bayesian}, and subsequently compare them with
appropriate criteria to select the number of states. 
However, in this case, the model needs to be fitted multiple times, and in the end, a single model is used for interpretation, but without accounting for the uncertainty in the model selection
process itself \citep{mclachlan2019finite}. {Alternatively, in a Bayesian framework, the number of latent states can be treated as an additional random variable, and hence reversible jump Markov chain Monte Carlo (RJMCMC) \citep{green1995reversible} methods can be employed to sample from the posterior distribution in this case where the model dimension is not fixed; indeed RJMCMC has been used extensively within an HMM context \citep{cappe2009inference, robert2000bayesian, cappe2003reversible,russo2022covariate}}. {We note that HMMs can equivalently be viewed from the perspective of dynamic mixture models \citep{spezia2020bayesian}, both in discrete and continuous space \citep{reynolds2009gaussian, bartolucci2011bayesian}, with the number of mixture components corresponding to the number of states. In this paper, we refer to HMMs and corresponding states, but the concepts equally apply to dynamic mixture models and corresponding components.}

Within a Bayesian framework, state allocation ($S_{t}$ in Figure \ref{fig:markov-observations}) can be sampled within the MCMC algorithm \citep{stephens1997bayesian}, so that the complete data likelihood is used \citep{king2014statistical}. This approach however can lead to a substantial number of sampled latent variables. Instead, state allocation can be marginalised out of the model, as is standard practice within the HMM machinery \citep{russo2022covariate}, so that the observed data likelihood is used for inference, and this is the approach we employ in this paper. 
However, in either case, HMMs with a variable number of states are prone to overfitting, and hence such algorithms can lead to an unnecessarily large number of similar states \citep{duan2018bayesian}. Recent advancements in the field of mixture modelling have introduced the use of repulsive priors, which promote parsimony in the model \citep{petralia2012repulsive,  quinlan2021class,natarajan2023cohesion}. These repulsive prior distributions serve to impose constraints on the proximity of state parameters, which discourages similar states from being created. Unavoidably, this particular form of penalty applied to the parameter space also affects the selection of the number of states \citep{natarajan2023cohesion}. An additional advantage associated with incorporating a repulsive prior into HMMs, whether with fixed or variable dimensions, is the mitigation of overfitting. In certain instances, conventional mixture models, and hence HMMs, may excessively fine-tune their components (states) to capture noise in the data, resulting in poor generalization of the obtained results. Extensive research, as documented in the literature \citep{petralia2012repulsive,  quinlan2021class,beraha2022mcmc}, has demonstrated that these issues can be effectively addressed by introducing repulsion constraints among distribution parameters within the model, thereby promoting dissimilarity among states. 

To introduce a repulsion constraint within an HMM framework we use interaction point processes, referred to as a repulsive prior in this paper, which is a class of distributions on a set of points that actively penalises cases where points (parameters) are close together (similar). Specifically, we consider a distribution belonging to the family of pairwise interaction point processes, the Strauss point process prior, first described in \cite{strauss1975model}, as a prior distribution on the state parameters of the emission distribution $f$ as illustrated in Figure \ref{fig:markov-observations}.
This approach enables us to achieve more effective and interpretable modelling without pre-specifying the number of states, thereby improving various aspects of analysis, including inference, model selection, and our overall understanding of system dynamics.  

HMMs have been extensively used in the ecological literature \citep{gimenez2009winbugs, schmidt2016ungulate, patterson2017statistical}, since they are well-suited to capturing the underlying latent state structure in ecological systems, enabling researchers to seamlessly integrate the observed data with the unobserved latent states \citep{glennie2023hidden}. In ecological systems, these latent states can correspond to life stages \citep{mcclintock2020uncovering} 
or behavioural states \citep{schmidt2016ungulate,nicol2023flywaynet}. 
In this paper, we demonstrate our approach using two ecological case studies: GPS data on muskox, \textit{Ovibos moschatus}, in Antarctica, also analysed in \citet{pohle2017selecting}, who used model selection criteria to select the number of states in their HMM, and acoustic data on Cape gannets, \textit{Morus capensis}, in South Africa, analysed in \citet{thiebault2021animal} where behavioural state classification was performed manually to train a subsequent model. {Our results demonstrate how our framework effectively explores the model space, defined by models with different latent state dimensions, while leading simpler models with latent states that are distinguished better and hence are more interpretable.} {Finally, our extensive simulation study demonstrates that, when the true model is fitted to the data, then the repulsive prior leads to inference, in terms of density estimation and state classification, that is equivalent to, or marginally better of, that of independent priors.} Therefore, for noisy, real data, where the model is typically a simplified version of the actual data-generating process, the repulsive prior avoids overfitting and leads to more parsimonious models, whilst in data simulated from the fitted model, the repulsive prior does not overpenalise, and leads to results that are on par to those obtained from an independent prior.


The article is structured as follows: Section \ref{sec:models} introduces the general concepts of HMMs and repulsive priors, and gives a broad overview of the model-fitting approach developed in this paper, with technical details provided in the Supplementary Material. Section \ref{sec:sim} discusses the results of our extensive simulation study, while Section \ref{sec:aplGPS} presents the results of modelling the muskox GPS data, and Section \ref{sec:aplAcoustic} the results of modelling the Cape gannet acoustic data. Finally, the paper concludes with a discussion in Section \ref{sec:disc}.

\section{Models}
\label{sec:models}

\subsection{Hidden Markov Models}
\label{subsec::hmm}

\par{
A first-order hidden Markov model is a stochastic process consisting of a set of hidden/latent states $S$ and observations $O$. The state process is assumed to be an N-state Markov chain $P(S_{t}|S_{1}, S_{2},..., S_{t-1}) = P(S_{t}|S_{t-1})$ with $S_{t}\in \left \{ 1,2,..., N \right \}$. The evolution of the hidden states across time is described by the transition matrix $P$, where $P_{ij}$ is the probability of transitioning from state $i$ to state $j$ for all $t$, i.e.,

\begin{equation*}
    P(S_{t}=j|S_{t-1}=i) = P_{ij}.
\end{equation*}

The probability of being in a particular state at the first time point can be modeled using an initial state distribution $\pi$ i.e. $P(S_1 = i) = \pi_i$. 
At each time step, we observe $O_{t}$, whose distribution only depends on the current value $S_t$, 

\begin{equation*}
    f(O_{t}|O_{1},...,O_{t-1},S_{1},...,S_{t}) = f(O_{t}|S_{t}).
\end{equation*}
Therefore, the model for a particular sequence of observations given the hidden states,
$f(O_{1},O_{2},...,O_{T}|S_{1},S_{2},...,S_{T})$, can be factorised as $\prod_{t=1}^{T}f(O_{t}|S_{t})$, and the joint model of a particular sequence of hidden states and observations is equal to 
\begin{equation}
\label{eq:hmm}
    P(O_{1}, O_{2},..., O_{T}, S_{1}, S_{2},..., S_{T}) = \pi_{S_{1}}f(O_{1}|S_{1}) \prod_{t=2}P_{S_{t-1},S_{t}}f(O_{t}|S_{t})
\end{equation}

\par{
{The emission distribution, $f$, which describes how the observations are generated conditional on the states, is a function of corresponding state-specific parameters, $\theta_i$, where $\theta_i$ can be a scalar or a vector of parameters, and it is on these parameters that we place the repulsive prior distributions proposed in this paper.}

\subsection{Repulsive prior}
\label{subsec:repuls}

We assume a pairwise interaction point processes on the parameters $\theta_{i}$ described in Section \ref{subsec::hmm}. Pairwise interaction point processes can be constructed by defining a point process with density of the form
\begin{equation}
    h(\theta_{1},\theta_{2},...,\theta_{N}|N,\xi_1, \xi_2) = \frac{1}{Z_{\xi}}g(\theta_{1}, \theta_{2}, ..., \theta_{N}|N,\xi_{1},\xi_{2}) =\frac{1}{Z_{\xi}}\prod_{i=1}^{N}\phi_{1}(\theta_{i}|\xi_1)\prod_{1\leq i < j \leq N}\phi_{2}(\theta_{i},\theta_{j}|\xi_2)
    \label{repeq}
\end{equation}
where $\xi = (\xi_1,\xi_2)$. More details on interaction point processes can be found in \citet{moller2003statistical}.  It is convenient to take $\phi_{1}(\theta_{i}| \xi_1) = \xi_1 \mathbb{I}[\theta_{i}\in R]$, where $\xi_1$ is the intensity of the points and $R$ is the region where $\theta$ is defined. In our case, we use the Strauss process \citep{strauss1975model}, which assumes   \[\phi_{2}(\theta_{i},\theta_{j}|\xi_2 = \left \{ a, d \right \}) = a^{\mathbb{I}[\left \| \theta_{i}-\theta_{j}<d \right \|]}.\]
This term denotes the interaction term between the locations $\theta_{i}, \theta_{j}$ for parameters $a, d$.
Parameter $a$ ranges from $0$ to $1$ and controls the penalty magnitude between the points $\theta_{i}$ and $\theta_{j}$; the smaller the $a$ the stronger the penalty. Parameter $d$ is the threshold such that if the distance (typically the Euclidean distance) between two components is less than $d$, the penalty applies. Lastly, the normalizing constant $Z_{\xi}$ of Equation~\eqref{repeq} is intractable, which makes inference on parameter $\xi$ challenging. If $a = 1$ there is no penalty, and we retrieve a point process (referred to as the independent point process for the rest of the article), with points being drawn from {an} independent Uniform distribution with $\phi_{1}(\theta_{i}|\xi_{1}) = \xi_1 \mathbb{I}[\theta_{i}\in R]$,
\begin{equation}
    h(\theta_{1}, \theta_{2}, ..., \theta_{N}|N,\xi_{1}) = \frac{1}{Z_{\xi_{1}}}\prod_{i=1}^{N}\phi_{1}(\theta_{i}|\xi_1),
    \label{standeq}
\end{equation}
In this case,  the normalizing constant is tractable and equal to $Z_{\xi} = \xi_1^{N}\left | R \right |^{N}$.



Finally, bringing together the concepts described in Sections \ref{subsec::hmm} and \ref{subsec:repuls}, the hierarchical representation of an HMM model with a random number of states and a repulsive prior on the state parameters is:

\begin{align}
& N \sim g(\cdot) \notag \\
    & O_{t} \sim f(O_{t}|\theta_{S_{t}}), \ t = 1, 2, ..., T \notag \\
     & \underline{\theta} = (\theta_{1}, \theta_{2}, ..., \theta_{N})|N \sim \text{StraussProcess}(\xi, a, d) \notag \\
      & P(S_{1} = i) = \pi_{i}, \ i = 1, 2, ..., N  \label{eq:fullhmm} \\    
    & P(S_{t}=j|S_{t-1}=i) = P_{ij}, \ i,j = 1, 2, ..., N, \ \  t = 2, 3, ..., T \notag \\
     & P_{i.} = (P_{i1}, P_{i2}, ..., P_{iN})|N \sim \text{Dirichlet}(a_{1}^{P}, a_{2}^{P}, ..., a_{N}^{P}), \ i = 1, 2, ..., N \notag \\
    &  \pi = (\pi_{1}, \pi_{2}, ..., \pi_{N})|N \sim \text{Dirichlet}(a_{1}^{\pi}, a_{2}^{\pi}, ..., a_{N}^{\pi})
    \notag
\end{align}

\subsection{Inference}
\label{subsec:infer}

\par{

Inference is made on the parameters $\underline{\theta}, \pi, P, \xi $ and $N$. Since the dimension of $\underline{\theta}, P, \pi$ changes according to $N$, we employ a RJMCMC sampling algorithm that allows us to move between models with different parameter dimensions. On each iteration of the algorithm, we implement a fixed and a variable dimension move. The fixed {dimension} move updates the model parameters $(\underline{\theta}, P, \pi)$ conditional on the number of states, and the variable {dimension}  move updates the dimension of the model. Finally, we update $\xi$ with the use of the exchange algorithm \citep{murray2012mcmc}, described in Section \ref{updt:xi}.


\par{
$\bullet $
{\textbf{Fixed dimension Moves}}
\par{
We update the model parameters $\pi, P, \underline{\theta}$, for a fixed value N, by sampling from the corresponding posterior distributions using a Metropolis Hastings  algorithm \citep{metropolis1953equation, hastings1970monte}, since the HMM likelihood with state allocation marginalised out is not conjugate {to} the prior distribution(s).}
}
}

\par{
$\bullet${\textbf{ Variable dimension moves}}

With probability $0.5$, we choose between the moves {Split/Combine} and {Birth/Death}. 


\par{
\textbf{ Split/Combine moves}

\par{
In each step, we choose whether to split or combine states with probability 0.5. In the split case, if we have a single state, with probability one we propose to split that state. If we have more than one states, we choose uniformly one of the $N$ states, denoted by $j_{*}$, which we propose to split into $j_{1}$ and $j_{2}$, therefore proposing to split $\pi_{j_{*}}, P_{j_{*},.}, P_{.j_{*}}, \theta_{j_{*}}$ into new model parameters $(\pi_{j_{1}}, P_{j_{1}.}, P_{.j_{1}}, \theta_{j_{1}})$ and $(\pi_{j_{2}}, P_{j_{2}.}, P_{.j_{2}}, \theta_{j_{2}})$.

The split move is accepted with probability ${\rm min}\left \{ 1,A \right \}$, where

\begin{align*}
    A & = \frac{f(\left \{ O_{t} \right \}_{t=1}^{T}|\left \{ \pi \right \}_{j=1}^{N+1},\left \{ P_{j.} \right \}_{j=1}^{N+1},\left \{ \theta_{j} \right \}_{j=1}^{N+1}) }{f(\left \{ O_{t} \right \}_{t=1}^{T}|\left \{ \pi \right \}_{j=1}^{N},\left \{ P_{j.} \right \}_{j=1}^{N},\left \{ \theta_{j} \right \}_{j=1}^{N})} 
     \frac{p(\left \{ \pi \right \}_{j=1}^{N+1},\left \{ P_{j.} \right \}_{j=1}^{N+1},\left \{ \theta_{j} \right \}_{j=1}^{N+1},N+1)}{p(\left \{ \pi \right \}_{j=1}^{N},\left \{ P_{j.} \right \}_{j=1}^{N},\left \{ \theta_{j} \right \}_{j=1}^{N},N) } \frac{q(N+1\rightarrow N) }{q(N\rightarrow N+1)}
\end{align*}
where {$p(\cdot)$ is the joint prior distribution of all parameters, and $q(N+1\rightarrow N)$ and $q(N\rightarrow N+1)$ are the proposal probabilities for the transdimensional moves with details given in Sections \ref{supsubsec:infergps} 
and \ref{supsubsec:model} 
in the Supplementary Material.} 

In the combine case, we choose the two states $j_{1}$ and $j_{2}$ whose distance is the smallest, and we propose to combine them to $j_{*}$. The combine move is accepted with probability $\text{min}\left \{ 1,A^{-1} \right \}.$
}
}
\\
\par{
\textbf{Birth/Death moves}

\par{
The Birth/Death move is performed similarly to the Split/Combine move. Specifically, if we have $N$ states, we choose with probability 0.5 to give birth to a new state or kill an existing one.

In the birth move, we propose a new state generated by sampling its parameters from the prior distribution. On the other hand, for the death move, we uniformly choose a state and propose to kill it. In this case, the acceptance probability of the birth move is again $\text{min}\left \{ 1,A \right \}$ whereas for the death move it is $\left \{1, A^{-1} \right \}$ with

\begin{align*}
   A & =  \frac{f(\left \{ O_{t} \right \}_{t=1}^{T}|\left \{ \pi \right \}_{j=1}^{N+1},\left \{ P_{j.} \right \}_{j=1}^{N+1},\left \{ \theta_{j} \right \}_{j=1}^{N+1})}{f(\left \{ O_{t} \right \}_{t=1}^{T}|\left \{ \pi \right \}_{j=1}^{N},\left \{ P_{j.} \right \}_{j=1}^{N},\left \{ \theta_{j} \right \}_{j=1}^{N})  } \frac{p(\left \{ \pi \right \}_{j=1}^{N+1},\left \{ P_{j.} \right \}_{j=1}^{N+1},\left \{ \theta_{j} \right \}_{j=1}^{N+1},N+1) }{p(\left \{ \pi \right \}_{j=1}^{N},\left \{ P_{j.} \right \}_{j=1}^{N},\left \{ \theta_{j} \right \}_{j=1}^{N},N)} \frac{q(N+1\rightarrow N) }{q(N\rightarrow N+1)}
\end{align*}
 where {$p(\cdot)$ is the joint prior distribution of all parameters, and $q(N+1\rightarrow N)$ and $q(N\rightarrow N+1)$ are the proposal probabilities for the transdimensional moves with details given in Sections \ref{supsubsec:infergps} 
and \ref{supsubsec:model} 
 in the Supplementary Material.} 

\subsection{Update $\xi$ }
\label{updt:xi}
{Parameter $\xi$ of the Strauss process prior is updated with a Metropolis Hastings algorithm. At each iteration, we propose $\xi_{*}$ from }
\begin{equation*}
    \xi_{*}  \sim q(\xi_{*}|\xi) = \text{LogNormal}(\text{log}(\xi),\tau_{\xi})
\end{equation*}

\noindent However, calculation of the  Metropolis Hastings acceptance ratio depends on the ratio of the corresponding densities (Equation \ref{repeq}) at $\xi$ and $\xi_{*}$,

\begin{equation}
    \frac{h(\theta_{1}, \theta_{2}, ..., \theta_{N}|N,\xi_{*}, a, d)}{h(\theta_{1}, \theta_{2}, ..., \theta_{N}|N,\xi, a, d)}  = \frac{\frac{1}{Z_{\xi_{*}}}\prod_{i=1}^{N}\xi_{*} \mathbb{I}[\theta_{i}\in R]}{\frac{1}{Z_{\xi}}\prod_{i=1}^{N}\xi \mathbb{I}[\theta_{i}\in R]} = \frac{Z_{\xi}}{Z_{\xi_{*}}}\frac{\prod_{i=1}^{N}\xi_{*} \mathbb{I}[\theta_{i}\in R]}{\prod_{i=1}^{N}\xi \mathbb{I}[\theta_{i}\in R]}
\end{equation}
which is intractable.
Therefore, we employ the exchange algorithm of \citet{murray2012mcmc}, and simulate {a new parameter}, $\theta_{aux}$, from the Strauss process (Equation \ref{repeq}) with parameter $\xi_{*}$ using the birth and death algorithm \citep{moller1994statistical} 
described in  Section \ref{alg:birthdeath} 
of the Supplementary Material, and accept $\xi_{*}$ with probability $\text{min}(1,A)$, where

\begin{equation}
    A = \frac{q(\xi|\xi_{*})p(\xi_{*})g(\theta_{1}, \theta_{2}, \ldots, \theta_{N}|N,\xi_{*},a,d)}{q(\xi_{*}|\xi)p(\xi)g(\theta_{1}, \theta_{2}, \ldots, \theta_{N}|N,\xi,a,d)}\frac{g(\underline{\theta_{aux}}|\left | \theta_{aux} \right |,\xi,a,d)}{g(\underline{\theta_{aux}}|\left | \theta_{aux} \right |,\xi_{*},a,d)}
    \label{acceptance:ratio:ex}
\end{equation}
{with $p(\xi)$ the prior distribution assigned on parameter $\xi$, $g(\cdot)$ the unormalised density of the Strauss process in Equation \eqref{repeq} and $\underline{\theta_{aux}}$ the simulated new parameters vector of size $\left | \theta_{aux} \right |$. }

In this paper, we choose $a$ and $d$ based on the recommendation of \citet{beraha2022mcmc}. The threshold is calculated as $d = \text{min}_{r>0}\left \{ r : \text{local minimum for} \ p(r) \right \}$ where $p(r)$ is the kernel density of all pairwise distances $r$ between the observations in the sample. {We present examples that demonstrate this process in Section \ref{subsec:penthresh} 
of the Supplementary Material.} The penalty $a$ is set equal to $\exp(-n^{*}\log(k_{s}))$, where $n^{*}$ corresponds to the minimum acceptable size of a state, for example 5\% of the sample size, {and set $\log(k_{s})=1$, so that $a$ is only a function of $n^{*}$.} Further details on the penalty choice are given in Section \ref{subsec:penthresh} 
of the Supplementary Material.

}
}

\subsection{Label Switching}
\label{sec:labelswitching}

{Inference for mixture models is usually complicated by label switching, which occurs because the labels of states are interchangeable, since reordering the states has no effect on resulting inference. Addressing label switching is important for interpreting the resulting states and corresponding parameters.

In this paper, we employ two approaches to {choose an}  ordering of states and hence deal with label switching.. The first, {employed in the case study of Section \ref{sec:aplGPS}}, involves imposing an ordering on one of the state parameters in cases when such ordering is meaningful, such as $\theta_{1} \leq \theta_{2} \leq ... \leq \theta_{N}$ , with $\theta_{i}$ the mean of the state $i$, as demonstrated by \citet{russo2022covariate}. During each iteration of the RJMCMC, the parameters are rearranged according to this predefined order, ensuring that the same state in different iterations maintains its label. The second, {employed in the case study of Section \ref{sec:aplAcoustic}}, is the post-processing method of \citet{bartolucci2011bayesian}, performed after the end of all MCMC iterations, which involves computing the posterior mode (MAP) and subsequently determining, for each distinct posterior sample, the permutation that minimizes the distance between the MAP estimate and the permuted posterior sample. 


\subsection{State Allocation}
\label{sec:stateallocation}
{As discussed in Section \ref{sec:models}, our approach relies on marginalising over the latent state allocation instead of sampling it as part of the MCMC inference. Therefore, when interest lies in interpreting state allocations, these can be obtained at a post-processing stage {by sampling from their posterior distribution} 
\begin{align*}
    P(S_{1}=j|\left \{ O_{t} \right \}_{t=1}^{T},\pi_{j},\theta_{j}) & \propto \pi_{j}f(O_{1};\theta_{j}), \qquad t = 1 \\
    P(S_{t} = j | S_{t-1} = i, \left \{ O_{t} \right \}_{t=1}^{T},P_{i,j},\theta_{j,l}) & \propto P_{i,j}f(O_{t};\theta_{j}), \qquad t > 2
\end{align*}


\section{Simulation Study}
\label{sec:sim}

{We conducted an extensive simulation study to compare a model with a repulsive prior   {with} a model with an independent prior. We simulated data from a model with five states, each described by a normal distribution with mean locations $\underline{\mu} = (-10, -5, 0, 5, 10)$ and standard deviations $\underline{\sigma} = (\sigma_{1}, \sigma_{2}, \sigma_{3}, \sigma_{4}, \sigma_{5})$ chosen so that there is increasing amount of overlap between the state distributions. Specifically, we took $\sigma_{1} = \sigma_{2} = \sigma_{3} = \sigma_{4} = \sigma_{5}$ and we chose them to have value 1.1408, 1.4726, 2.5709 and 4.2319, such the overlap between consecutive {mixture} is equal to 3\%, 9\%, 33\% and 55\%, respectively, which corresponds to  5\%, 15\%, 50\%, and 75\% overall overlap. The consecutive index overlap is calculated by integrating the area where the two density functions overlap, which is equal to  $\int_{\mathbb{R}} \min\left[ \text{Normal}(x;\mu_{i},\sigma_{i}),\text{Normal}(x;\mu_{i+1},\sigma_{i+1})\right ]dx$ for $i = 1, 2, ..., 4$ 
whereas the overall overlap is based on the overlap index described in \citet{pastore2019measuring}, 
We plot the corresponding state distributions in each case in Section \ref{supsec:sim} of the Supplementary Material.} {For each degree of overlap, we varied the sample size $n=\left \{ 50,100 \right \}$ and number of time points 
$T=\left \{ 5,10 \right \}$.} {The initial probability distribution $\pi$ and transition probability matrix $P$ were chosen to have  {all elements equal to 1/5}, ensuring equal state sizes across all time points, allowing us to focus on the effect of state overlap on inference.}

The repulsive parameters, penalty $a$ and threshold $d$, were chosen according to \citet{beraha2022mcmc} as described in Section \ref{updt:xi}. We considered two values of $n^{*}$: $n_{2.5}^{*}$, corresponding to penalties that consider cluster of sizes less than $2.5\%$ and $n_{5}^{*}$, corresponding to penalties that consider cluster of sizes less than $5\%$.  The results are given in Tables \ref{tab:rafsim_2.5} and \ref{tab:rafsim} of Section \ref{supsec:sim} 
of the Supplementary Material}. {For each simulation scenario, we placed repulsive (Equation \ref{repeq}) and independent (Equation \ref{standeq}) priors on the location parameters $\mu$. Finally, we placed Dirichlet priors with parameters equal to 1 on the initial and transition probabilities $\pi$ and $P$ and a uniform distribution on standard deviations $\sigma$ with lower and upper bound $0$ and $2\times$ 90\% quantile of the observations, respectively.} 

We report the Kullback-Liebler (KL) divergence between the true state distributions and the estimated distributions, together with the misclassification rate between the true state allocation of observations and the inferred allocation, averaged across MCMC iterations, time points, and $100$ replications for each scenario. Details about the calculation of KL divergence and miscassification rate are
given in Section \ref{supsec:sim} 
of the Supplementary Material. 

{We employed the ordering constraint described in Section \ref{sec:labelswitching}, with $\mu_{i} \leq \mu_{i+1}$, for $i = 1, 2, ..., N$ to deal with label switching and we calculate the misclassification rate in each case computing state allocation as described in Section \ref{sec:stateallocation}.}} {Finally, we summarise the posterior mode of the distribution for the number of states in each scenario, averaged across all replications.}


{The results for case with $n^{*}_{2.5}$ and $n^{*}_{5}$} are presented in Table \ref{tab:rafsim_2.5} and \ref{tab:rafsim}, respectively, of Section \ref{supsec:sim} in the Supplementary Material.  The results demonstrate that the two models have very similar performance in terms of selected number of states, density estimation and misclassification but with a small advantage of the RP. For penalty $n^{*}_{2.5}$, 13/16 cases and 10/16, in Table \ref{tab:rafsim_2.5}, KL and misclassification error is lower for RP compared to ID. On the other hand, for penalty $n^{*}_{5}$, 14/16 and 10/16, in Table \ref{tab:rafsim}, KL and misclassification error is lower for RP compared to ID. As the number of time points $T$ or sample size $n$ increases, KL divergence and misclassification error decrease for both models in general. As expected, as the amount of overlap increases, the posterior mode of the number of states decreases for both models. In this case, there is no obvious differences in the effects of the two chosen penalties.

\section{Case study 1: Muskox GPS data}
\label{sec:aplGPS}
We consider data on muskox movement in east Greenland  analysed in \citet{pohle2017selecting}. The data consist of 25103 hourly GPS locations, covering a period of roughly three years, giving information on the step length, $L_{t}$, which represents the distance in meters between time points $t-1$ and $t$, and the turning angle between time points $t-2$ and $t$, $A_{t}$, as is standard practice in GPS tracking data \citep{zucchini2009hidden,langrock2012flexible,patterson2017statistical}.

We model the step-length at time $t$, $L_{t}$, using a 0-inflated Gamma distribution to account for the number of 0s in the data (0.58$\%$ or 145): 
\begin{equation}
    f(L_{t}|S_{t}) = z_{S_{t}}\delta_{L_{t}}(0) + (1-z_{S_{t}}) {\rm Gamma}(L_{t};\mu_{S_{t}},\sigma_{S_{t}}) \label{eq:length}
\end{equation}}

\noindent where $z_{S_{t}}$ represents the probability of individuals being stationary given their corresponding state at time $t$, with $\delta_{L_{t}}(0)$ being a Dirac measure at step-length 0, and $\mu_{S_{t}}$ and $\sigma_{S_{t}}$ denote the mean and standard deviation of the Gamma distribution governing the step length, conditional on state.

We model the turning angle between time points $t-2$ and $t$, $A_{t}$, 
using a \text{vonMises} distribution with location and concentration deviation parameters $m_{S_{t}}$ and $k_{S_{t}}$ 
, respectively 
\begin{equation}
     f(A_{t}|S_{t}) = \text{vonMises}(A_{t};m_{S_{t}},k_{S_{t}}) \label{eq:angle}
\end{equation}

Therefore, the observation at time $t$, given state at time $t$, is modelled as 
\begin{equation}
        f(O_{t}|S_{t}) = f(L_{t}|S_{t})f(A_{t}|S_{t}) \label{eq:gpsobs}
\end{equation}

{We choose to set a repulsive prior on the mean step length, 
$\mu_{1}, \mu_{2}, ..., \mu_{N}$}
\begin{equation}
   \underline{\mu} = (\mu_{1}, \mu_{2},...,\mu_{N})|N  \sim \text{StraussProcess}(\mu_{1}, \mu_{2},...,\mu_{N};\xi, a, d) \label{eq:straussgps} 
\end{equation}
 and for comparison purposes also present results considering an independent, prior distribution 
 \begin{equation}
   \underline{\mu} = (\mu_{1}, \mu_{2},...,\mu_{N})|N   \sim \text{IndependentProcess}(\mu_{1}, \mu_{2},...,\mu_{N};\xi)\label{eq:poissgps} 
\end{equation}
as described in Section \ref{subsec:repuls}.
Details on Equations \eqref{eq:straussgps} and \eqref{eq:poissgps} are given in Section \ref{supsubsec:modelgps}  in the Supplementary Material.

We also place the following prior distributions on the remaining model parameters (with more details on the prior distribution choices and inference given in Section \ref{supsec:smgps} of the Supplementary Material.) 
\begin{align*}
    N & \sim \text{Uniform}\left \{ 1, 2, ..., N_{max} \right \} \\
    \pi = (\pi_{1}, \pi_{2}, ..., \pi_{N})|N & \sim \text{Dirichlet}(a^{\pi}_{1}, a^{\pi}_{2}, ..., a^{\pi}_{N}) \\
    P_{i.} = (P_{i,1}, P_{i,2}, ..., P_{i,N})|N & \sim \text{Dirichlet}(a^{P}_{1}, a^{P}_{2}, ..., a^{P}_{N}), \ \ i = 1, 2, ..., N  \\
    z_{i} & \sim \text{Beta}(a^{z},b^{z}), \ \ i = 1, 2, ..., N \\
    k_{i} & \sim \text{Uniform}(a^{k},b^{k}), \ \ i = 1, 2, ..., N \\
    m_{i} & \sim \text{Uniform}(a^{m},b^{m}), \ \ i = 1, 2, ..., N \\
    \sigma_{i} & \sim \text{Uniform}(a^{\sigma},b^{\sigma}), \ \ i = 1, 2, ..., N
\end{align*}

We run a RJMCMC algorithm for 500,000 iterations with 50,000 burn-in iterations, imposing the ordering constraint $\mu_{i}\leq \mu_{i+1}$ for $i = 1, 2, ..., N$. We fit the model with the repulsive prior of Equation \eqref{eq:straussgps} and the independent prior of Equation \eqref{eq:poissgps} and compare our results to those obtained by \citet{pohle2017selecting}. {The penalty parameter $a$ and threshold $d$ were chosen as described in Section \ref{sec:models}, with $a = \exp(-n^{*}_{2.5})$ and $n^{*}_{2.5} = 627$, with results presented in Figure \ref{fig:cs1}, or $a = \exp(-n^{*}_{5})$ with $n^{*}_{5} = 1255$, with results presented in Figure 2 of Section \ref{supsubsec:resgps}  in the Supplementary Material.} To select which value of the penalty $a$ is the most appropriate, we computed the Bayes factor between the two models using the two different settings. The log Bayes Factor is $1077$ in favour of $a=\exp(-n^{*}_{2.5})$. Therefore, we conclude that the model with $a = \exp(-n^{*}_{2.5})$ better supports the data and the value $a=\exp(-n^{*}_{5})$ overpenalizes the number of states. Moreover, the value $a=\exp(-n^{*}_{5})$ tends to point towards two states, which also contradict the findings of \cite{pohle2017selecting}. 


{The repulsive prior of Equation \eqref{eq:straussgps} leads to a posterior mode of four states, {with posterior distribution on the number of explored states $p(2) = 0.009$, 
 $p(3) = 0.211$, $p(4) = 0.480$, 
 $p(5) = 0.295$, $p(6)=0.004$, $p(7) = 0.001$ with $\sum_{i=2}^{8}p(i)=1$},  whereas the independent prior of Equation \eqref{eq:poissgps} leads to a posterior mode of seven states. \citet{pohle2017selecting} considered models with up to five states and selected the model with four states according to the integrated completed likelihood \citep[ICL][]{biernacki2000assessing}  criterion. However, we note that the model with seven states actually leads to a smaller ICL (see Table \ref{tab:sim1:ICL}  in Section  \ref{supsubsec:resgps} of the Supplementary Material), agreeing with our results in the case of a independent prior.

We plot the resulting distributions of step length and angle of the last time point, conditional on four states, from our model with a repulsive prior and those obtained by \citet{pohle2017selecting} in Figure \ref{fig:cs1}. {The corresponding  distribution results for the independent prior are given in Section  \ref{supsubsec:resgps} of the Supplementary Material.}


\begin{figure}[htp]
    \includegraphics[height = 6cm, width=\textwidth]{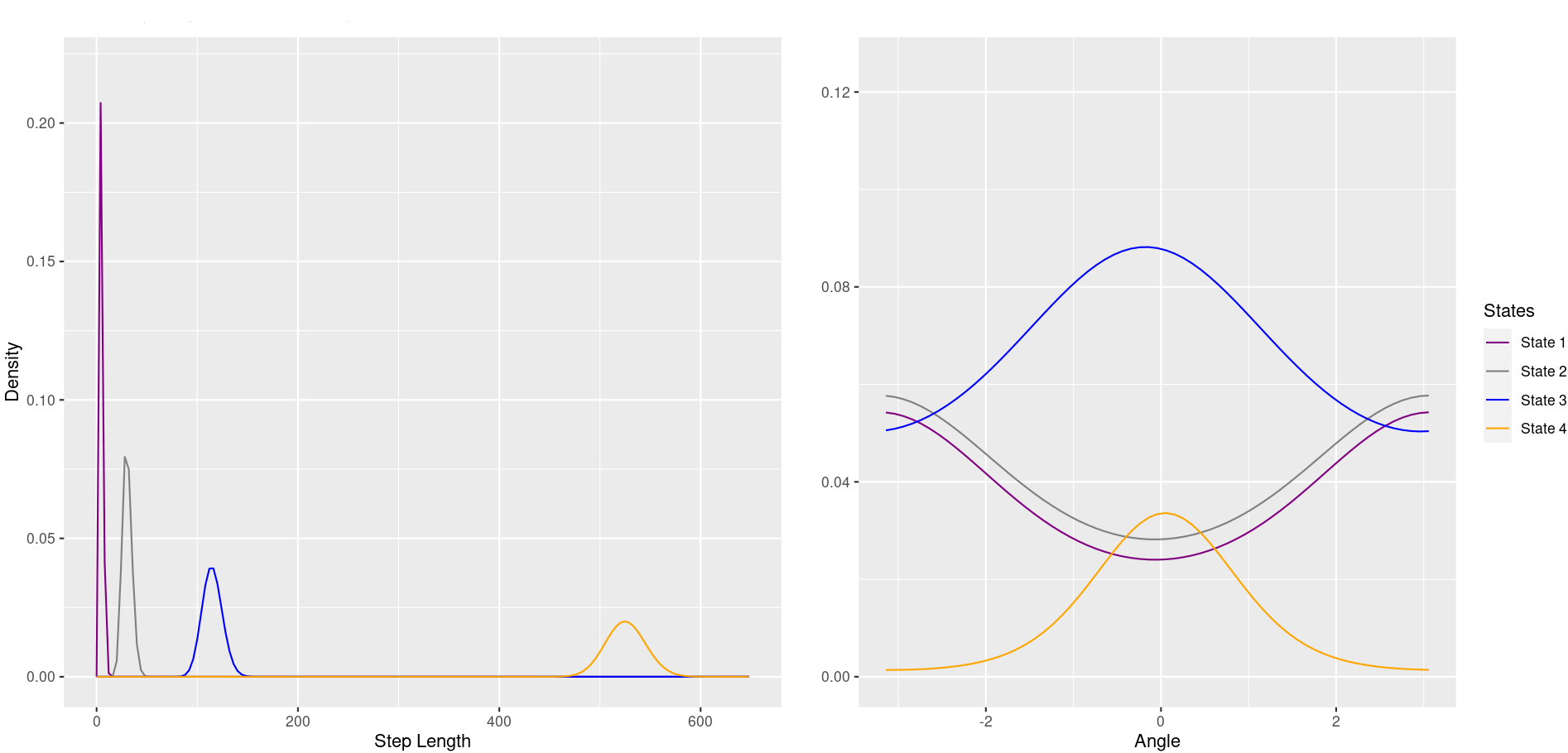}
        \\[\smallskipamount]
    \includegraphics[height = 6cm, width=\textwidth]{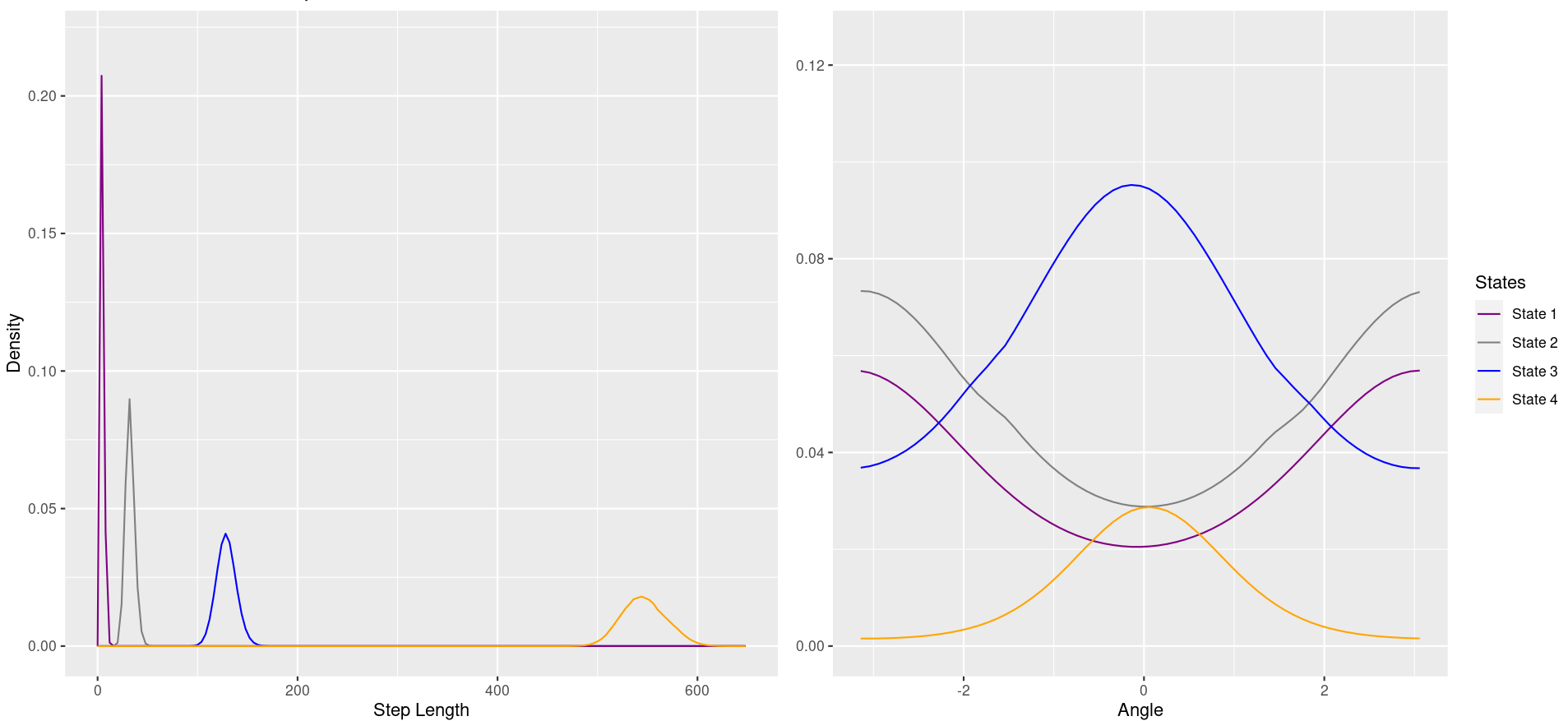}
     \caption{The first row illustrates the distribution of step length (left) and angle (right) for the last time point as inferred by \citet{pohle2017selecting}. The second row illustrates the posterior distribution of the step length (left) and angle (right) as inferred by our model, conditional on the posterior mode of four states, with a repulsive prior distribution.}
 \label{fig:cs1}
\end{figure}


Similarly to \citet{pohle2017selecting}, we identify four types of step length, corresponding to hardly any movement (state 1), small movement (state 2), moving (state 3), and traveling (state 4). 
Additionally, states 3 and 4 have a much more directed movement compared to states 1 and 2 as observed from Figure \ref{fig:cs1} and discussed in \citet{pohle2017selecting}. In contrast, when we consider a independent prior on the mean step length parameters $\mu_{1}, \mu_{2}, ..., \mu_{N}$, we observe common issues with models of this type, namely state distributions that are almost completely overlapping and state distributions that are assigned very small weights (Figure  \ref{sup:Fig:standpriorgps} in  Section  \ref{supsubsec:resgps} of the Supplementary Material).

\clearpage

\section{Case study 2 : Cape gannet acoustic data}
\label{sec:aplAcoustic}

We consider data on Cape gannets in South Africa, comprising  \ADtext{of} 3078.1 seconds of acoustic time points using animal-borne devices, analyzed in \citet{thiebault2021animal}. The data were recorded at 22.05kHz sampling frequency and were pre-processed by downsampling the audio at 12 kHz and with a high-pass filter above 10 Hz before being segmented into 2179 intervals of 1.4 seconds \citep{thiebault2021animal}. For each time segment of length 1.4 seconds we extracted 12 acoustic features based on the Mel-frequency cepstral coefficients with $n$ measurements each \citep{cheng2010call}, which is standard practise in acoustic data analysis \citep{cheng2010call,ramirez2018comparative,noda2019acoustic,chalmers2021modelling}. 
However, these 12 features are correlated with each other, and so we employ principal component analysis (PCA) to obtain a set of uncorrelated components as model inputs, instead of modelling the 12 features directly, as described in \citet{trang2014proposed}. We consider the first two principal components (2-PC) that explain 70\% of the variability of the original Mel-frequency cepstral coefficients.

We model the 2-PC at time $t$, $ \underline{E_{t,c}} = ({E_{t,c,1}}, {E_{t,c,2}})$, {with $c = 1, 2, ..., C$ the index of Mel-frequency cepstral coefficients measurements.}  
using a Multivariate Normal distribution: 
\begin{equation*}
    \underline{E_{t,c}} \sim \text{Normal}_{2}(\underline{\mu_{S_{t}}},\Sigma_{S_{t}})
\end{equation*}
\noindent where $\underline{\mu_{S_{t}}}$ corresponds to the mean vector of the 2-PC for the latent state $S_{t}$ and similarly {the} $\Sigma_{S_{t}}$ is the covariance matrix for the 2-PC under the latent state $S_{t}$.

We considered a repulsive prior on the mean parameters $\underline{\mu_{1}},\ldots, \underline{\mu_{N}}$ and, for comparison, the independent prior described in Section \ref{subsec:repuls}. Details can be found in Section \ref{supsubsec:model} of the Supplementary Material.

The prior distributions placed on the rest of parameters of the model, such as initial probabilities, transition probabilities and covariance matrices are the following 

\begin{align*}
    N & \sim \text{Uniform}\left \{ 1, 2, ..., N_{max} \right \} \\
    \pi = (\pi_{1}, \pi_{2}, ..., \pi_{N})|N & \sim \text{Dirichlet}(a^{\pi}_{1}, a^{\pi}_{2}, ..., a^{\pi}_{N}) \\
    P_{i} = (P_{i,1}, P_{i,2}, ..., P_{i,N}) |N& \sim \text{Dirichlet}(a^{P}_{1}, a^{P}_{2}, ..., a^{P}_{N}), \ \ i = 1, 2, ..., N  \\
    \Sigma_{i} & \sim \text{Wishart}(n^{\Sigma},\Sigma_{0}), \ \ i = 1, 2, ..., N
\end{align*}
{The penalty parameter $a$ was chosen as $a =\exp(-n^{*}_{2.5})$ with $n^{*}_{2.5}=54$ with $n^{*}_{2.5}=54$, with results presented in Section \ref{sec:aplGPS}, or $a = \exp(-n^{*}_{5})$ with $n^{*}_{5}=108$, with results presented in {Section \ref{supsubsec:res} of the Supplementary Material}, and $d = 21$.} To select which value of the penalty is the most appropriate for the data, we used the Bayes factor between the two different settings for $a$. The value $a=\exp(-n^{*}_{2.5})$ is supported with a log Bayes Factor of $20043$. Hence, the model with $a = \exp(-n^{*}_{2.5})$ is preferred from the data, compared to $a=\exp(-n^{*}_{5})$ even though both models give similar results. We run a RJMCMC algorithm for 100,000 iterations with 10,000 burn-in  iterations. Details of the prior distribution choices and inference are displayed in Sections \ref{supsubsec:model}, \ref{supsubsec:res} of the Supplementary Material. We use the post-processing technique described in Section \ref{sec:labelswitching} to label the states and the state allocation method described in Section \ref{sec:stateallocation} to obtain the posterior distribution of state allocation for each observation.} We fit the model with the repulsive prior and the independent
prior and show the posterior distributions in Section \ref{supsubsec:res} in of the Supplementary Material.

{For each model, we sample the allocation state of each observation at each time point as described in Section \ref{supsubsec:model} of the Supplementary Material}. In \citet{thiebault2021animal}, state allocation is conducted manually with the assistance of experts by listening to the audio. The results of state allocation across time within our Bayesian framework are compared to the manual allocation of \citet{thiebault2021animal} in Figure \ref{two_three_state_comp}. We focus our interpretation on the model with three states, which is the posterior mode of the distribution on the number of states {(the corresponding results for states such as two and four, for the repulsive case are given in Section \ref{supsubsec:res} of the Supplementary Material)}. 

\begin{figure}[htp]
    \includegraphics[height = 6cm, width=\textwidth]{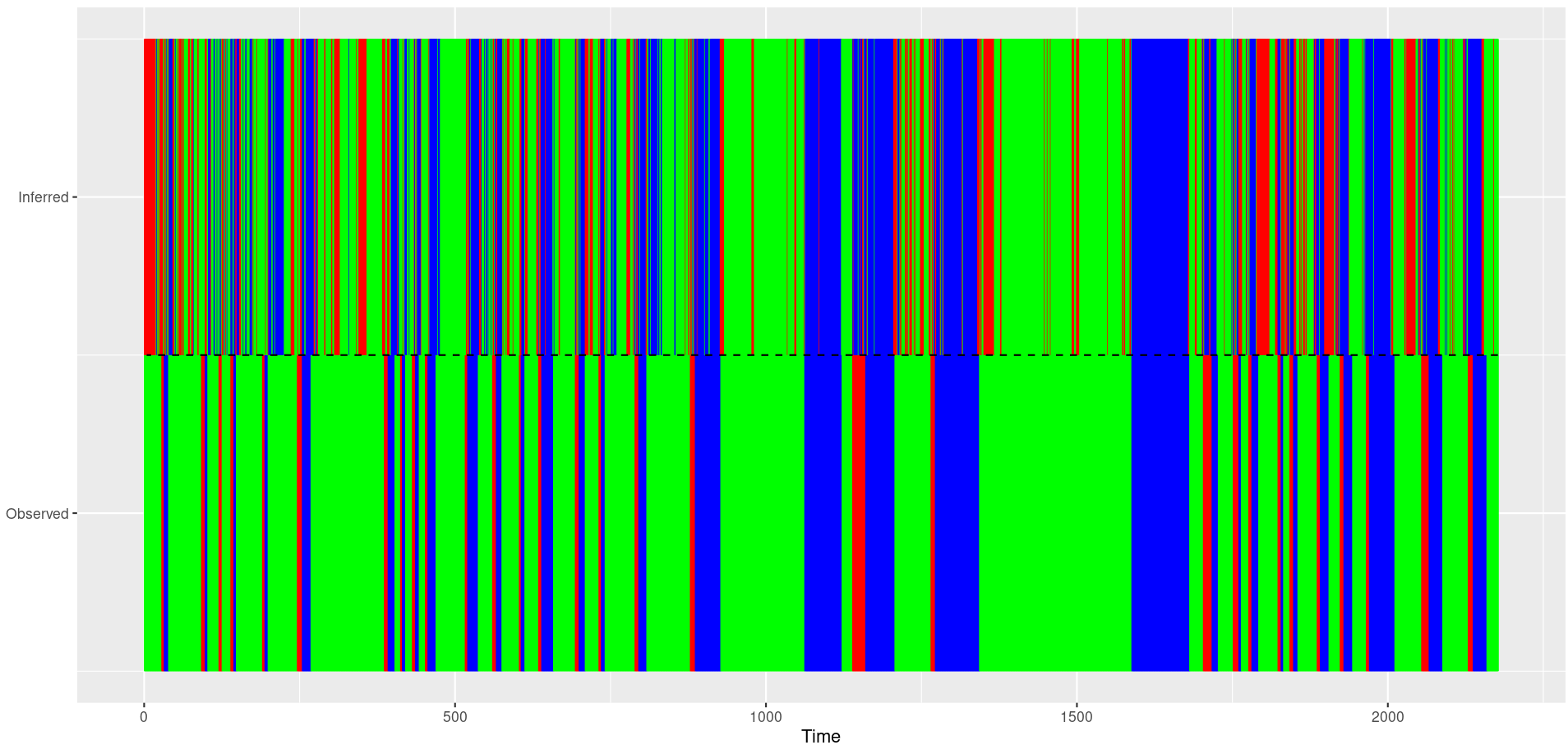}
     \caption{Comparison of the posterior classification of our model (top half ) with the manual classification of \citet{thiebault2021animal} (bottom half). Based on the manual classification of \citet{thiebault2021animal}, the states are: floating on water (blue), flying (green) and diving (red). }
 \label{two_three_state_comp}
\end{figure}

\citet{thiebault2021animal} identified three states, flying (green), floating on water (blue) and diving (red), as indicated in Figure \ref{two_three_state_comp}. There is general agreement in state allocation between our model, which is completely unsupervised, and the manual allocation by \citet{thiebault2021animal},
in particular for the most common states of water (blue) and flying (green). However, our model classifies more time points to the third (red) state than the manual classification. According to expert knowledge and based on the recording, the time points allocated to this third state correspond to sounds made mainly after the take-off of the species with vigorous flapping, and general alterations in flying behavior, such as changes in flying direction or speed, coupled with variations in wind speed. Hence, in our  three state model the third state corresponds to diving, alongside nuisance acoustic occurrences from the device or instantaneous changes in the individual's flying behaviour. 

In Section \ref{supsubsec:res}  of the Supplementary Material in Figure 4 we display the uncertainty of the classification inferred by our Bayesian framework, by plotting the posterior allocation probabilities for each state across time.  All time points are in fairly high proportion of the time ($\sim 25\%$) allocated to states other than their modal state, indicating that state allocation has a high degree of uncertainty in this case, which is expected given the noisy and multivariate nature of the data and the very short time span of 1.4 seconds between time points. Nevertheless, the model successfully manages to allocate the majority of points to a state that agrees with expert knowledge, demonstrating the potential of the approach even when modelling noisy and multidimensional data. 

In Figure \ref{fig:biplot} we display the {observations} with points coloured according to their modal state allocation on the domain defined by the 2-PC.
The first PC is dominated by the 1st Mel-frequency cepstral coefficient, which has a negative coefficient, whereas the second PC is a contrast between a weighted average of 5th, 8th, 10th and 11th Mel-frequency cepstral coefficients and 2nd Mel-frequency cepstral coefficient. {Those PC contrasts can be helpful to acoustic experts since they provide means of identifying correlations between the different type of sounds that each Mel-frequency cepstral coefficient captures. }


Based on Figure \ref{fig:biplot}, flying (green) is characterized by large scores for the first PC, whereas floating on the water (blue) is characterized by small scores. Diving with the instantaneous nuisance acoustics (red) is associated with increasing PC2 scores, widening the range of PC1 scores, albeit always remaining mid-range. This suggests that when there is no strong support for allocating a time point to flying or floating on the water, then it is allocated to the third state.

\begin{figure}[htp]
    \includegraphics[height = 6cm, width=\textwidth]{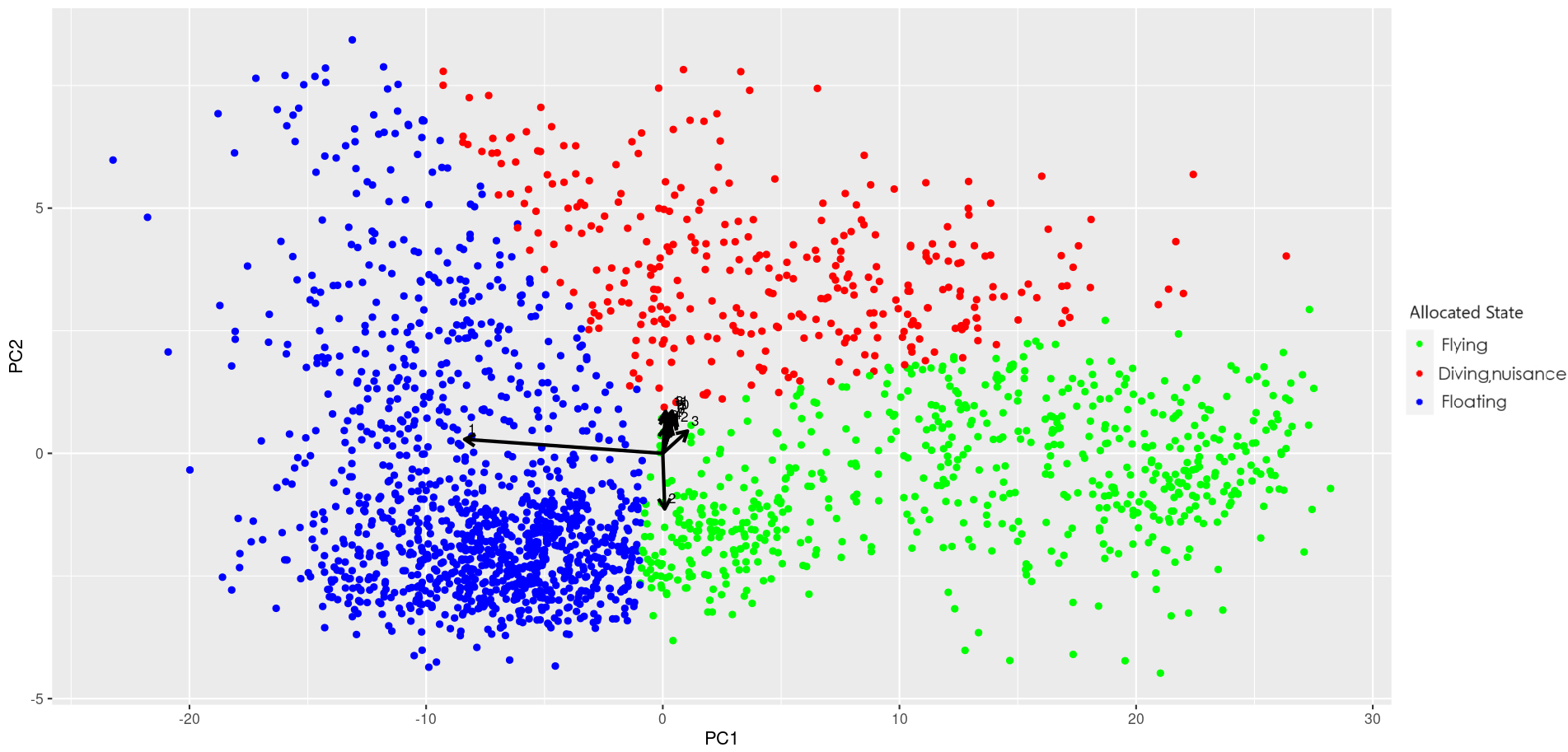}
     \caption{Biplot, with observations coloured according to their modal state allocation, in the case of three states, plotted on the domain of the first two PC. {Based on the manual classification of \citet{thiebault2021animal} blue corresponds to floating on water, green to flying and red to diving.}}
 \label{fig:biplot}
\end{figure}



Finally, when the independent prior is used instead, Equation (\ref{standeq}), the posterior distribution of the number of states is very diffuse, with 2, 3, 4, ..., 42 states almost equally supported, with the results given in Figure 3, Section \ref{supsubsec:res} of the Supplementary Material, suggesting clear overfitting, particularly evident in this case because of the noisy nature of acoustic data. 

\clearpage

\section{Conclusion}
\label{sec:disc}


In this paper, we developed a new modeling framework for inferring the number of states and the corresponding distribution parameters in HMMs. In particular, we placed a repulsive prior on the location parameters of the HMM distributions in order to penalize small differences between the state parameters, and we treated the number of latent states as random and inferred it directly from the model without making any subjective decisions. We accomplished this using an RJMCMC algorithm, which samples the entire Bayesian model space. 

We demonstrated the model using two interesting and challenging types of ecological applications using GPS and acoustic data. The results demonstrated the ability of our framework to yield parsimonious models with good state allocation ability in a completely unsupervised modelling framework. The case studies showcase the effectiveness and practicality of our framework, with the repulsive prior penalizing the number of underlying states, leading to simpler models, while effectively exploring the model sample space. 
{Additionally, we conducted an extensive simulation study, which demonstrated that, when the generating model is fitted to the data, the repulsive prior model and the independent prior model yield practically indistinguishable results in terms of chosen number of states, density estimation and state allocation, with a small benefit in some cases when using the repulsive prior. Future work could explore simulation scenarios where the fitted model is only an approximation or simplification of the data-generating process, as is typically the case for real data.}

The repulsive prior distribution is a function of a threshold and penalty parameter, which are not being inferred in the model and instead they have to be subjectively chosen. Our chosen approach, based on the work by \citet{beraha2022mcmc}, builds on a systematic way of choosing the penalty and threshold based on the minimum cluster size expected. 
Alternatively, different repulsive models that avoid the use of threshold and penalty parameters can be used, as discussed in \citet{petralia2012repulsive}.

{In this case, we only placed repulsive prior distributions on one set of state parameters, the mean step length and mean vectors of the PC}. Potential avenues for future research would be to apply joint repulsion priors {on} different parameters, such as step length and angle, or apply repulsion solely on the variance matrix for the acoustic coefficient. Such research might lead to interesting combinations of parameter components for the HMMs,  {giving} a finer resolution of state dynamics. 

Overall, the {approach} of fitting HMMs to ecological data within a dynamic mixture modelling framework with a repulsive prior on the latent number of components provide a valuable new point of view for this widely used class of models, which can be applied to a variety of disciplines such as in finance, biology, social science, medicine and ecology among others  making it generic framework for statisticians and practitioners to use on their corresponding disciplines.

\title{\textbf{Supporting Information for\\ ``Hidden Markov models with an unknown number of states and a repulsive prior on the state parameters'' \\ by Ioannis Rotous, Alex Diana, Alessio Farcomeni, Eleni Matechou and Andréa Thiebault }}

\maketitle

\section{Birth and Death Algorithm }
\label{alg:birthdeath}

A brief description of the Birth hand Death algorithm is

\begin{enumerate}
    \item We initialize a point pattern $\theta_{1}$, choose birth and death probabilities $q_{\text{birth}}$ and $q_{\text{death}}$, specify the number of iterations $M$, and proposal parameters $\eta$.
    \item At iteration $it$, if the cardinality of $\theta_{it}$ is one, then with probability $1$ we choose to give birth to a new point and add it to $\theta_{it}$ and form the $\theta^{*}$, sampled from a proposal distribution with proposal parameters $\eta$. In any other case, we either propose to give birth to a new point and add it to $\theta_{it}$ to form $\theta^{*}$, sampled from a proposal distribution with probability $q_{\text{birth}}$ or give death to a randomly uniformly chosen point from $\theta_{it}$ with probability $q_{\text{death}}$ and form $\theta^{*}$.
    \item The acceptance ratio is 
    \begin{equation*}
        A = \frac{g(\underline{\theta^{*}}|\left | \underline{\theta^{*}} \right |,\xi_{*},a,d)}{g(\underline{\theta_{it}}|\left | \underline{\theta_{it}} \right |,\xi_{*},a,d)}\frac{q(\theta_{i}\rightarrow \theta^{*})}{q(\theta^{*}\rightarrow \theta_{i})}
    \end{equation*}
    where $q(\cdot\rightarrow \cdot)$ corresponds to the proposal distribution times the death or birth probability.
    \item We repeat this process for $M$ iterations. The resulting $\underline{\theta_{\text{aux}}} = \underline{\theta_{M}}$.
\end{enumerate}

\clearpage

\section{Penalty \& Threshold}
\label{subsec:penthresh}

{The density over distances 
$r$ is expected to have more than one mode, as the existence of only one mode suggests the presence of only one mixture component. Local minima of 
$p(r)$ can be likened to ``valleys'' between ``peaks'' of higher density, indicating distances where fewer observations are present. By choosing the minimum over the local minimum for $p(r)$, we select a distance that is not too large to affect the density estimation severely. Examples, can be found in Section \ref{sec:thres} of the Supplementary Material. }

{The penalty $a$ as explained in \citet{beraha2022mcmc} is chosen as follows: suppose we have two components with location parameters $\theta_{h}$ and $\theta_{h^{'}}$ such that their distance is $\left \| \theta_{h}-\theta_{h^{'}} \right \|\leq d$, while the pairwise distances between $\theta_{h}, \theta_{h^{'}}$ and the  rest of the component parameters are larger than $d$. Hence, the conditional likelihood of the component $\theta_{h}$ is proportional to the penalty $a$ (since we have only one pair of distance less than the threshold $d$) times the likelihood function for the observations assigned to the $h$th mixture component, denoted as $F_{h}$. Therefore, $p(\theta_{h}) \propto a F_{h}$. Now, in the case where the cardinality of the cluster $h$ is small, we want the penalization to determine whether we keep the mixture component $h$ or not. Therefore, $a$ has to be chosen such that $a F_{h}$ is small. In that case we define what we assume of being a small cluster size $n^{*}$, and then as they mention in \citet{beraha2022mcmc} we take a ``guess'' of the value of $F_{h}$ denoted as $k_{s}$. Then an estimate of $a = \exp(-n^{*}\log(k_{s}))$. This choice of $a$ regulates how small $a F_{h}$ is going to be since, we can rewrite it as $\exp(-n^{*}\log(k_{s}))F_{h} = k_{s}^{-n^{*}}F_{h} = F_{h}/k_{s}^{n^{*}}$ making the conditional likelihood of the component $\theta_{h}$ sufficiently small, hence in that case the repulsion prevails. }

\clearpage

\section{Threshold}
\label{sec:thres}

To understand the derivation of the threshold $d$, we will provide a few examples. The first example is a mixture of two normal distributions with means 0 and 4, and standard deviations 1, respectively, each with mixture weights of 0.5. In Figure \ref{2_comp_thresh} we display the kernel density estimate for the observations of the mixture, and the kernel density estimate of the distances between the observations. From Figure \ref{2_comp_thresh}, we observe that the kernel density estimation of the distances has only one local minimum, corresponding to the value of $2.8304$, which is an ideal threshold. Since the true means of the normal distributions are at $0$ and $4$, which have larger distance than $2.8304$, hence mixture components with smaller distance will be penalized. The second example corresponds to a mixture of three normal distributions with means -5, 0, and 5, and a standard deviation equal to 1, respectively. Each distribution has a mixture weight of $1/3$.  In Figure \ref{3_comp_thresh}, we display the kernel density estimate for the observations of the mixture and the kernel density estimate of the distances between the observations. As we can see, there are two local minimum. However, we consider the minimum local minimum indicated with a red dot, corresponding to the value $2.7279$. The other local minimum has a value of $7.9296$. Choosing the latter one would severely affect the estimation of the true mixture density of the observations, as the mixture components have distances less than $7.9296$. Therefore, we choose the minimum local minimum to avoid overpenalization of the mixture components.

\begin{figure}[htp]
    \includegraphics[height = 6cm, width=\textwidth]{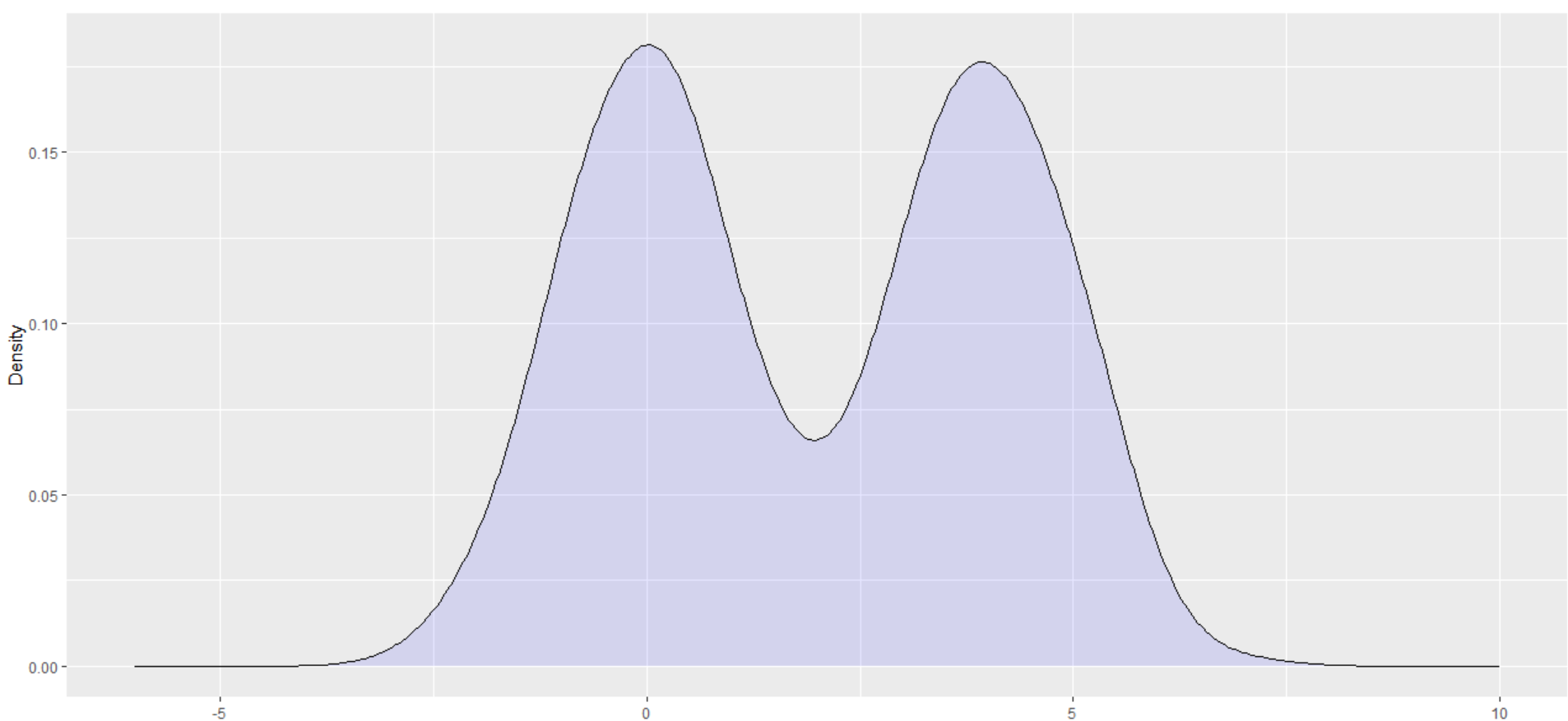}
        \\[\smallskipamount]
    \includegraphics[height = 6cm, width=\textwidth]{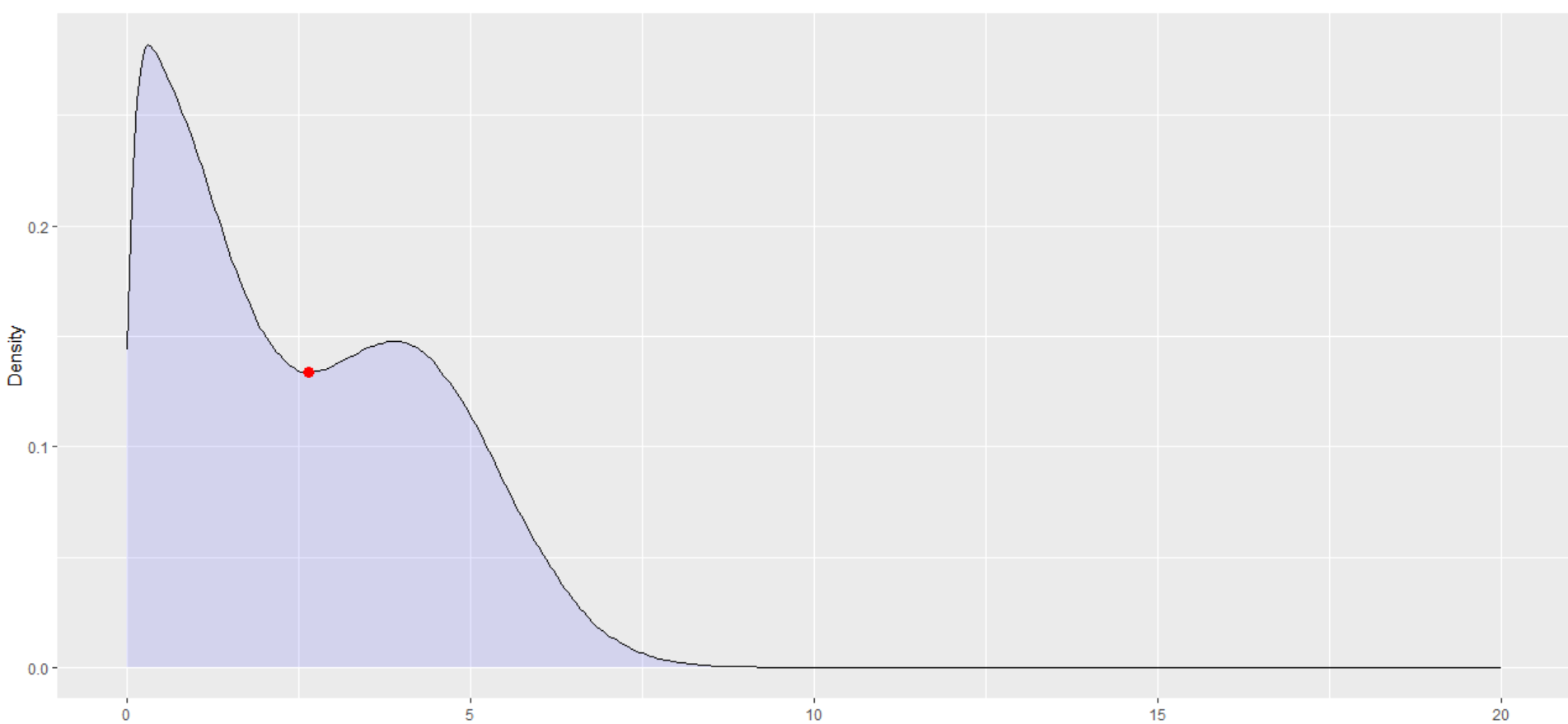}
     \caption{The top row corresponds to kernel density estimation of the observations, while the bottom row corresponds to the kernel density estimate of the distances. The red dot indicates the local minimum of the kernel density estimate of the distances..}
 \label{2_comp_thresh}
\end{figure}

\begin{figure}[htp]
    \includegraphics[height = 6cm, width=\textwidth]{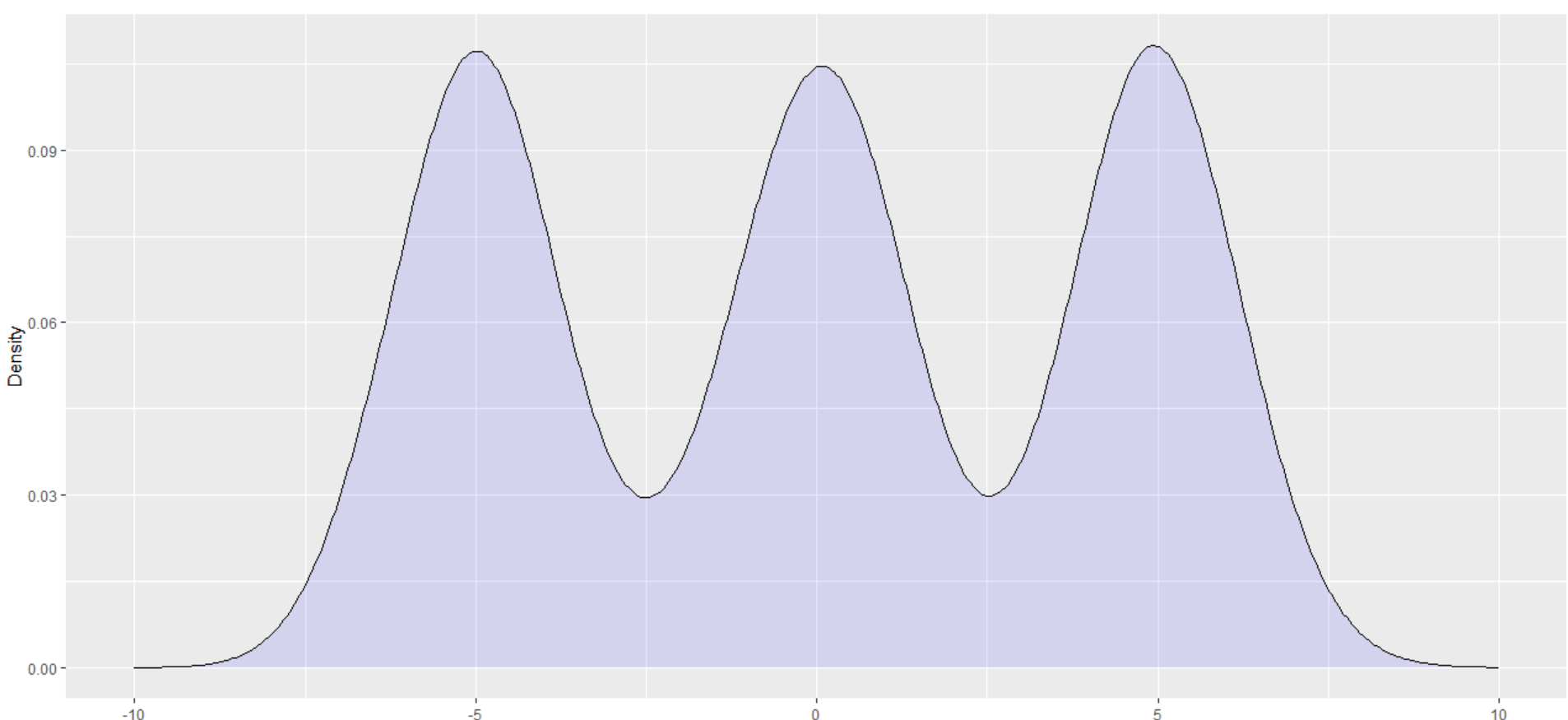}
        \\[\smallskipamount]
    \includegraphics[height = 6cm, width=\textwidth]{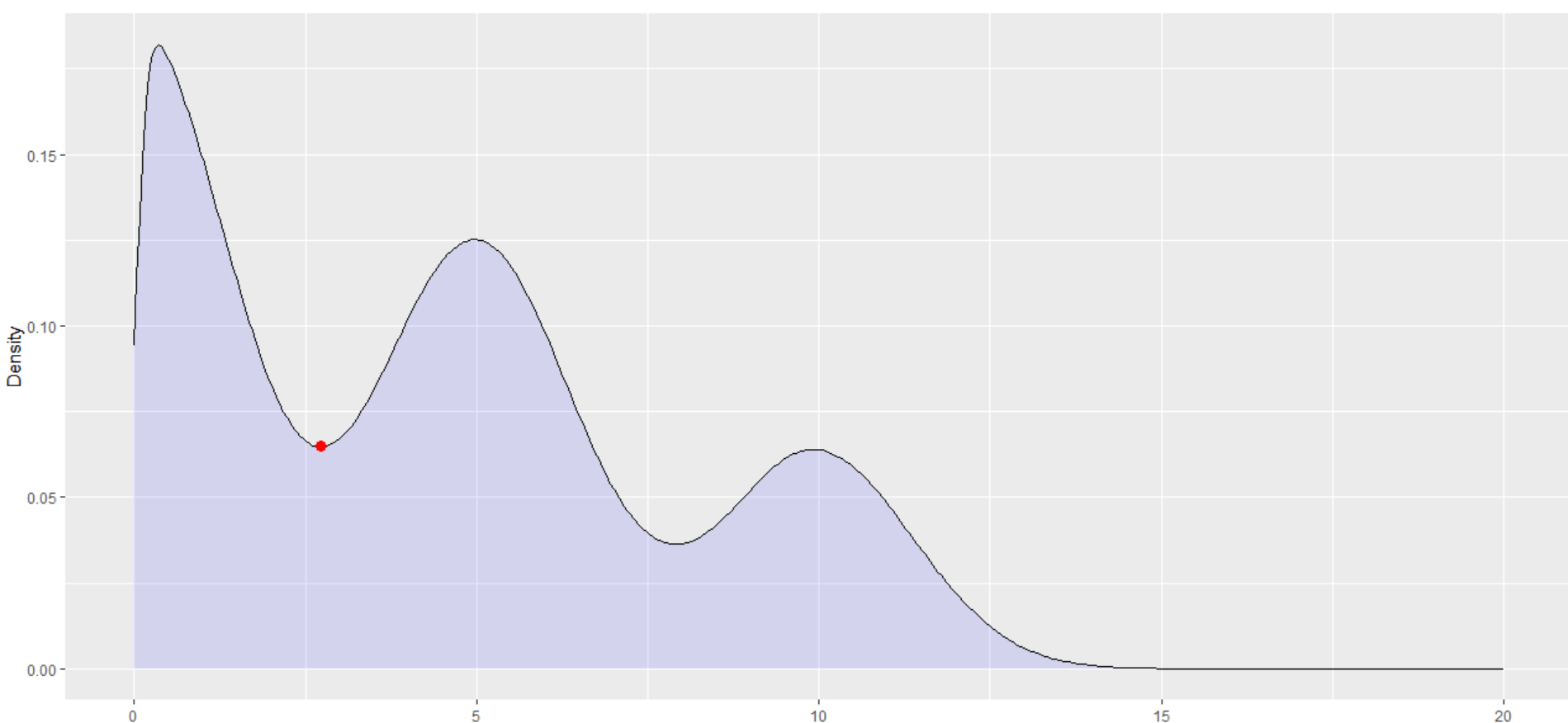}
     \caption{The top row corresponds to kernel density estimation of the observations, while the bottom row corresponds to the kernel density estimate of the distances. The red dot indicates the minimum of the local minimum of the kernel density estimate of the distances..}
 \label{3_comp_thresh}
\end{figure}

\clearpage

\section{Simulations}
\label{supsec:sim}

We conducted a simulation study varying the sample size
$n=\left \{ 50,100 \right \}$ and number of time points 
$T=\left \{ 5,10 \right \}$ for four cases of{overlap between consecutive states distributions equal to 3\%, 9\%, 33\% and 55\% or equivalently 
this can be viewed as 5\%, 15\%, 50\%, and 75\% overall overlap}. The results were averaged across 100 replications. The Normal distributions have mean $(-10, -5, 0, 5, 10)$ {and standard deviations $\sigma_{1} = \sigma_{2} = ... = \sigma_{5}$ where for each degree of overlap are 1.1408, 1.4726, 2.5709 and 4.2319}. The initial probability distribution $\pi$ and transition probability matrix $P$ were chosen to have  {all elements equal to 1/5}, ensuring equal state sizes across all time points, allowing us to focus on the effect of state overlap on inference. The corresponding mixture in each case are displayed in Figure \ref{fig:density_plot}.

\begin{figure}[ht]
    \centering
    \includegraphics[width=\linewidth]{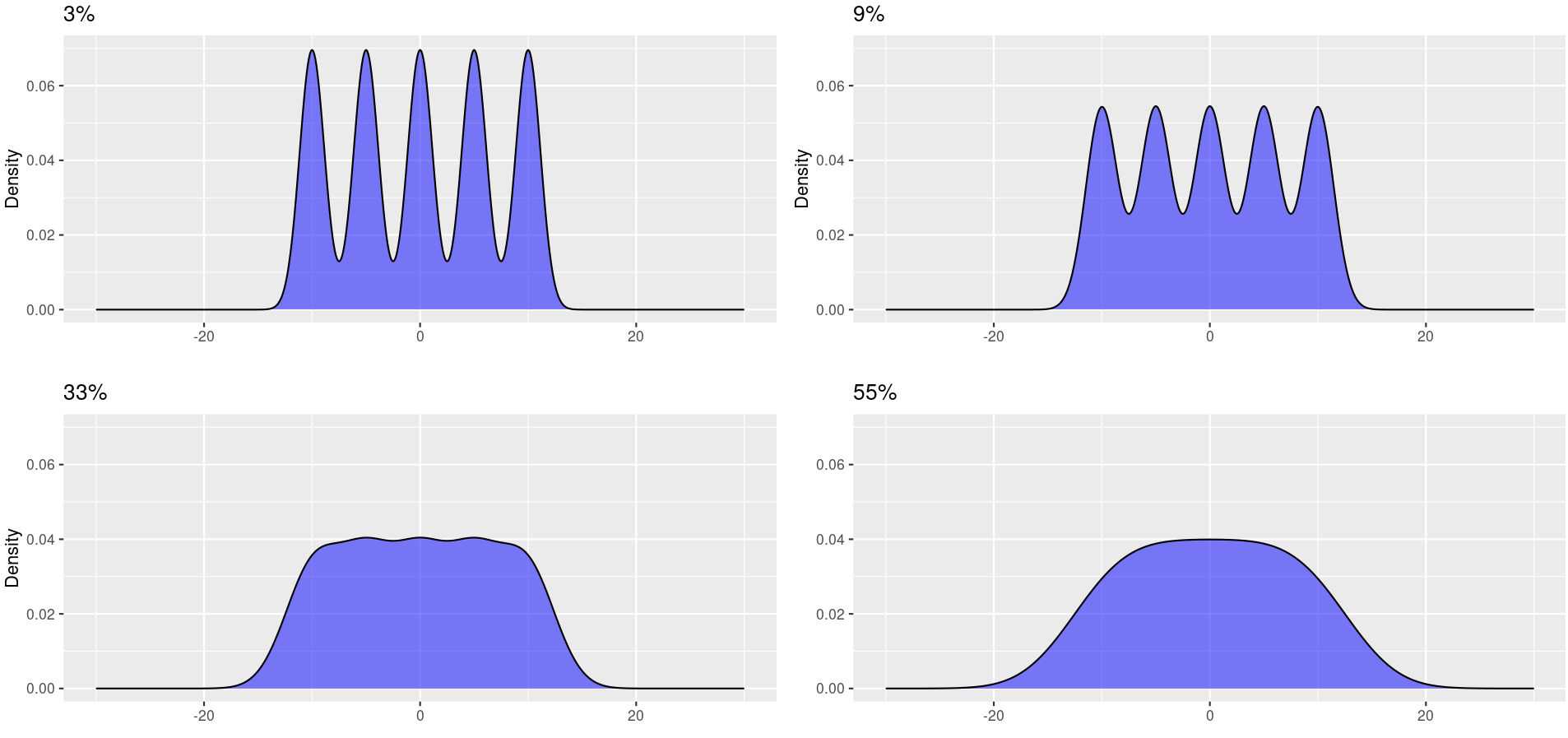}
    \caption{Density mixture distributions under the different consecutive overlaps of 3\%, 9\%, 33\%, and 55\%.}
    \label{fig:density_plot}
\end{figure}


To begin with, we provide details about the KL divergence statistics utilized in our simulation study. At each time point $t$,  we have access to the true distribution $p^{0}_{t}$, from which we generated our observations. For a replication $r$ we compute the KL divergence for each time point 
$t$ and posterior sample $l = 1, 2, ..., L$, as we have estimated the posterior density $p_{l,t,r}$ as $\text{KL}^{l,r}_{t}(p^{0}_{t}|p_{l,t,r}) = \int p^{0}_{t}(x)log\frac{p^{0}_{t}(x)}{p_{l,t,r}(x)}$, Subsequently, we average across posterior samples to obtain $\bar{\text{KL}}^{r}_{t} = \frac{1}{L}\sum_{l=1}^{L}\text{KL}^{l,r}_{t}(p^{0}_{t}|p_{l,t,r})$ and across time points to acquire $\bar{\text{KL}}^{r} = \frac{1}{T}\sum_{t=1}^{T}\bar{\text{KL}}^{r}_{t}$ and across replications $\bar{\text{KL}} = \frac{1}{100}\sum_{r=1}^{100}\bar{\text{KL}}^{r}$ for which we have calculated also their $95\%$ credible intervals. Moreover, for the misclassification statistics we define the true similarity matrix for time point $t$ and replication $r$, $S_{t,r}$ of dimensions $n\times n$, defined such that entries where $(i,j)$ belong to the same component take the value 1, otherwise zero for time point $t$, for all $i,j = 1, 2, ..., n$ and we compare it with the posterior sample similarity matrix $\hat{S_{l,t,r}}$ for each posterior sample $l$, each time point $t$ and replication $r$. We report the misclassification error as described in \cite{petralia2012repulsive} averaged across time points for replication $r$. $\bar{MS_{r}} = \frac{1}{T}\sum_{t=1}^{T}\sum_{l=1}^{L}\frac{1}{\frac{n(n-1)}{2}}\sum_{i=1}^{n}\sum_{j=i+1}^{n}1(\hat{S_{l,t,r}}(i,j)\neq S_{t,r}(i,j))$. Then we average across replications and derive $\bar{MS} = \frac{1}{100}\sum_{r=1}^{100}\bar{MS_{r}}$ alongside with their $95\%$ credible intervals.

 We employed RJMCMC for 10,000 iterations, of which we discarded the first 1000 as burn-in. {To ensure identifiability, we use the ordering constraint $\mu_{i}\leq \mu_{i+1}$, for $i = 1, 2, ..., N$}. For each scenario, we replicated and averaged our results over 100 iterations. {Table \ref{tab:rafsim_2.5} displays the mean and 95\% credible interval of the KL divergence and misclassification error for the independent (ID) and repulsive (RP) priors for the penalty case $n^{*}_{2.5}$.}

\begin{center}
\begin{table}
\caption{Comparison of different degrees of overlaps 3\%, 9\%, 33\% and 55\% between independent and repulsive priors based on measurements of Kullback–Leibler (KL) and misclassification error (Miscl) when considering $a=\exp(-n^{*}_{2.5})$ single.}
\resizebox{\textwidth}{!}{
\begin{tabular}{ r r  r r r r r r}
\multicolumn{6}{c}{\multirow{2}{*}{}} \\
\multicolumn{6}{c}{} \\[-6ex]
\hline
$n$ & $T$ & ID:N &ID:KL & ID:Miscl & RP:N & RP:KL & RP:Miscl  \\
\hline
\multicolumn{6}{c}{\textbf{Overlap 3\%}} \\
50& 5  &2 &0.2124(0.1402,0.2364) & 0.5098(0.3229,0.6332) &2 &0.1986(0.1451,0.2336) &  0.5000(0.2975,0.6207) \\
50 & 10  & 3&0.1285(0.0535,0.2059) &0.3603(0.1047,0.5540)   &4 &0.1238(0.0449,0.2008)  & 0.3373(0.0424,0.5937)     \\
100 &  5 & 4& 0.1381(0.0489,0.2057) &  0.3570(0.1095,0.6077) & 4&0.1268(0.0376,0.2003) &  0.3386(0.0359,0.5696)  \\
100 &  10 & 5&0.0503(0.0146,0.1142) & 0.1437(0.0278,0.3512)  & 4& 0.0652(0.0161,0.1169)  & 0.1839(0.0282,0.3812)   \\
\hline
\multicolumn{6}{c}{\textbf{Overlap 9\%}} \\
50 & 5  & 2&0.1181(0.1026,0.1441) & 0.4991(0.3450,0.6518) & 2&0.1167(0.0843,0.1481)  & 0.5087(0.3573,0.6746) \\
50 &  10 & 3&0.0905(0.0571,0.1189) &  0.4423(0.3058,0.6099) & 3&0.0856(0.0563,0.1141)& 0.3999(0.3086,0.5976) \\
100 & 5  & 3& 0.0889(0.0547,0.1106) &  0.4185(0.3031,0.5853)&3& 0.0854(0.0536,0.1114)  &  0.4359(0.3107,0.5965) \\
100 & 10 & 3&0.0542(0.0304,0.0961) & 0.3642(0.2741,0.5214)  & 3&0.0500(0.0310,0.0920)  &  0.3393(0.2462,0.5349)    \\
\hline
\multicolumn{6}{c}{\textbf{Overlap 33\%}} \\
50 &  5 & 2 &0.0533(0.0319,0.0771) &0.5629(0.3682,0.7706)   & 2&0.0559(0.0303,0.0742)  &0.6186(0.3785,0.8018)   \\
50 &  10 & 2&0.0291(0.0210,0.0487) &  0.4442(0.3699,0.6199)&2&0.0282(0.0198,0.0665) &0.4259(0.3704,0.8015)   \\
100 & 5  & 2&0.0315(0.0208,0.0634)  & 0.4630(0.3700,0.7221)  & 2&0.0279(0.0202,0.0680)& 0.4420(0.3670,0.8015)  \\
100 &  10 & 2&0.0200(0.0120,0.0273) &   0.3981(0.3327,0.4825)  & 2&0.0195(0.0118,0.0319)    &  0.3922(0.3671,0.5069)  \\
\hline
\multicolumn{6}{c}{\textbf{Overlap 55\%}} \\
50 & 5  & 1 &0.0321(0.0269,0.0396) & 0.7639(0.5412,0.7835)  & 1&0.0291(0.0259,0.0385) & 0.7714(0.5861,0.8043) \\
50 &  10 & 1 &0.0249(0.0110,0.0299) & 0.7500(0.4294,0.7842)&1&0.0228(0.0110,0.0303) & 0.7102(0.4237,0.8035) \\
100 &  5 &  1& 0.0251(0.0123,0.0319) &0.7446(0.4304,0.7854) & 1&0.0240(0.0129,0.0294)   &0.7483(0.4479,0.8029)  \\
100 &  10 & 2& 0.0134(0.0057,0.0246) & 0.5671(0.4108,0.7823) &2&0.0103(0.0054,0.0245)  & 0.5082(0.4058,0.8016)    \\
\hline
\end{tabular}
}
\label{tab:rafsim_2.5}
\end{table}
\end{center}

Next, we display the simulation results for the penalty case $n^{*}_{5}$ for the mean and 95\% credible interval of the KL divergence and misclassification error for the independent (ID) and repulsive (RP) priors in Table \ref{tab:rafsim}. All the input and tuning parameters for the simulation and RJMCMC were kept the same.

\begin{center}
\begin{table}
\caption{Comparison of different degrees of consecutive overlaps 3\%, 9\%, 33\% and 55\% between independent and repulsive priors based on measurements of Kullback–Leibler (KL) and misclassification error (Miscl) when considering $a=\exp(-n^{*}_{5})$ single.}
\resizebox{\textwidth}{!}{
\begin{tabular}{ r r  r r r r r r}
\multicolumn{6}{c}{\multirow{2}{*}{}} \\
\multicolumn{6}{c}{} \\[-6ex]
\hline
$n$ & $T$ & ID:N &ID:KL & ID:Miscl & RP:N & RP:KL & RP:Miscl  \\
\hline
\multicolumn{6}{c}{\textbf{Overlap 3\%}} \\
50& 5  &2 &0.2124(0.1402,0.2364) & 0.5098(0.3229,0.6332) &2 &0.1902(0.1385,0.2298) &  
0.4784(0.2851,0.6215)
\\
50 & 10  & 3&0.1285(0.0535,0.2059)
&0.3603(0.1047,0.5540)   &3 &
0.1280(0.0514,0.2032)
& 0.3579(0.1035,0.6044)
\\
100 &  5 & 4& 0.1381(0.0489,0.2057) &   0.3570(0.1095,0.6077) & 3&
0.1319(0.0356,0.2039)
&  0.3635(0.0325,0.5808)
\\
100 &  10 & 5&
0.0503(0.0146,0.1142) & 
0.1437(0.0278,0.3512)
& 4& 
0.0626(0.0161,0.1334)
& 
 0.1710(0.0269,0.3778)
\\
\hline
\multicolumn{6}{c}{\textbf{Overlap 9\%}} \\
50 & 5  & 2&
0.1181(0.1026,0.1441)
&
0.4991(0.3450,0.6518)
& 2&
0.1163(0.1003,0.1681)
&
0.5069(0.3624,0.7920)
\\
50 &  10 & 3&0.0905(0.0571,0.1189) & 
0.4423(0.3058,0.6099)
& 2&
 0.0888(0.0578,0.1156)
& 
 0.4058(0.2984,0.6142)
\\
100 & 5  & 3& 
0.0889(0.0547,0.1106)

&  
0.4185(0.3031,0.5853)

&3& 
0.0822(0.0567,0.1169)
&  
 0.4383(0.3138,0.5970)
\\
100 & 10 & 3&
0.0542(0.0304,0.0961)
& 
 0.3642(0.2741,0.5214)
& 3&
0.0514(0.0300,0.0971)
&  
0.3488(0.2673,0.5213)
\\
\hline
\multicolumn{6}{c}{\textbf{Overlap 33\%}} \\
50 &  5 & 2 &
0.0533(0.0319,0.0771)
&0.5629(0.3682,0.7706)   & 2&
0.0576(0.0318,0.0737)
&
0.6433(0.3926,0.8058)
\\
50 &  10 & 2&
0.0291(0.0210,0.0487)
&  
0.4442(0.3699,0.6199)
&2&
0.0271(0.0210,0.0683)

&
0.4302(0.3662,0.8016)
\\
100 & 5  & 2&
 0.0315(0.0208,0.0634)
& 0.4630(0.3700,0.7221)  & 2&
0.0308(0.0205,0.0688)
&
0.4513(0.3693,0.8027)
\\
100 &  10 & 2&
0.0200(0.0120,0.0273)
&   0.3981(0.3327,0.4825)  & 2&
0.0188(0.0098,0.0280)
&  0.3901(0.3548,0.4988)  \\
\hline
\multicolumn{6}{c}{\textbf{Overlap 55\%}} \\
50 & 5  & 1 &0.0321(0.0269,0.0396) & 0.7639(0.5412,0.7835)  & 1&
0.0283(0.0258,0.0387)
& 0.7815(0.6505,0.8080) \\
50 &  10 & 1 &
 0.0249(0.0110,0.0299)
& 0.7500(0.4294,0.7842)&1&
0.0224(0.0104,0.0295)
& 0.7033(0.4222,0.8034) \\
100 &  5 &  1& 0.0251(0.0123,0.0319) &0.7446(0.4304,0.7854) & 1&
0.0229(0.0113,0.0279)
&0.7222(0.4359,0.8037)  \\
100 &  10 & 2& 0.0134(0.0057,0.0246) & 0.5671(0.4108,0.7823) &2&
0.0103(0.0048,0.0246)
& 0.4980(0.4087,0.8020)    \\
\hline
\end{tabular}
}
\label{tab:rafsim}
\end{table}
\end{center}

\clearpage
\section{GPS Muskox Application}
\label{supsec:smgps}

In this section we present Supplementary information for the first case study of Section 4.
\subsection{Model}
\label{supsubsec:modelgps}
In our case study, GPS technology is used to monitor individual  location over time. As a result, the data gathered include both step lengths, $L_{t}$ and turning angles, $A_{t}$ between time points. The model is described as follows:
\begin{align*}
    O_{t} & = (L_{t},A_{t}) \\
    S_{t} &\in \left \{ 1,2,...,N \right \}\\
    f(O_{t}|S_{t}) &= f(L_{t}|S_{t})f(A_{t}|S_{t}) \\
    f(L_{t}|S_{t}) &= z_{S_{t}}\delta_{L_{t}}(0) + (1-z_{S_{t}}) \text{Gamma}(L_{t};\mu_{S_{t}},\sigma_{S_{t}}) \\
    f(A_{t}|S_{t}) &= \text{vonMises}(A_{t};m_{S_{t}},k_{S_{t}}) = \frac{e^{k_{S_{t}}cos(A_{t}-m_{S_{t}})}}{2\pi I_{0}(k_{S_{t}})}, \  I_{0} \ \text{ Bessel function of order 0} 
\end{align*}
for $t = 1, 2, ..., T$ and
the specified priors are:
\begin{align*}
    N & \sim \text{Uniform}\left \{ 1, 2, ..., 80 \right \} \\
    \pi = (\pi_{1}, \pi_{2}, ..., \pi_{N}) & = (\frac{\lambda_{1}}{\sum_{i=1}^{N}\lambda_{i}},\frac{\lambda_{2}}{\sum_{i=1}^{N}\lambda_{i}},...,\frac{\lambda_{N}}{\sum_{i=1}^{N}\lambda_{i}}) \Rightarrow \lambda_{i} \sim \text{Gamma}(1,1), \ \ i = 1, 2, ..., N \\
    P_{i.} = (P_{i,1}, P_{i,2}, ..., P_{i,N})  & = (\frac{\Lambda_{i1}}{\sum_{j=1}^{N}\Lambda_{ij}},\frac{\Lambda_{i2}}{\sum_{j=1}^{N}\Lambda_{ij}},...,\frac{\Lambda_{iN}}{\sum_{j=1}^{N}\Lambda_{ij}})  \Rightarrow \Lambda_{ij} \sim \text{Gamma}(1,1), \ \ i,j = 1, 2, ..., N  \\
    z_{i} & \sim \text{Beta}(1, 100), \ \ i = 1, 2, ..., N \\
    k_{i} & \sim \text{Uniform}(0.5,2), \ \ i = 1, 2, ..., N \\
    m_{i} & \sim \text{Uniform}(-\pi,\pi), \ \ i = 1, 2, ..., N \\
    \sigma_{i} & \sim \text{Uniform}(0.5\left \{ l_{t}:\mathbb{P}(L_{t}\leq l_{t})=0.1 \right \},2\left \{ l_{t}:\mathbb{P}(L_{t}\leq l_{t})=0.9 \right \}), \ \ i = 1, 2, ..., N
\end{align*}
the corresponding priors on the mean parameters of the step length for the repulsive case:
\begin{align*}
    \underline{\mu} = (\mu_{1}, \mu_{2},...,\mu_{N})|N & \sim \text{StraussProcess}(\mu_{1}, \mu_{2},...,\mu_{N};\xi, a, d) = h(\mu_{1}, \mu_{2},...,\mu_{N}|N,\xi, a, d) \\
   & \propto \left[\prod_{i=1}^{N}\xi \mathbb{I}[\mu_{i} \in R] \right]a^{\sum_{1\leq i \leq j \leq N}\mathbb{I}[\left \| \mu_{i}-\mu_{j} \right \|<d]} 
\end{align*}
for the independent case:
\begin{align*}
    \underline{\mu} = (\mu_{1}, \mu_{2},...,\mu_{N})|N &  \sim \text{IndependentProcess}(\mu_{1}, \mu_{2},...,\mu_{N};\xi) = h(\mu_{1}, \mu_{2},...,\mu_{N}|N,\xi) \\
    & = \frac{1}{\xi^{N}\left | R \right |^{N}}\left[\prod_{i=1}^{N}\xi \mathbb{I}[\mu_{i} \in R]\right] = \prod_{i=1}^{N} \frac{\mathbb{I}[\mu_{i}\in R]}{\left | R \right |}
\end{align*}
which corresponds to the product of $N$ Uniform distributions in the region $R$. The point process parameter $\xi$ for either cause of Strauss Process or Independent Process has 
\begin{equation*}
    \xi \sim \text{Uniform}(|R|^{-1},80|R|^{-1})
\end{equation*}
The unknown number of states, denoted as $N$, follows a Uniform prior distribution with an upper bound of $80$. This choice reflects our prior belief that there is not good reason to consider more than $80$ behavioral states. It is important to note that this upper bound is subjective; although we could have opted for a smaller value, it should not be selected close to the expected number of behavioral states. This precaution is taken to allow the algorithm to independently identify the latent states without introducing bias to the results through the prior distribution. 

For the initial and transition probability distribution we used a Dirichlet prior distribution, with parameters equal to 1, and use their Gamma decomposition equivalence explained in \cite{argiento2022infinity} for obtaining a much more efficient mixing for the RJMCMC algorithm. Next, for the probability parameter $z_{i}$ of the zero-inflated Gamma distribution, we selected a 
Beta prior distribution to express the proportion of zeros in the step length data. The prior distributions for the parameters $k_{i}$ a Uniform within $[0.5,2]$, $m_{i}$ a Uniform within $[-\pi,\pi]$ and $\sigma_{i}$ a Uniform within $[0.5\left \{ l_{t}:\mathbb{P}(L_{t}\leq l_{t})=0.1 \right \},2\left \{ l_{t}:\mathbb{P}(L_{t}\leq l_{t})=0.9 \right \}]$ which are the same distributions chosen in \cite{pohle2017selecting} for initializing the parameters values for their likelihood optimization. 
We made these choices for the prior distributions of the previously mentioned parameters, in order to be a ground of comparison between the methods proposed in \cite{pohle2017selecting} and our method, i.e. our goal was to observe how the RJMCMC alongside with the Strauss point process are affecting the inference, by minimizing the effect of prior distributions. 

Moreover, for the repulsive prior we choose as norm measure $\left \| \cdot \right \|$ the euclidean distance, for the penalty $a$ and threshold $d$ we can choose them based on the method explained in \cite{beraha2022mcmc} which are equal to $a=\exp(-n^{*}_{2.5})$ with $n^{*}_{2.5}=627$, $a=\exp(-n^{*}_{5})$ with $n^{*}_{5} = 1255$ and $d = 98$.


\subsection{Inference}
\label{supsubsec:infergps}
\par{
$\bullet$ \text{\textbf{Fixed dimension Moves}}
\par{
In the first step of the algorithm, we update the model parameters, for a fixed value N, by sampling from the corresponding posterior distributions. We sequentially update each parameter using a Metropolis Hastings algorithm. The proposal steps are of the following form
\begin{enumerate}
    \item $\mu^{*}_{S_{t}}=\mu_{S_{t}} + \epsilon_{\mu}, $ \ \ \ $\epsilon_{\mu}\sim \text{LogNormal}(0,0.01),$ \ \ $S_{t}=1,2,...,N$
    \item $\sigma^{*}_{S_{t}}=\sigma_{S_{t}} + \epsilon_{\sigma}, $ \ \ \ $\epsilon_{\sigma}\sim \text{LogNormal}(0,0.03),$ \ \ $S_{t}=1,2,...,N$
    \item $k^{*}_{S_{t}}=k_{S_{t}} + \epsilon_{k}, $ \ \ \ $\epsilon_{k}\sim \text{LogNormal}(0,0.08),$ \ \ $S_{t}=1,2,...,N$
    \item $m^{*}_{S_{t}}=m_{S_{t}} + \epsilon_{m}, $ \ \ \ $\epsilon_{m}\sim \text{Normal}(0,0.08),$ \ \ $S_{t}=1,2,...,N$
    \item $\Lambda^{*}_{S_{t}S_{t+1}}=\Lambda_{S_{t}S_{t+1}} + \epsilon_{L}, $ \ \ \ $\epsilon_{L}\sim \text{LogNormal}(0,0.05),$ \ \ $S_{t},S_{t+1}=1,2,...,N$
    \item $\lambda^{*}_{S_{1}}=\lambda_{S_{1}} + \epsilon_{\lambda}, $ \ \ \ $\epsilon_{\lambda}\sim \text{LogNormal}(0,0.07),$ \ \ $S_{1}=1,2,...,N$
    \item $\text{logit}(z^{*}_{S_{t}})=\text{logit}(z_{S_{t}}) + \epsilon_{z}, $ \ \ \ $\epsilon_{z}\sim \text{Normal}(0,0.5),$ \ \ $S_{t}=1,2,...,N$ $\Leftrightarrow  z^{*}_{S_{t}} \sim \text{LogNormal}(\text{logit}(z_{S_{t}}),\tau_{z}),$ 
    \item $\xi$ is sampled from each full conditional with a Metropolis Hastings algorithm and the steps are described in Section 2.4 in the main text.
\end{enumerate}
Care must be taken when calculating the Metropolis-Hastings ratio since most of the proposed moves are not symmetric, which must be accounted for. The acceptance probabilities of the proposed values, for both versions, include
the Jacobian that arises because we work with a logit and log scale transformation. The proposal distribution were chosen such the acceptance ratio were close to 0.25. 
}
}
\\
\par{
$\bullet$ \text{\textbf{ Variable dimension moves}}

With probability $0.5$, we choose between the moves {Split/Combine} and {Birth/Death}. The Split/Combine move splits or combines two existing components; in particular the choice of components to me combined is based on how similar they are, the similarity measure can be found later on. In the Birth/Death case, we kill or give birth to a new component by sampling from the corresponding proposal distributions.
\\
\par{
\text{\textbf{ Split/combine moves}}

\par{
 In this step, we choose whether to split or combine components with probability 0.5. If we only have a single component, then with probability one, we split. In the split move, we choose uniformly one of the $N$ components, denoted as $j_{*}$ which we decide to split it to $j_{1}$ and $j_{2}$. Then the corresponding parameters split as follows

\begin{enumerate}
    \item $\lambda_{j_{1}} = \rho \lambda_{j_{*}},$ \ $\lambda_{j_{2}} = (1-\rho)\lambda_{j_{*}},$ \ $\rho \sim \text{Beta}(2,2)$
    \item $z_{j_{1}} = z_{j_{*}}-u_{z},$ \ \ $z_{j_{2}} = z_{j_{*}} + u_{z},$ \  $u_{z}\sim \text{Uniform}(0,\text{min}(z_{j_{*}},1-z_{j_{*}}))$
    \item $\Lambda_{jj_{1}} = \Lambda_{jj_{*}}\rho_{j},$ \ $\Lambda_{jj_{2}} = \Lambda_{jj_{*}}(1-\rho_{j}),$ \ $\rho_{j}\sim \text{Beta}(2,2)$, \ $j\neq j_{*}$
    \item $\Lambda_{j_{1}j} = \Lambda_{j_{*}j}\theta_{j},$ \ $\Lambda_{j_{2}j} = \Lambda_{j_{*}j}/\theta_{j},$ \ $\theta_{j}\sim \text{Gamma}(1,3)$, \ $j\neq j_{*}$
    \item \begin{align*}
        \Lambda_{j_{1}j_{1}} & = \Lambda_{j_{*}j_{*}}\rho_{j_{*}}\theta_{j_{1}}, \ \Lambda_{j_{1}j_{2}} = \Lambda_{j_{*}j_{*}}(1-\rho_{j_{*}})\theta_{j_{2}} \\
        \Lambda_{j_{2}j_{1}} & = \Lambda_{j_{*}j_{*}}\rho_{j_{*}}/\theta_{j_{1}}, \ \Lambda_{j_{2}j_{2}}=\Lambda_{j_{*}j_{*}}(1-\rho_{j_{*}})/\theta_{j_{2}}
    \end{align*}
    with $\rho_{j_{*}}\sim \text{Beta}(2,2)$ and $\theta_{j_{1}},\theta_{j_{2}}\sim \text{Gamma}(1,3)$
    \item $\mu_{j_{1}} = \mu_{j_{*}}-\theta_{\mu}, \ \mu_{j_{2}} = \mu_{j_{*}}+\theta_{\mu}, \ \theta_{\mu}\sim \text{Uniform}(0,\mu_{j_{*}}-\mu_{j_{*}}/2)$ 
    \item $\sigma_{j_{1}} = \sigma_{j_{*}}-\theta_{\sigma}, \ \sigma_{j_{2}} = \sigma_{j_{*}}+\theta_{\sigma}, \ \theta_{\sigma}\sim \text{Uniform}(0,\sigma_{j_{*}}-\sigma_{j_{*}}/2)$ 
    \item $k_{j_{1}} = k_{j_{*}}-\theta_{k}, \ k_{j_{2}} = k_{j_{*}}+\theta_{k}, \ \theta_{k}\sim \text{Uniform}(0,k_{j_{*}}-k_{j_{*}}/2)$ 
    \item $m_{j_{1}} = m_{j_{*}}+\epsilon_{m}, \ m_{j_{2}} = m_{j_{*}}-\epsilon_{m}, \ \epsilon_{m}\sim \text{Normal}(0,2)$ 
\end{enumerate}
In the reverse move, we merge the most similar components $j_{1}$ and $j_{2}$ and to $j_{*}$.
\begin{enumerate}
    \item $\lambda_{j_{*}} = \lambda_{j_{1}} + \lambda_{j_{2}}$
    \item $z_{j_{*}} = \frac{z_{j_{1}}+z_{j_{2}}}{2}$
    \item $\Lambda_{jj_{*}} = \Lambda_{jj_{1}} + \Lambda_{jj_{2}}$ \ $j\neq j_{*}$
    \item $\Lambda_{j_{*}j}=(\Lambda_{j_{1}j}\Lambda_{j_{2}j})^{0.5},$ \ $j\neq j_{*}$
    \item $\Lambda_{j_{*}j_{*}}=(\Lambda_{j_{1}j_{1}}\Lambda_{j_{2}j_{1}})^{0.5}+(\Lambda_{j_{1}j_{2}}\Lambda_{j_{2}j_{2}})^{0.5}$
    \item $\mu_{j_{*}} = \frac{\mu_{j_{1}} + \mu_{j_{2}}}{2}$ 
    \item $\sigma_{j_{*}} = \frac{\sigma_{j_{1}} + \sigma_{j_{2}}}{2}$
    \item $k_{j_{*}} = \frac{k_{j_{1}} + k_{j_{2}}}{2}$
    \item $m_{j_{*}} = \frac{m_{j_{1}}+m_{j_{2}}}{2}$ 
\end{enumerate}
The split is accepted with probability $\text{min}\left \{ 1,A \right \}$ whereas in the combine move we accepted with $\text{min}\left \{ 1,A^{-1} \right \}$.

\begin{align*}
    A  = & \frac{f(\left \{ O_{t} \right \}_{t=1}^{T}|\left \{ \pi \right \}_{j=1}^{N+1},\left \{ P_{j} \right \}_{j=1}^{N+1},\left \{ \mu_{j} \right \}_{j=1}^{N+1},\left \{ \sigma_{j} \right \}_{j=1}^{N+1},\left \{ m_{j} \right \}_{j=1}^{N+1},\left \{ k_{j} \right \}_{j=1}^{N+1}) }{f(\left \{ O_{t} \right \}_{t=1}^{T}|\left \{ \pi \right \}_{j=1}^{N},\left \{ P_{j} \right \}_{j=1}^{N},\left \{ \mu_{j} \right \}_{j=1}^{N},\left \{ \sigma_{j} \right \}_{j=1}^{N},\left \{ m_{j} \right \}_{j=1}^{N},\left \{ k_{j} \right \}_{j=1}^{N}) } \\
& \frac{p(\left \{ \pi \right \}_{j=1}^{N+1},\left \{ P_{j} \right \}_{j=1}^{N+1},\left \{ \mu_{j} \right \}_{j=1}^{N+1},\left \{ \sigma_{j} \right \}_{j=1}^{N+1},\left \{ m_{j} \right \}_{j=1}^{N+1},\left \{ k_{j} \right \}_{j=1}^{N+1})p(N+1)}{p(\left \{ \pi \right \}_{j=1}^{N},\left \{ P_{j} \right \}_{j=1}^{N},\left \{ \mu_{j} \right \}_{j=1}^{N},\left \{ \sigma_{j} \right \}_{j=1}^{N},\left \{ m_{j} \right \}_{j=1}^{N},\left \{ k_{j} \right \}_{j=1}^{N}){p(N)}}\\ 
    &\underset{q(N+1\rightarrow N)/q(N\rightarrow N+1)}{\underbrace{ \frac{(N+1)!}{N!} \frac{P_{c}(N+1)/[2(N+1)]}{P_{s}(N)/N} \frac{\left | J \right |}{p(\theta_{j_{1}})p(\theta_{j_{2}})\prod_{j}p(\theta_{j})p(\theta_{\mu})p(\theta_{\sigma})p(\theta_{k})p(\epsilon_{m})p(\rho)\prod_{j}p(\rho_{j})p(z_{j_{*}})}}}
\end{align*}
The Jacobian, $|J|$ of the transformation from $N$ to $N+1$, is equal to
\begin{equation*}
    |J| = 2\lambda_{j_{*}}k_{j_{*}}\mu_{j_{*}}\sigma_{j_{*}}4\rho_{j_{*}}(1-\rho_{j_{*}})\frac{L_{j_{*}j_{*}}^{3}}{\theta_{j_{1}}\theta_{j_{2}}}\prod_{j}\Lambda_{jj_{*}} 2^{N-1} \prod_{j}\frac{\Lambda_{j_{*}}j}{\theta_{j}}
\end{equation*}
Also, the probabilities $P_{c}$ and $P_{s}$ correspond to the probabilities of making a combine or split movie, which in our case will cancel out. The split and merge moves are the ones described in \cite{bartolucci2012latent}.

Lastly the similarity measure for combining two components, is calculated as 
\begin{equation*}
    d_{i} = \sum_{j \neq i}\sqrt{(\mu_{i}-\mu_{j})^{2} + (\sigma_{i}- \sigma_{j})^{2} + (m_{i}-m_{j})^{2} + (k_{i}-k_{j})^{2}}
\end{equation*}
for each component $i$. Then we choose the two components that have the smallest $d_{i}'s$ values, and we combine them.
}
}
\\
\par{
\text{\textbf{Birth/death moves}}

\par{
The Birth/Death move is performed similarly to the Split/Combine. Likewise, if we have $N$ components, we choose with probability 0.5 to give birth to a new component or death to an existing one. In the birth move, we generate new component parameters from the prior distribution, and the rest of the components are simply copied. On the other hand, for the death move, we uniformly choose a component and kill it. In this case, the acceptance probability of birth move is again $\text{min}\left \{ 1,A \right \}$ whereas for the death move $\text{min}\left \{1, A^{-1} \right \}$ with 
\begin{align*}
   A  = & \frac{f(\left \{ O_{t} \right \}_{t=1}^{T}|\left \{ \pi \right \}_{j=1}^{N+1},\left \{ P_{j} \right \}_{j=1}^{N+1},\left \{ \mu_{j} \right \}_{j=1}^{N+1},\left \{ \sigma_{j} \right \}_{j=1}^{N+1},\left \{ m_{j} \right \}_{j=1}^{N+1},\left \{ k_{j} \right \}_{j=1}^{N+1}) }{f(\left \{ O_{t} \right \}_{t=1}^{T}|\left \{ \pi \right \}_{j=1}^{N},\left \{ P_{j} \right \}_{j=1}^{N},\left \{ \mu_{j} \right \}_{j=1}^{N},\left \{ \sigma_{j} \right \}_{j=1}^{N},\left \{ m_{j} \right \}_{j=1}^{N},\left \{ k_{j} \right \}_{j=1}^{N}) } \\
& \frac{p(\left \{ \pi \right \}_{j=1}^{N+1},\left \{ P_{j} \right \}_{j=1}^{N+1},\left \{ \mu_{j} \right \}_{j=1}^{N+1},\left \{ \sigma_{j} \right \}_{j=1}^{N+1},\left \{ m_{j} \right \}_{j=1}^{N+1},\left \{ k_{j} \right \}_{j=1}^{N+1})p(N+1)}{p(\left \{ \pi \right \}_{j=1}^{N},\left \{ P_{j} \right \}_{j=1}^{N},\left \{ \mu_{j} \right \}_{j=1}^{N},\left \{ \sigma_{j} \right \}_{j=1}^{N},\left \{ m_{j} \right \}_{j=1}^{N},\left \{ k_{j} \right \}_{j=1}^{N}){p(N)}}\\ 
  & \underset{q(N+1\rightarrow N)/q(N\rightarrow N+1)}{\underbrace{ \frac{(N+1)!}{N!} \frac{P_{d}(N+1)/N}{P_{b}(N)/(N+1)}\frac{|J|}{p(\lambda_{j_{*}})p(z_{j_{*}})p(\mu_{j_{*}})p(\sigma_{j_{*}})p(k_{j_{*}})p(m_{j_{*}})p(\Lambda_{j_{*}j_{*}})\prod_{j}p(\Lambda_{jj_{*}})p(\Lambda_{j_{*}j})}}}
\end{align*}
Since the parameters are drawn from their respective priors, the Jacobian term $|J|$ will equal one.
}



\subsection{Results}
\label{supsubsec:resgps}
We display the results of the ICL information criterion for model comparison as described in \cite{pohle2017selecting}, in Table \ref{tab:sim1:ICL}.

\begin{center}
\begin{table}[!ht]
\caption{Values of model selection criterion ICL for the different numbers of components 2, 3, 4, 5, 6 and 7, computed with the algorithm displayed in \citet{pohle2017selecting}}
\label{tab:sim1:ICL}
\centering
 \text{Number of components}
\\
\begin{tabular}{ |c|c|c|c|c|c|c| } 
\hline
 & 2 & 3 & 4 & 5 & 6 & 7\\
\hline
ICL & 354,829 &  351,544 &350,159 &  351,247 &   354,701 & 350,051\\
\hline
\end{tabular}
\end{table}
\end{center}

Then for the last time point of the time series we display the averaged posterior mixture distributions of the step length and angle for the independent prior model, in Figure \ref{sup:Fig:standpriorgps}.

\begin{figure}[htp]
    \centering \includegraphics[height=10cm,width=15cm]{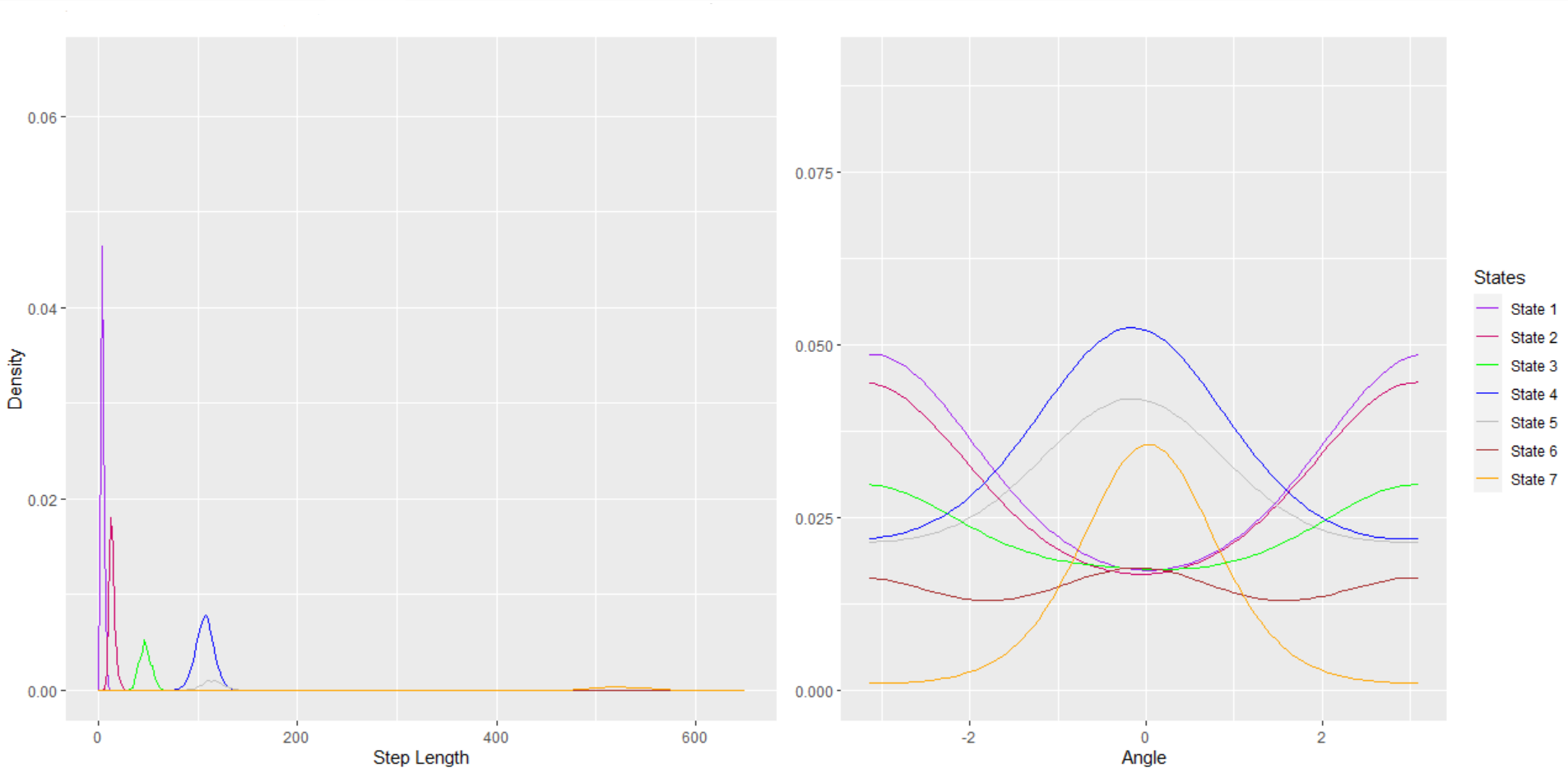}
    \caption{Averaged posterior mixture distribution of step length (left) and angle (right)
for the last time point of the time series, for the independent prior model.}
\label{sup:Fig:standpriorgps}
\end{figure}

We also present the averaged posterior mixture distributions of the step length and angle for the repulsive prior model, with $a=\exp(-n^{*}_{5})$ with $n^{*}_{5}=1255$, displayed in Figure \ref{K_2_musk}.

\begin{figure}[htp]
    \centering \includegraphics[height=10cm,width=15cm]{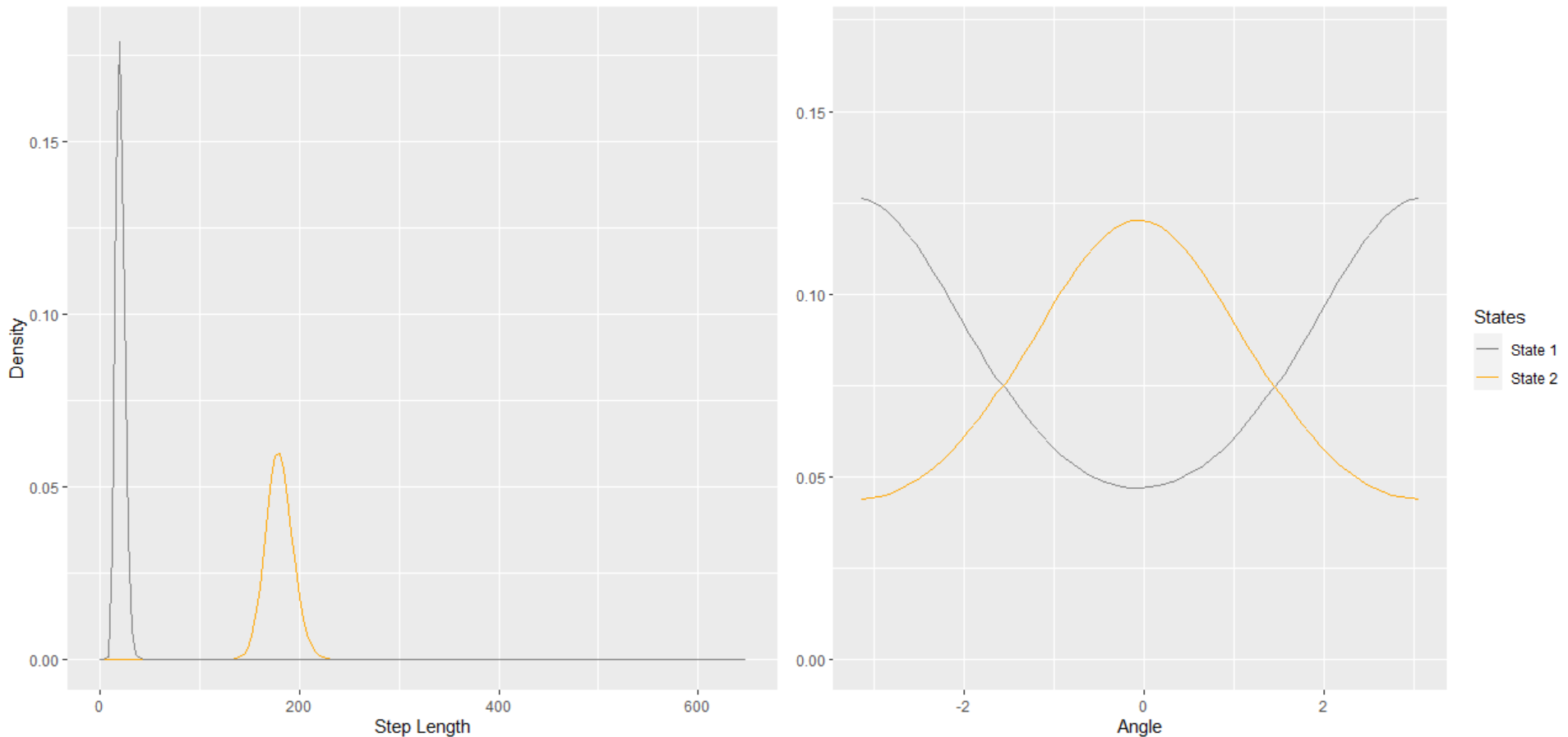}
    \caption{Averaged posterior mixture distribution of step length (left) and angle (right)
for the last time point of the time series, for the repulsive prior model, when accounting for penalty $a=\exp(-n^{*}_{5})$.}
\label{K_2_musk}

\end{figure}

Since, we have used a smaller $a$ compared to the case of $2.5\%$ we expect to infer a small number of clusters which is evident from the Figure \ref{K_2_musk} where we identify two clusters, one corresponding to small undirected steps (State 1) and one corresponding to big directed steps (State 2). Interestingly, if we look at the posterior distribution on the number of states $N$, $p(2) = 0.4307, p(3) = 0.1265, p(4) = 0.2357, p(5) = 0.2045, p(6) = 0.0026$, we can observe that even though the mode of $N$ is on $2$ there is some significant mass on the number of states $4$ and $5$ showing evidence that the choice of $5\%$ might lead to overpenalization.

\clearpage
\section{Acoustic Application}
\label{supsec:acoustic}
In this section we present Supplementary information for the second case study of Section 5.

\subsection{Model}
\label{supsubsec:model}
In our case study, airborne  devices are used to collect acoustic data over time. As a result, the data we gathered are acoustic features based on the Mel-frequency cepstral coefficient measured for each segment time point. However, those features are highly correlated for which reason we applied a PCA and kelp only the first two PC. The model is described as follows:
\begin{align*}
    O_{t,c} & = \underline{E_{t,c}} = \left \{ E_{c,t,1},E_{c,t,1} \right \}, \ c = 1, 2, ..., C\\
    S_{t} & \in \left \{ 1,2,...,N \right \} \\
    f(O_{t,c}|S_{t}) &= f(\underline{E_{t,c}}|S_{t}) \\   f(\underline{E_{t,c}}|S_{t}) &= \text{Normal}_{2}(\underline{E_{t,c}};\underline{\mu_{S_{t}}},\Sigma_{S_{t}}) 
\end{align*}
the specified priors are:

\begin{align*}
    N & \sim \text{Uniform}\left \{ 1, 2, ..., 120 \right \} \\
    \pi = (\pi_{1}, \pi_{2}, ..., \pi_{N}) & = (\frac{\lambda_{1}}{\sum_{i=1}^{N}\lambda_{i}},\frac{\lambda_{2}}{\sum_{i=1}^{N}\lambda_{i}},...,\frac{\lambda_{N}}{\sum_{i=1}^{N}\lambda_{i}}) \Rightarrow \lambda_{i} \sim \text{Gamma}(1,1), \ \ i = 1, 2, ..., N \\
    P_{i.} = (P_{i,1}, P_{i,2}, ..., P_{i,N})  & = (\frac{\Lambda_{i1}}{\sum_{j=1}^{N}\Lambda_{ij}},\frac{\Lambda_{i2}}{\sum_{j=1}^{N}\Lambda_{ij}},...,\frac{\Lambda_{iN}}{\sum_{j=1}^{N}\Lambda_{ij}})  \Rightarrow \Lambda_{ij} \sim \text{Gamma}(1,1), \ \ i,j = 1, 2, ..., N  \\
    \Sigma_{i} & \sim \text{Wishart}(20,\hat{\Sigma}_{0}/10), \ \ i = 1, 2, ..., N
\end{align*}
and the corresponding priors on the vector means are:
\begin{align*}
    \underline{\mu} = (\underline{\mu_{1}}, \underline{\mu_{2}},...,\underline{\mu_{N}})|N & \sim \text{StraussProcess}(\underline{\mu_{1}}, \underline{\mu_{2}},...,\underline{\mu_{N}};\xi, a, d) = h(\underline{\mu_{1}}, \underline{\mu_{2}},...,\underline{\mu_{N}}|N,\xi, a, d) \\
   & \propto \left[\prod_{i=1}^{N}\xi \mathbb{I}[\underline{\mu_{i}} \in R^{2}] \right]a^{\sum_{1\leq i \leq j \leq N}\mathbb{I}[\left \| \underline{\mu_{i}}-\underline{\mu_{j}} \right \|<d]} 
\end{align*}
for the independent prior case:
\begin{align*}
    \underline{\mu} = (\underline{\mu_{1}}, \underline{\mu_{2}},...,\underline{\mu_{N}})|N &  \sim \text{IndependentProcess}(\underline{\mu_{1}}, \underline{\mu_{2}},...,\underline{\mu_{N}};\xi) = h(\underline{\mu_{1}}, \underline{\mu_{2}},...,\underline{\mu_{N}}|N,\xi) \\
    & = \left[\prod_{i=1}^{N}\xi \mathbb{I}[\underline{\mu_{i}} \in R^{2}]\right] \frac{1}{\xi^{N}\left | R^{2} \right |^{N}} = \prod_{i=1}^{N} \frac{\mathbb{I}[\underline{\mu_{i}}\in R^{2}]}{\left | R^{2} \right |}
\end{align*}
which corresponds to the product of $N$ Uniform distributions in the region $R^{2}$.
The point process parameter $\xi$ for either cause of Strauss Process or Strandard
Process has 
\begin{equation*}
    \xi \sim \text{Uniform}(|R|^{-2},120|R|^{-2})
\end{equation*}

The unknown number of states, denoted as $N$, follows a Uniform prior distribution with an upper bound of $120$. This choice reflects our prior belief that there should be no more than $120$ behavioral states. It is important to note that this upper bound is subjective; although we could have opted for a smaller value, it should not be selected close to the expected number of behavioral states. This precaution is taken to allow the algorithm to independently identify the latent states without introducing bias to the results through the prior distribution. 

For the initial and transition probability distribution we used Dirichlet prior distributions with parameters equal to 1, and use their Gamma decomposition equivalence explained in \cite{argiento2022infinity} for obtaining a much more efficient mixing for the RJMCMC algorithm. Next, for the covariance matrix we allow for a Wishart prior distribution with $\hat{\Sigma_{0}}$ the sampled covariance matrix of all 
 segment time points cosidered together. Lastly, if $R$ is the range of the transformed acoustic features on the first two PC, then we have the range of the product space $R^{2}$ as $|R|^{2}$.
 
  The penalty parameters of the Strauss point process $a,$ $d$ are fixed  and the prior distribution for $\xi$ are defined, based on the rules outlined in \cite{beraha2022mcmc}. 
  Moreover, for the repulsive prior we choose as norm measure $\left \| \cdot \right \|$ the euclidean distance, for the penalty $a$ and threshold $d$ we can choose them based on the method explained in \cite{beraha2022mcmc} which are equal to $a=\exp(-n^{*}_{2.5})$ with $n^{*}_{2.5}=54$, $a=\exp(-n^{*}_{5})$ with $n^{*}_{5} = 108$ and $d = 21$.
\\
\par{
$\bullet$ \text{\textbf{Fixed dimension Moves}}
\par{
In the first step of the algorithm, we update the model parameters, for a fixed value N, by sampling from the corresponding posterior distributions. We sequentially update each parameter using a Metropolis Hastings algorithm. The steps are of the following form
\begin{enumerate}
    \item $\underline{\mu}^{*}_{S_{t}}=\mu_{S_{t}} + \epsilon_{\mu}, $ \ \ \ $\epsilon_{\mu}\sim \text{Normal}(0,0.3),$ \ \ $S_{t}=1,2,...,N$
    \item $\Sigma^{*}_{S_{t}}\sim \text{Wishart}(1200,\Sigma_{S_{t}}/1200),$ \ \ $S_{t}=1,2,...,N$
    \item $\Lambda^{*}_{S_{t}S_{t+1}}=\Lambda_{S_{t}S_{t+1}} + \epsilon_{L}, $ \ \ \ $\epsilon_{L}\sim \text{LogNormal}(0,1),$ \ \ $S_{t},S_{t+1}=1,2,...,N$
    \item $\lambda^{*}_{S_{1}}=\lambda_{S_{1}} + \epsilon_{\lambda}, $ \ \ \ $\epsilon_{\lambda}\sim \text{LogNormal}(0,1.5),$ \ \ $S_{1}=1,2,...,N$
    \item $\xi$ is sampled from each full conditional with a Metropolis Hastings algorithm and the steps are described in Section 2.4 in the main text.
\end{enumerate}
Care must be taken when calculating the Metropolis-Hastings ratio since most of the proposed moves are not symmetric, which must be accounted for. The acceptance probabilities of the proposed values, for both versions, include
the Jacobian that arises because we work with a logit and log scale transformation. The proposal distribution were chosen such the acceptance ratio were close to 0.25. 
}
}
\\
\par{
$\bullet$ \text{\textbf{ Variable dimension moves}}

With probability $0.5$, we choose between the moves {Split/Combine} and {Birth/Death}. The Split/Combine move splits or combines two existing components; in particular the choice of components to me combined is based on how similar they are, the similarity measure can be found later on. In the Birth/Death case, we kill or give birth to a new component by sampling from the corresponding proposal distributions.
\\
\\
\par{
\text{\textbf{ Split/combine moves}}

\par{
In this step, we choose whether to split or combine components with probability 0.5. If we only have a single component, then with probability one, we split. In the split move, we choose uniformly one of the $N$ components, denoted as $j_{*}$ which we decide to split it to $j_{1}$ and $j_{2}$. Based on the split/combine moves described in \cite{zhang2004learning, bartolucci2012latent} the corresponding parameters split are as follows

\begin{enumerate}
    \item $\lambda_{j_{1}} = \rho \lambda_{j_{*}},$ \ $\lambda_{j_{2}} = (1-\rho)\lambda_{j_{*}},$ \ $\rho $
    \item $\Lambda_{jj_{1}} = \Lambda_{jj_{*}}\rho_{j},$ \ $\Lambda_{jj_{2}} = \Lambda_{jj_{*}}(1-\rho_{j}),$ \ $\rho_{j}\sim \text{Beta}(2,2)$, \ $j\neq j_{*}$
    \item $\Lambda_{j_{1}j} = \Lambda_{j_{*}j}\theta_{j},$ \ $\Lambda_{j_{2}j} = \Lambda_{j_{*}j}/\theta_{j},$ \ $\theta_{j}\sim \text{Gamma}(1,3)$, \ $j\neq j_{*}$
    \item \begin{align*}
        \Lambda_{j_{1}j_{1}} & = \Lambda_{j_{*}j_{*}}\rho_{j_{*}}\theta_{j_{1}}, \ \Lambda_{j_{1}j_{2}} = \Lambda_{j_{*}j_{*}}(1-\rho_{j_{*}})\theta_{j_{2}} \\
        \Lambda_{j_{2}j_{1}} & = \Lambda_{j_{*}j_{*}}\rho_{j_{*}}/\theta_{j_{1}}, \ \Lambda_{j_{2}j_{2}}=\Lambda_{j_{*}j_{*}}(1-\rho_{j_{*}})/\theta_{j_{2}}
    \end{align*}
    with $\rho_{j_{*}}\sim \text{Beta}(2,2)$ and $\theta_{j_{1}},\theta_{j_{2}}\sim \text{Gamma}(1,3)$
    \item $\underline{\mu_{j_{1}}} = \underline{\mu_{*}} - \sqrt{\frac{\pi_{j_{2}}}{\pi_{j_{1}}}} \sum_{d=1}^{12}r_{*d}^{\frac{1}{2}}u_{d}\underline{a_{d}},$ \ \ $\underline{\mu_{j_{2}}} = \underline{\mu_{*}} + \sqrt{\frac{\pi_{j_{1}}}{\pi_{j_{2}}}} \sum_{d=1}^{12}r_{*d}^{\frac{1}{2}}u_{d}\underline{a_{d}},$ \ \ $u_{d}\sim \text{Beta}(2,2)$
    \\
    the $r_{*d}$ is the $d$th eigenvalue of the covariance matrix $\Sigma_{*}$ and the $a_{d}$ is the $d$th eigenvector of the sampled covariance matrix $\hat{\Sigma_{0}}.$
    \item $r_{j_{1}d} = \beta_{d}(1-u_{d}^{2})\frac{\pi_{*}}{\pi_{j_{1}}}r_{*d},$ \ \ $r_{j_{2}d}=(1-\beta_{d})(1-u_{d}^{2})\frac{\pi_{*}}{\pi_{j_{2}}}r_{*d},$ \ $\beta_{d} \sim \text{Beta}(1,1),$ \ $d = 1, 2, ..., 12$
    \\
    we employ this specific eigenvalue decomposition split move, as described in \cite{zhang2004learning}, to ensure that the resulting covariance matrices for components $j_{1}$ and $j_{2}$ are both positive-definite and symmetric. This guarantees that they meet the criteria for being valid covariance matrices.
    \end{enumerate}
In the reverse move, we merge the most similar components $j_{1}$ and $j_{2}$ and to $j_{*}$.
\begin{enumerate}
    \item $\lambda_{j_{*}} = \lambda_{j_{1}} + \lambda_{j_{2}}$
    \item $\Lambda_{jj_{*}} = \Lambda_{jj_{1}} + \Lambda_{jj_{2}}$ \ $j\neq j_{*}$
    \item $\Lambda_{j_{*}j}=(\Lambda_{j_{1}j}\Lambda_{j_{2}j})^{0.5},$ \ $j\neq j_{*}$
    \item $\Lambda_{j_{*}j_{*}}=(\Lambda_{j_{1}j_{1}}\Lambda_{j_{2}j_{1}})^{0.5}+(\Lambda_{j_{1}j_{2}}\Lambda_{j_{2}j_{2}})^{0.5}$
    \item $\mu_{j_{*}} = \mu_{j_{1}}\frac{\pi_{j_{1}}}{\pi_{j_{*}}} + \mu_{j_{2}}\frac{\pi_{j_{2}}}{\pi_{j_{*}}}$
    \item $\lambda_{j_{*}} = \frac{\pi_{j_{1}}}{\pi_{j_{*}}}r_{j_{1}d} + \frac{\pi_{j_{2}}}{\pi_{j_{*}}}r_{j_{2}d} + \frac{\pi_{j_{1}}\pi_{j_{2}}}{\pi_{j_{*}}}(\mu_{j_{1}d}-\mu_{j_{2}d})^{2},$ \ \ $d = 1, 2, ..., d$
\end{enumerate}
The split is accepted with probability $\text{min}\left \{ 1,A \right \}$ whereas in the combine move we accepted with $\text{min}\left \{ 1,A^{-1} \right \}$.

\begin{align*}
    A  = & \frac{f(\left \{ O_{t} \right \}_{t=1}^{T}|\left \{ \pi \right \}_{j=1}^{N+1},\left \{ P_{j} \right \}_{j=1}^{N+1},\left \{ \underline{\mu_{j}} \right \}_{j=1}^{N+1},\left \{ \Sigma_{j} \right \}_{j=1}^{N+1}) }{f(\left \{ O_{t} \right \}_{t=1}^{T}|\left \{ \pi \right \}_{j=1}^{N},\left \{ P_{j} \right \}_{j=1}^{N},\left \{ \underline{\mu_{j}} \right \}_{j=1}^{N},\left \{ \Sigma_{j} \right \}_{j=1}^{N}) } \\
& \frac{p(\left \{ \pi \right \}_{j=1}^{N+1},\left \{ P_{j} \right \}_{j=1}^{N+1},\left \{ \underline{\mu_{j}} \right \}_{j=1}^{N+1},\left \{ \Sigma_{j} \right \}_{j=1}^{N+1})p(N+1)}{p(\left \{ \pi \right \}_{j=1}^{N},\left \{ P_{j} \right \}_{j=1}^{N},\left \{ \underline{\mu_{j}} \right \}_{j=1}^{N},\left \{ \Sigma_{j} \right \}_{j=1}^{N}){p(N)}}\\ 
     & \underset{q(N+1\rightarrow N)/q(N\rightarrow N+1)}{\underbrace{\frac{(N+1)!}{N!} \frac{P_{c}(N+1)/[2(N+1)]}{P_{s}(N)/N} \frac{\left | J \right |}{p(\theta_{j_{1}})p(\theta_{j_{2}})\prod_{j}p(\theta_{j})p(\rho)\prod_{j}p(\rho_{j})\prod_{d}p(\beta_{d})p(u_{d})}}}
\end{align*}
The Jacobian, $|J|$ of the transformation from $N$ to $N+1$, is equal to
\begin{equation*}
    |J| = 4\rho_{j_{*}}(1-\rho_{j_{*}})\frac{L_{j_{*}j_{*}}^{3}}{\theta_{j_{1}}\theta_{j_{2}}}\prod_{j}\Lambda_{jj_{*}} 2^{N-1} \prod_{j}\frac{\Lambda_{j_{*}}j}{\theta_{j}} \\
    \frac{\pi_{j_{*}}^{3*12+1}}{(\pi_{j_{1}}\pi_{j_{2}})^\frac{3*12}{2}}\sum r_{kd}^{\frac{3}{2}}(1-u_{d}^{2})
\end{equation*}
Also, the probabilities $P_{c}$ and $P_{s}$ correspond to the probabilities of making a combine or split movie, which in our case will cancel out. The split and merge moves are the ones described in \cite{bartolucci2012latent}.

Lastly the similarity measure for combining two components, is calculated as 
\begin{equation*}
    d_{i} = \sum_{j \neq i}\left \|  \underline{\mu_{i}}-\underline{\mu_{j}}\right \|_{2} 
\end{equation*}
for each component $i$. Then we choose the two components that have the smallest $d_{i}'s$ values, and we combine them. The $\left \| \cdot \right \|$ corresponds to the euclidean distance.
}
}
\\
\par{
\text{\textbf{Birth/death moves}}

\par{
The Birth/death move is performed similarly to the Split/Combine. Likewise, if we have $N$ components, we choose with probability 0.5 to give birth to a new component or death to an existing one. In the birth move, we generate new component parameters from the prior distribution, and the rest of the components are simply copied. On the other hand, for the death move, we uniformly choose a component and kill it. In this case, the acceptance probability of birth move is again $\text{min}\left \{ 1,A \right \}$ whereas for the death move $\text{min}\left \{1, A^{-1} \right \}$ with 
\begin{align*}
   A  = & \frac{f(\left \{ O_{t} \right \}_{t=1}^{T}|\left \{ \pi \right \}_{j=1}^{N+1},\left \{ P_{j} \right \}_{j=1}^{N+1},\left \{ \underline{\mu_{j}} \right \}_{j=1}^{N+1},\left \{ \Sigma_{j} \right \}_{j=1}^{N+1}) }{f(\left \{ O_{t} \right \}_{t=1}^{T}|\left \{ \pi \right \}_{j=1}^{N},\left \{ P_{j} \right \}_{j=1}^{N},\left \{ \underline{\mu_{j}} \right \}_{j=1}^{N},\left \{ \Sigma_{j} \right \}_{j=1}^{N}) } \\
& \frac{p(\left \{ \pi \right \}_{j=1}^{N+1},\left \{ P_{j} \right \}_{j=1}^{N+1},\left \{ \underline{\mu_{j}} \right \}_{j=1}^{N+1},\left \{ \Sigma_{j} \right \}_{j=1}^{N+1})p(N+1)}{p(\left \{ \pi \right \}_{j=1}^{N},\left \{ P_{j} \right \}_{j=1}^{N},\left \{ \underline{\mu_{j}} \right \}_{j=1}^{N},\left \{ \Sigma_{j} \right \}_{j=1}^{N}){p(N)}}\\ 
  &\underset{q(N+1\rightarrow N)/q(N\rightarrow N+1)}{\underbrace{ \frac{(N+1)!}{N!} \frac{P_{d}(N+1)/N}{P_{b}(N)/(N+1)}\frac{|J|}{p(\lambda_{j_{*}})p(\Lambda_{j_{*}j_{*}})\prod_{j}p(\Lambda_{jj_{*}})p(\Lambda_{j_{*}j})\prod_{d}p(\beta_{d})p(u_{d})}}}
\end{align*}
Since the parameters are drawn from their respective priors, the Jacobian term $|J|$ will equal one.
}

\subsection{Results}
\label{supsubsec:res}
Posterior distribution on the number of states $N$ with repulsive prior for $a = \exp(-n^{*}_{2.5})$, is $p(2) = 0.08764, p(3) = 0.12862, p(4) = 0.11209,  p(5) = 0.10901, p(6) = 0.10076, p(7) = 0.09890 , ..., p(25) = 0.00008$ with $\sum_{i=2}^{25}p(i) = 1$ and for the independent prior  we display the posterior distribution of the allocation in Figure \ref{fig:acoust:postdist}.

\begin{figure}[htp]
    \includegraphics[height = 6cm, width=\textwidth]{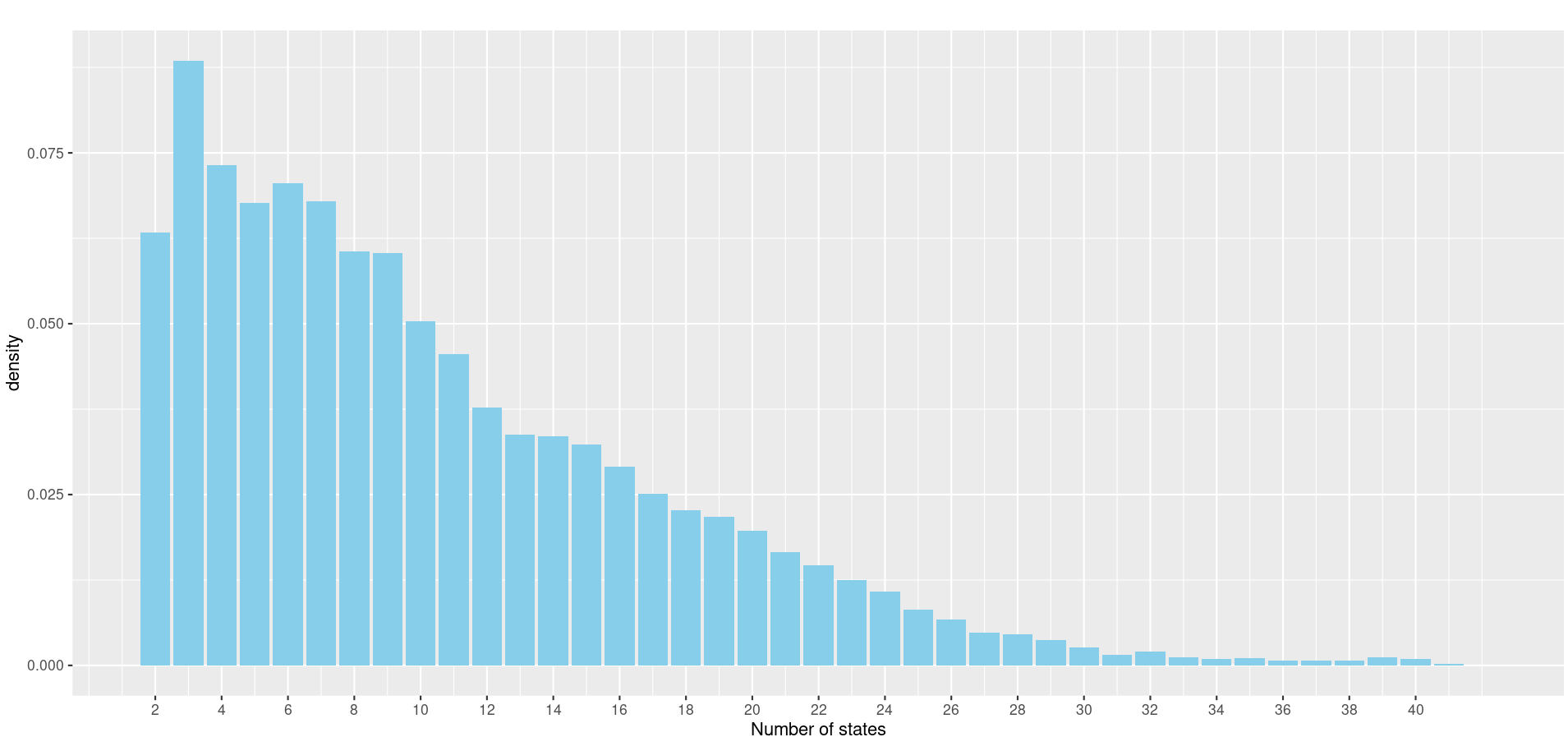}
     \caption{Posterior distribution across the states 2 to 41 for the independent prior model.}
     \label{fig:acoust:postdist}
\end{figure}

We give the posterior uncertainty probabilities of classification for the models with two and three and four mixture states, in Figure \ref{two_three_state_uncert}.
\begin{figure}[htp]
    \includegraphics[height = 6cm, width=\textwidth]{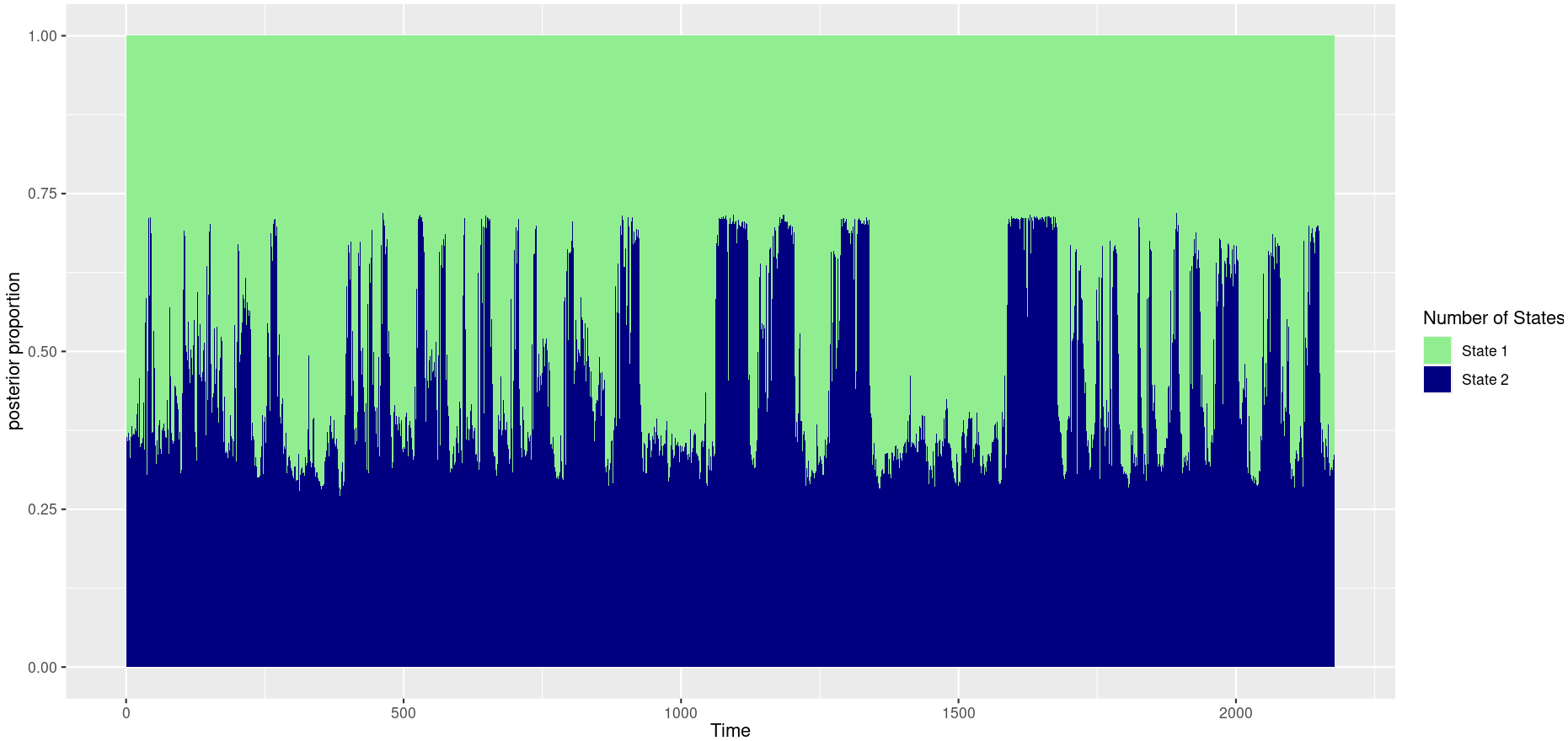}
         \\[\smallskipamount]
    \includegraphics[height = 6cm, width=\textwidth]{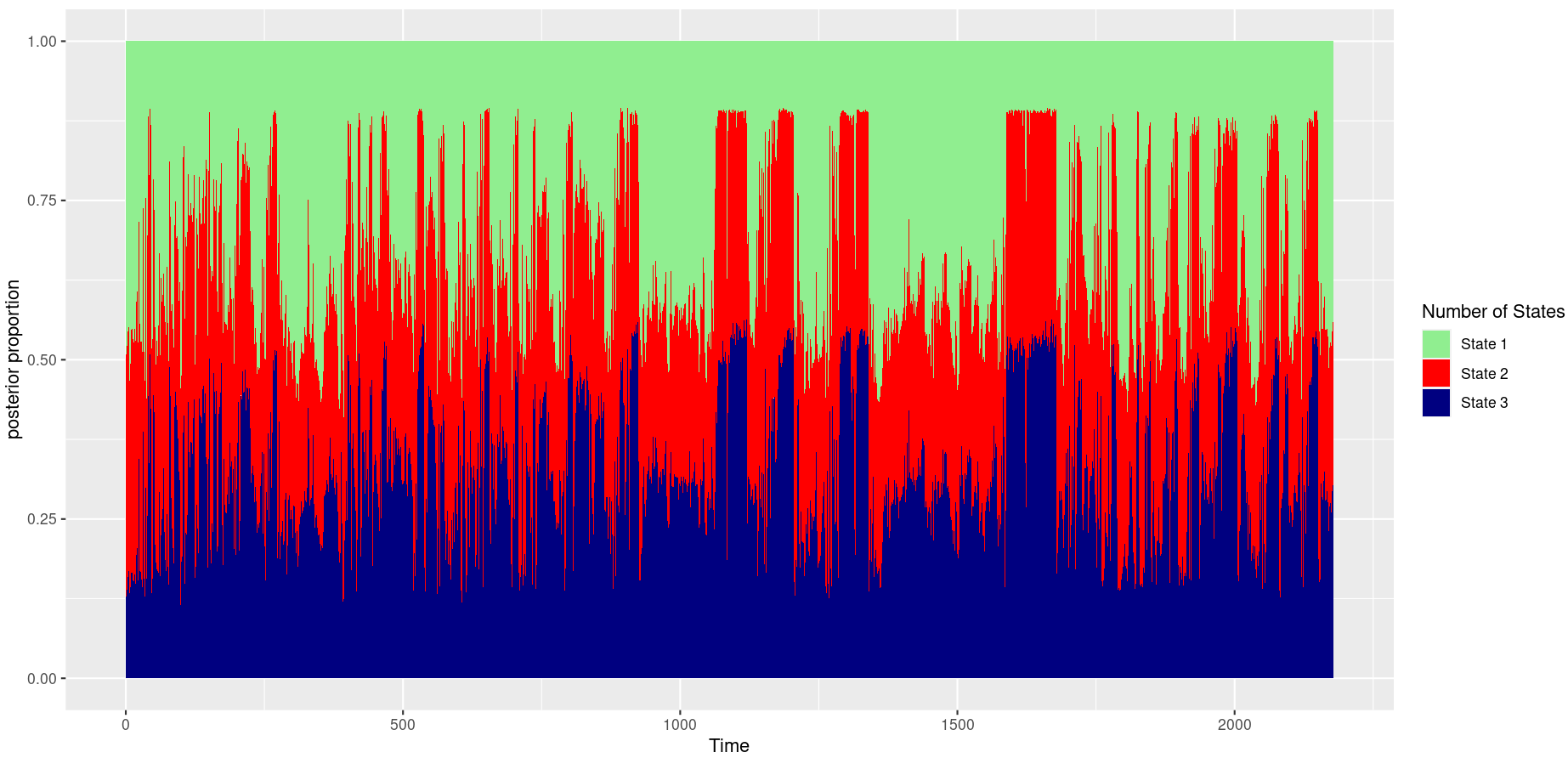}\\[\smallskipamount]
    \includegraphics[height = 6cm, width=\textwidth]{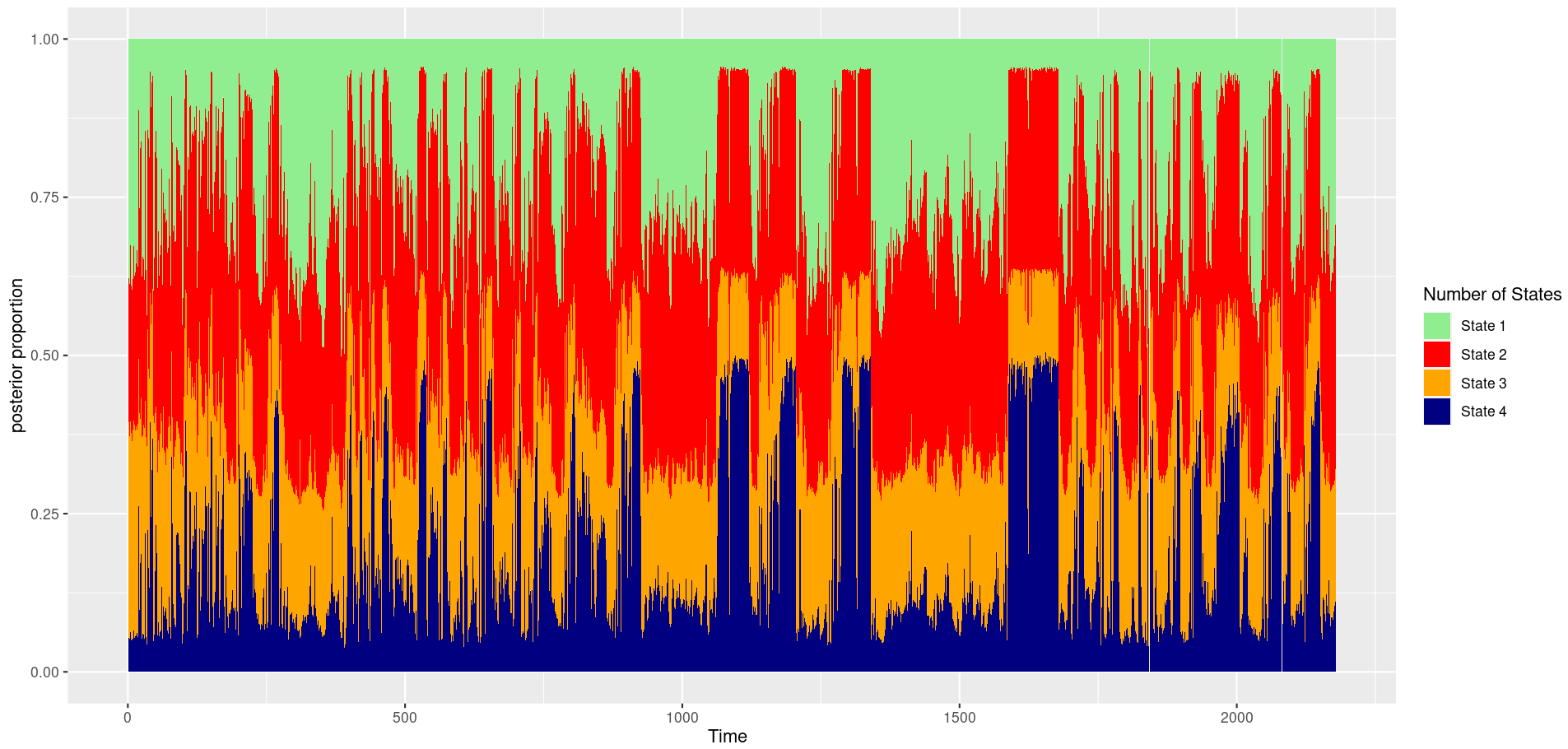}
     \caption{Posterior uncertainty probabilities of classification for the two, three and four state mixture models. On top row we have two states corresponding to (blue) floating on the water and (green) flying. On the middle row we have three states corresponding to (blue) floating on the water, (green) flying and (red) diving/nuisance. Lastly, on the bottom row, we have (blue) floating on the water, (green) flying, (red) diving/nuisance and (orange) floating/flying/nuisance.}
 \label{two_three_state_uncert}
\end{figure}

Next, on Figure \ref{fig:biplot_sup} we display the observations with points coloured according to their modal state allocation on the domain defined by the 2-PC.

\begin{figure}[htp]
    \includegraphics[height = 6cm, width=\textwidth]{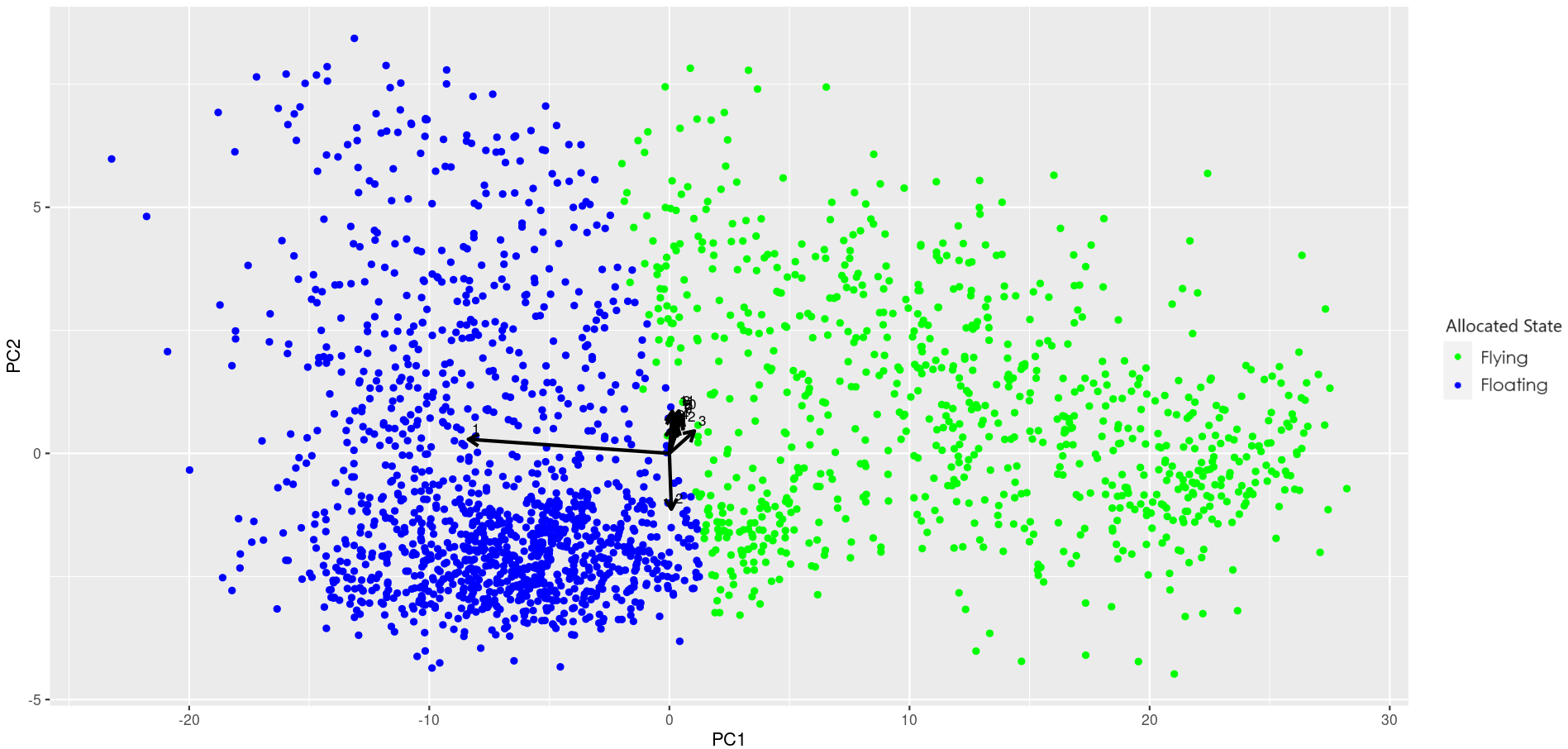}
        \\[\smallskipamount]
    \includegraphics[height = 6cm, width=1.055\textwidth]{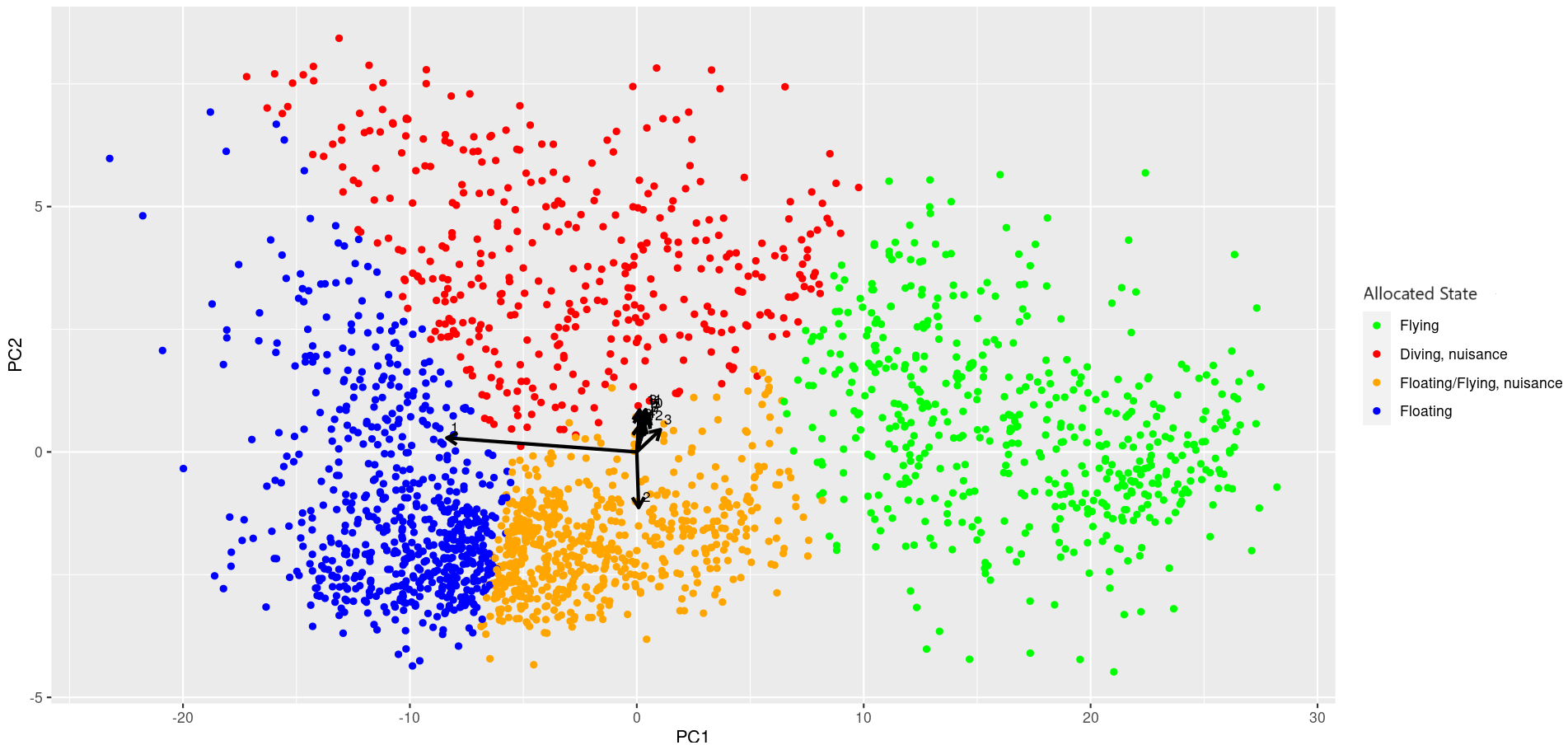}
     \caption{Biplot, with observations coloured according to their modal state allocation, in the case of two states  (top row) and four states (bottom row), plotted on the domain of the first two PC.}
 \label{fig:biplot_sup}
\end{figure}

\begin{figure}[htp]
    \includegraphics[height = 6cm, width=\textwidth]{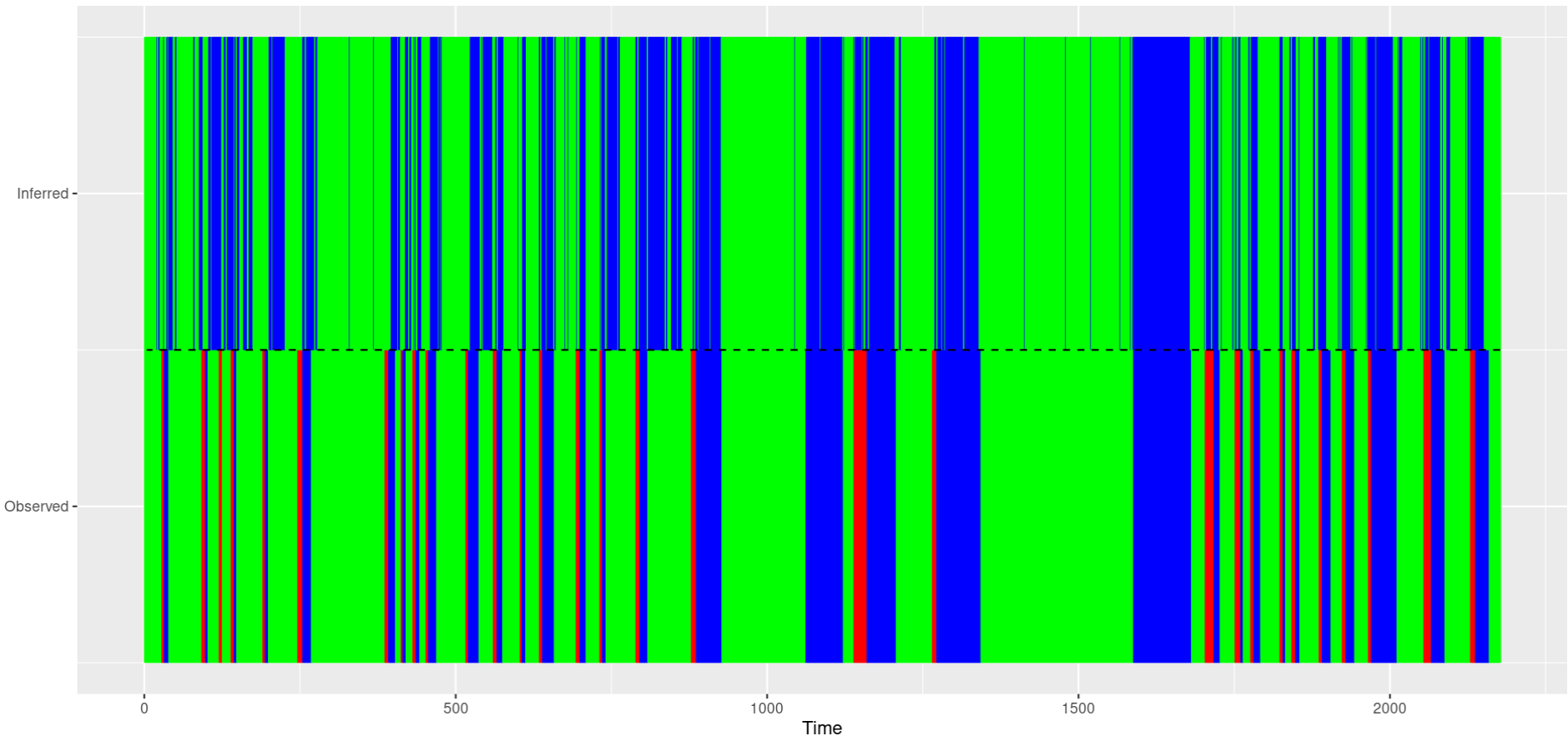}
        \\[\smallskipamount]
    \includegraphics[height = 6cm, width=\textwidth]{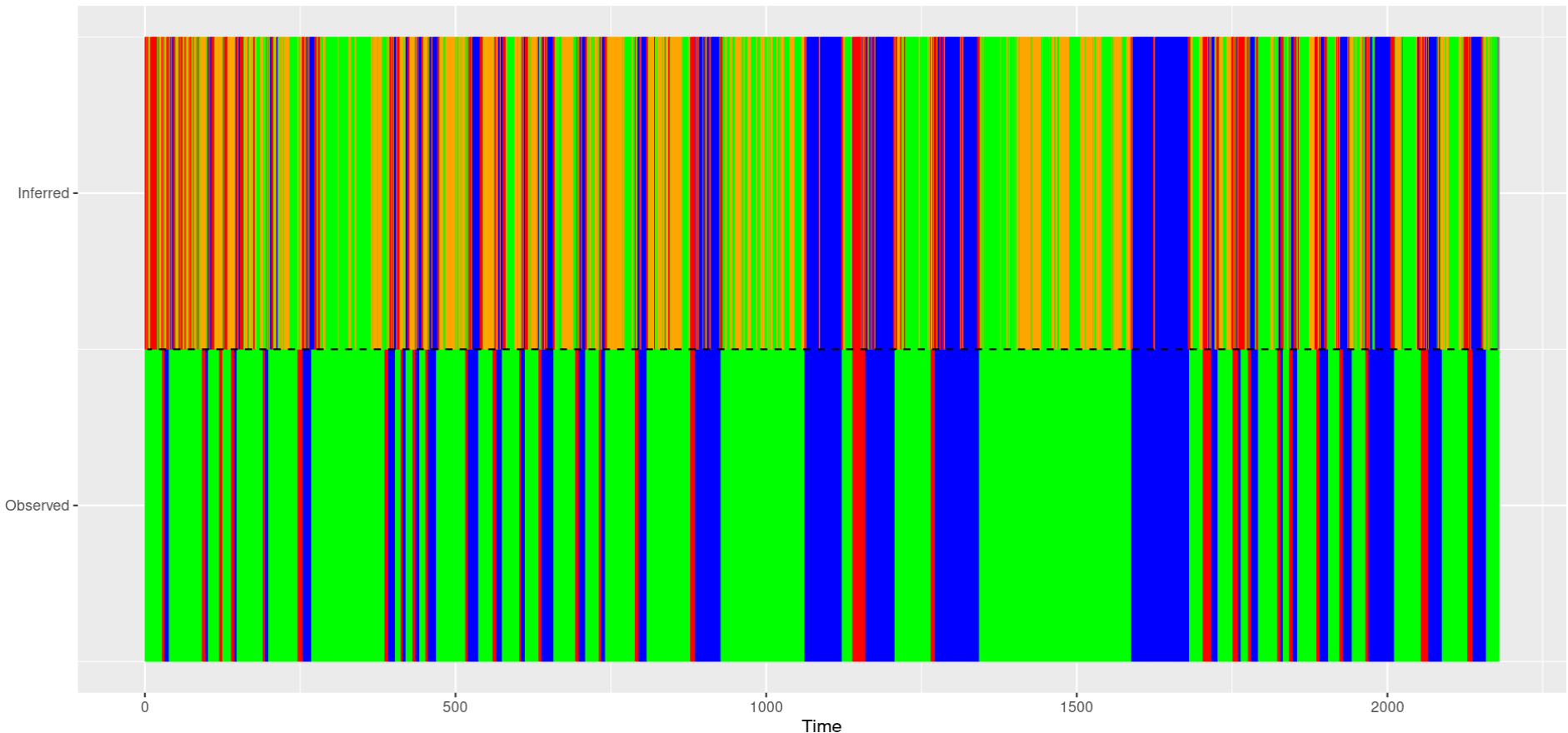}
     \caption{Comparison of the posterior classification of our model, for two states (top row) and four states (bottom row) with the manual
classification of Thiebault et al. (2021) (on the bottom half of each row). Based on the manual classification of
Thiebault et al. (2021), the states are: floating on water (blue), flying (green) and diving
(red). In our model the states are: (blue) floating on the water, (green) flying, (red) diving/nuisance, (orange) floating/flying/nuisance.}
 \label{fig:real_sup}
\end{figure}

We also, present results for the case of $a = \exp(-n^{*}_{5})$ in Figures \ref{fig:real_sup_5}, \ref{two_three_state_uncert_5} and \ref{fig:biplot_sup_5}.

\begin{figure}[htp]
    \includegraphics[height = 6cm, width=\textwidth]{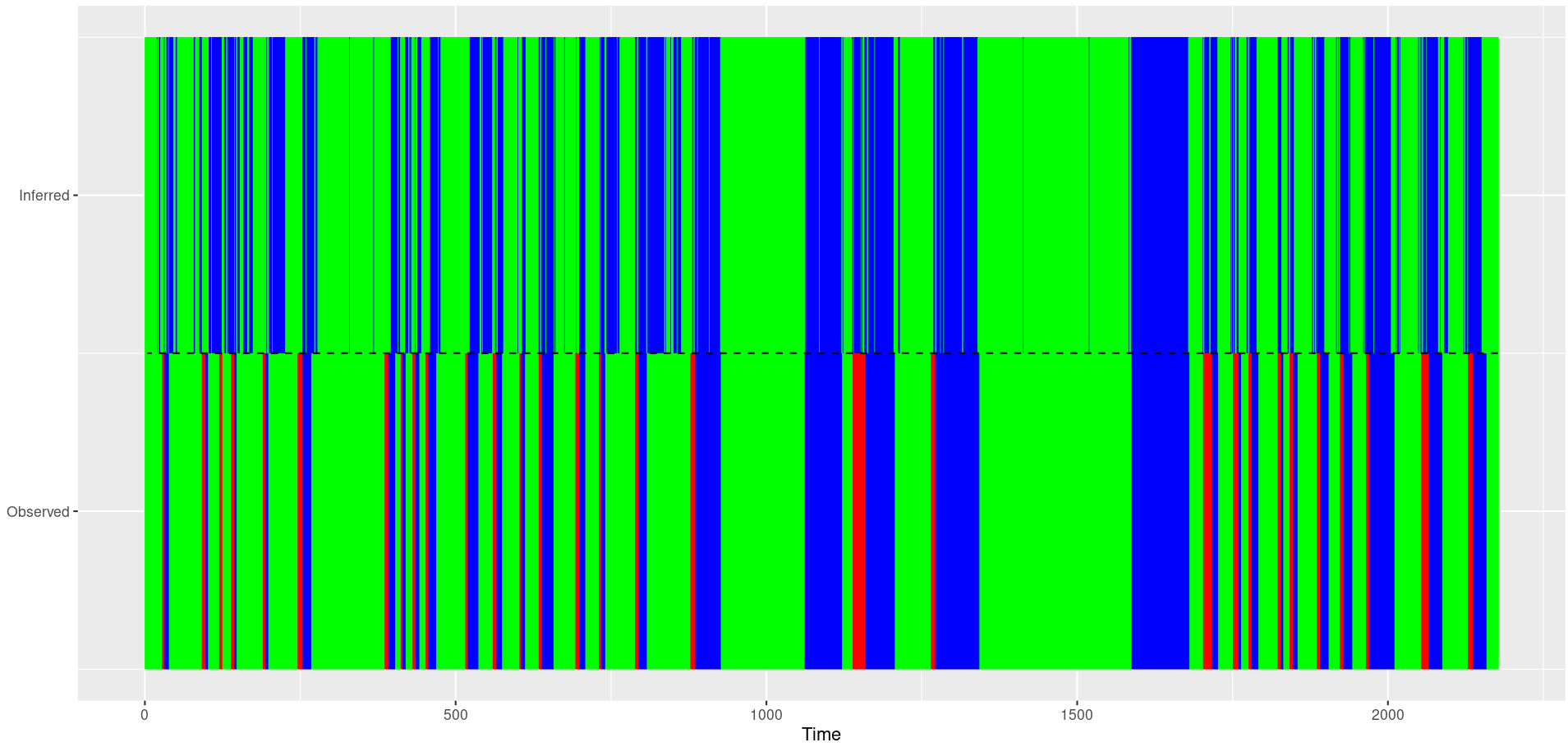}
        \\[\smallskipamount]
    \includegraphics[height = 6cm, width=\textwidth]{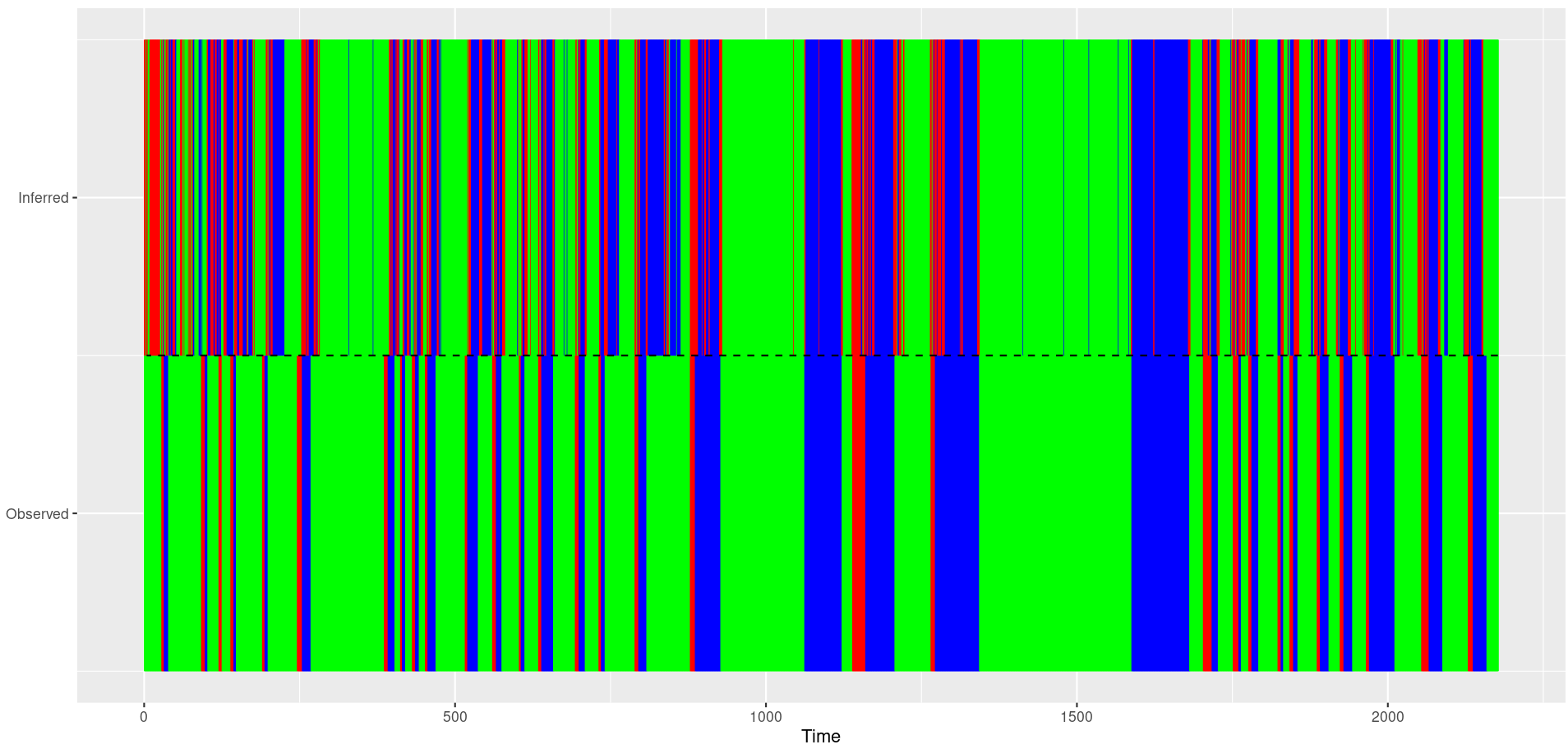}
        \\[\smallskipamount]
    \includegraphics[height = 6cm, width=\textwidth]{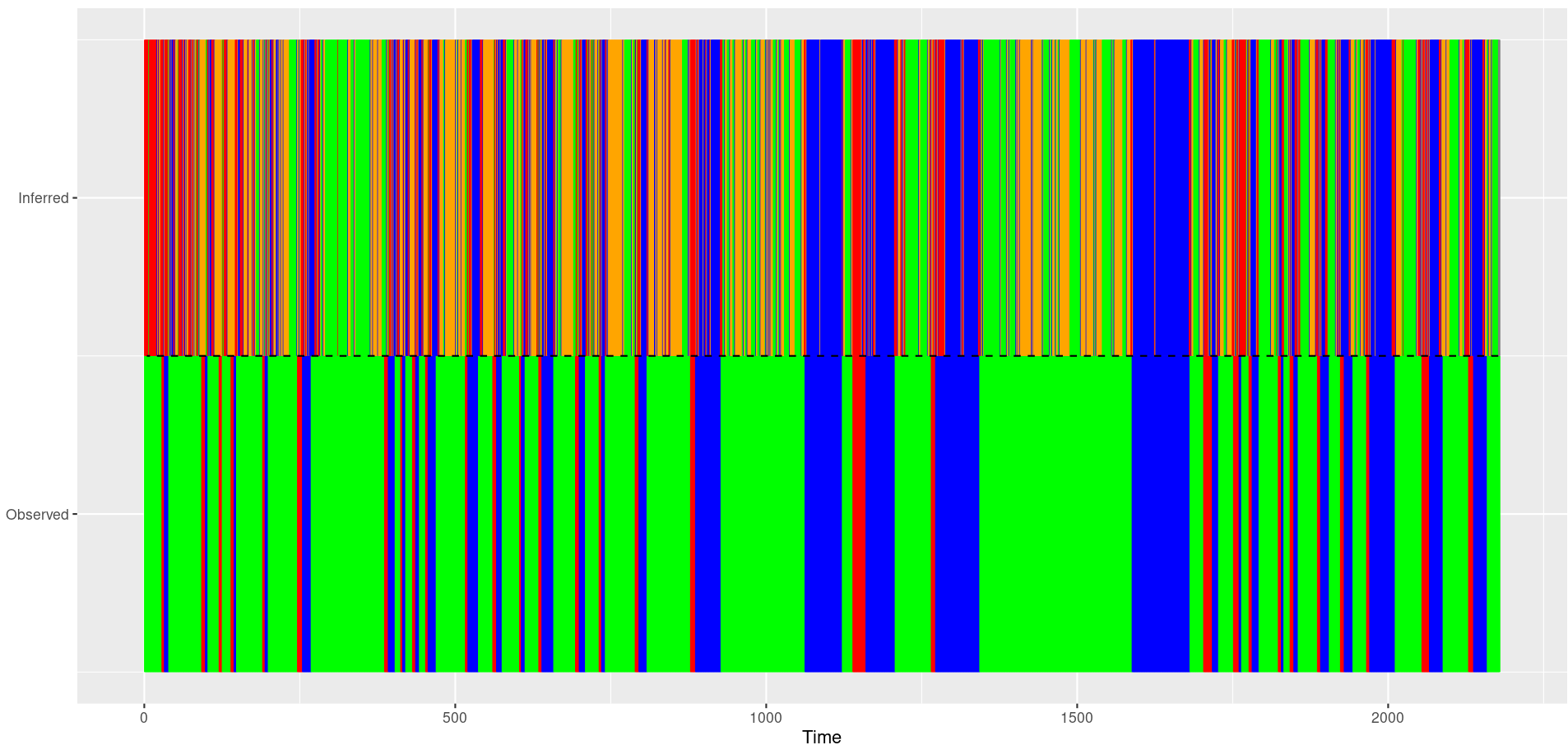}
     \caption{Comparison of the posterior classification of our model, for two states (top row) and four states (bottom row) with the manual
classification of Thiebault et al. (2021) (on the bottom half of each row). Based on the manual classification of
Thiebault et al. (2021), the states are: floating on water (blue), flying (green) and diving
(red). In our model the states are: (blue) floating on the water, (green) flying, (red) diving/nuisance, (orange) floating/flying/nuisance.}
 \label{fig:real_sup_5}
\end{figure}

\begin{figure}[htp]
    \includegraphics[height = 6cm, width=\textwidth]{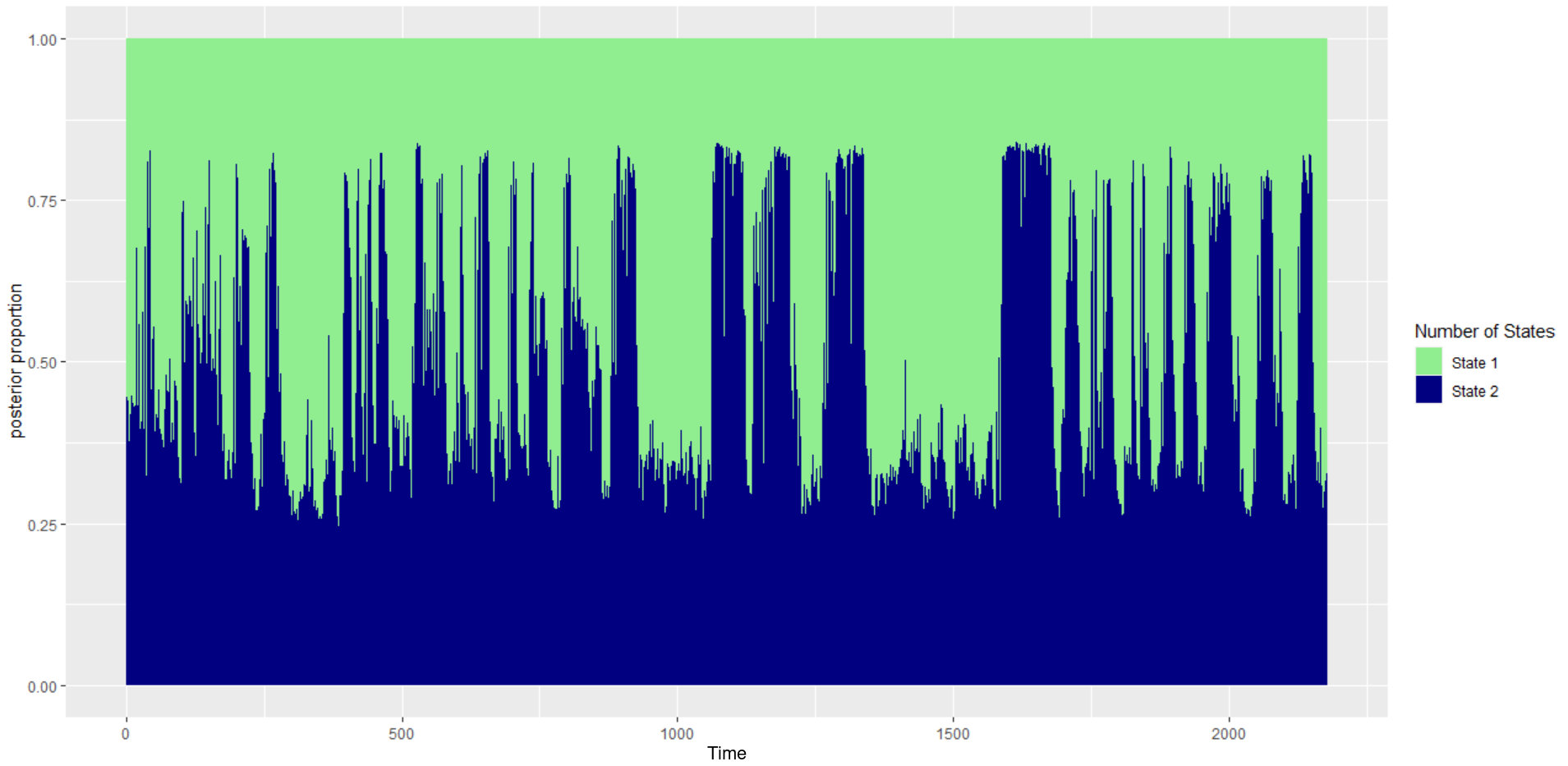}
         \\[\smallskipamount]
    \includegraphics[height = 6cm, width=\textwidth]{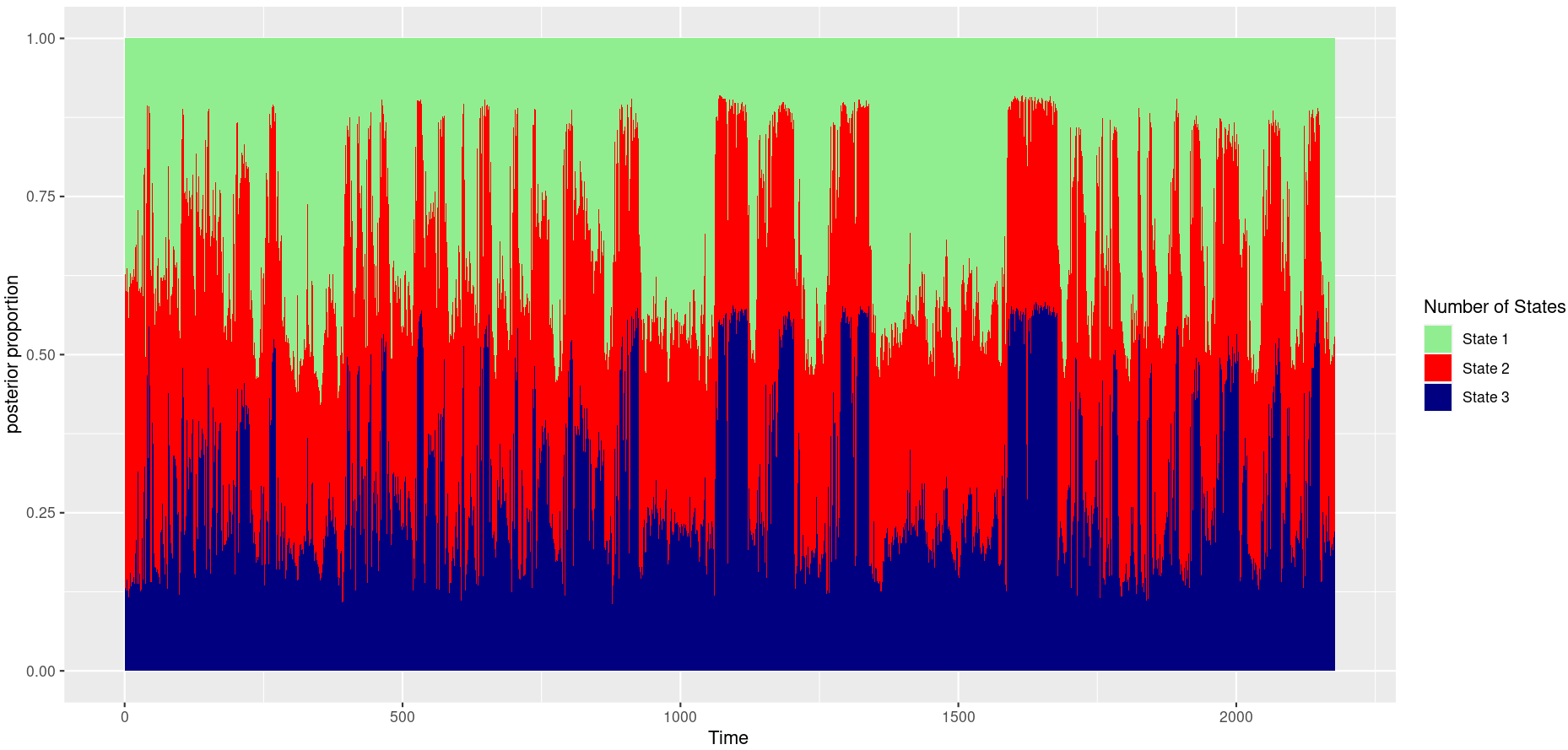}\\[\smallskipamount]
    \includegraphics[height = 6cm, width=\textwidth]{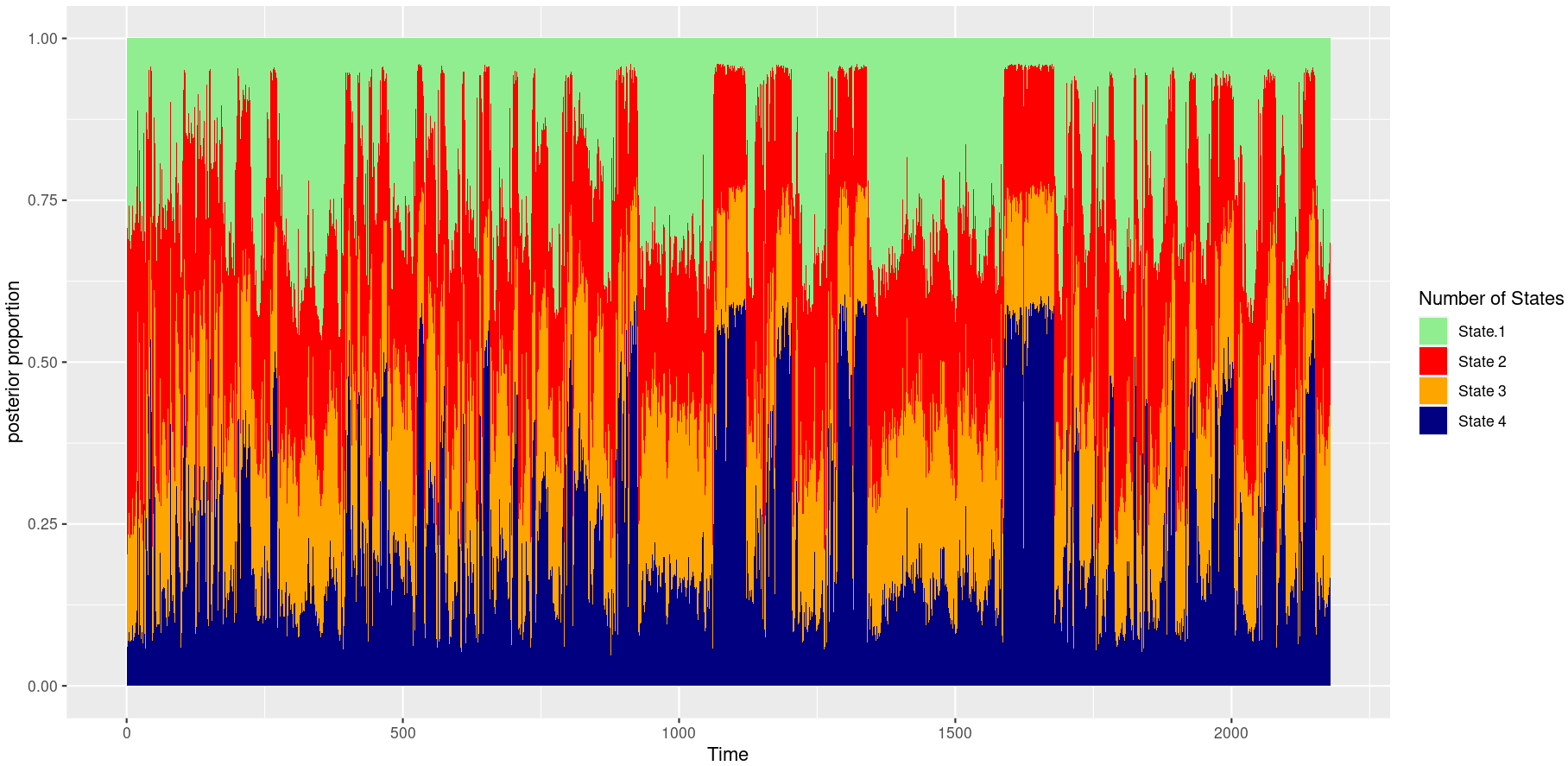}
     \caption{Posterior uncertainty probabilities of classification for the two, three and four state mixture models. On top row we have two states corresponding to (blue) floating on the water and (green) flying. On the middle row we have three states corresponding to (blue) floating on the water, (green) flying and (red) diving/nuisance. Lastly, on the bottom row, we have (blue) floating on the water, (green) flying, (red) diving/nuisance and (orange) floating/flying/nuisance.}
 \label{two_three_state_uncert_5}
\end{figure}

\begin{figure}[htp]
    \includegraphics[height = 6cm, width=\textwidth]{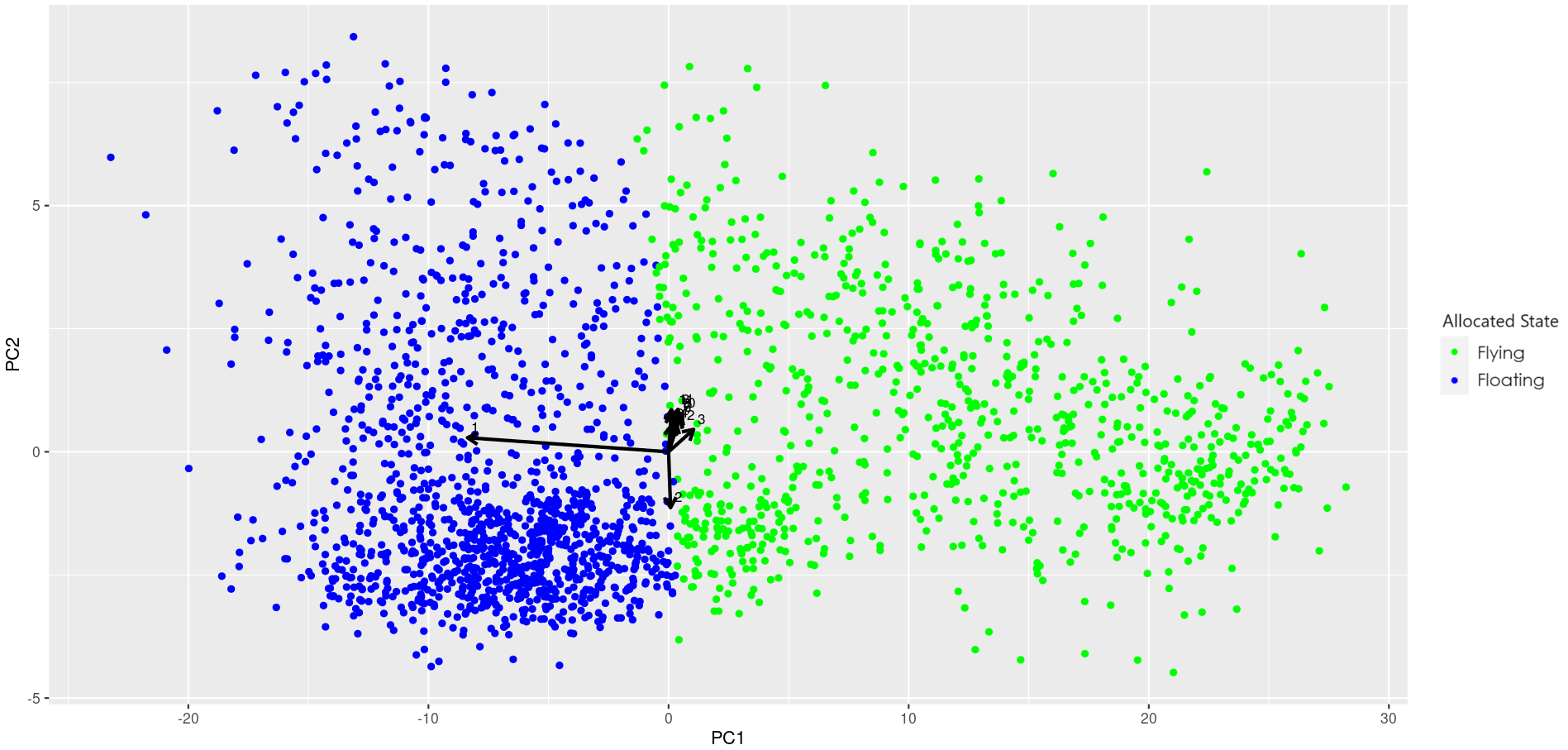}
        \\[\smallskipamount]
    \includegraphics[height = 6cm, width=1.015\textwidth]{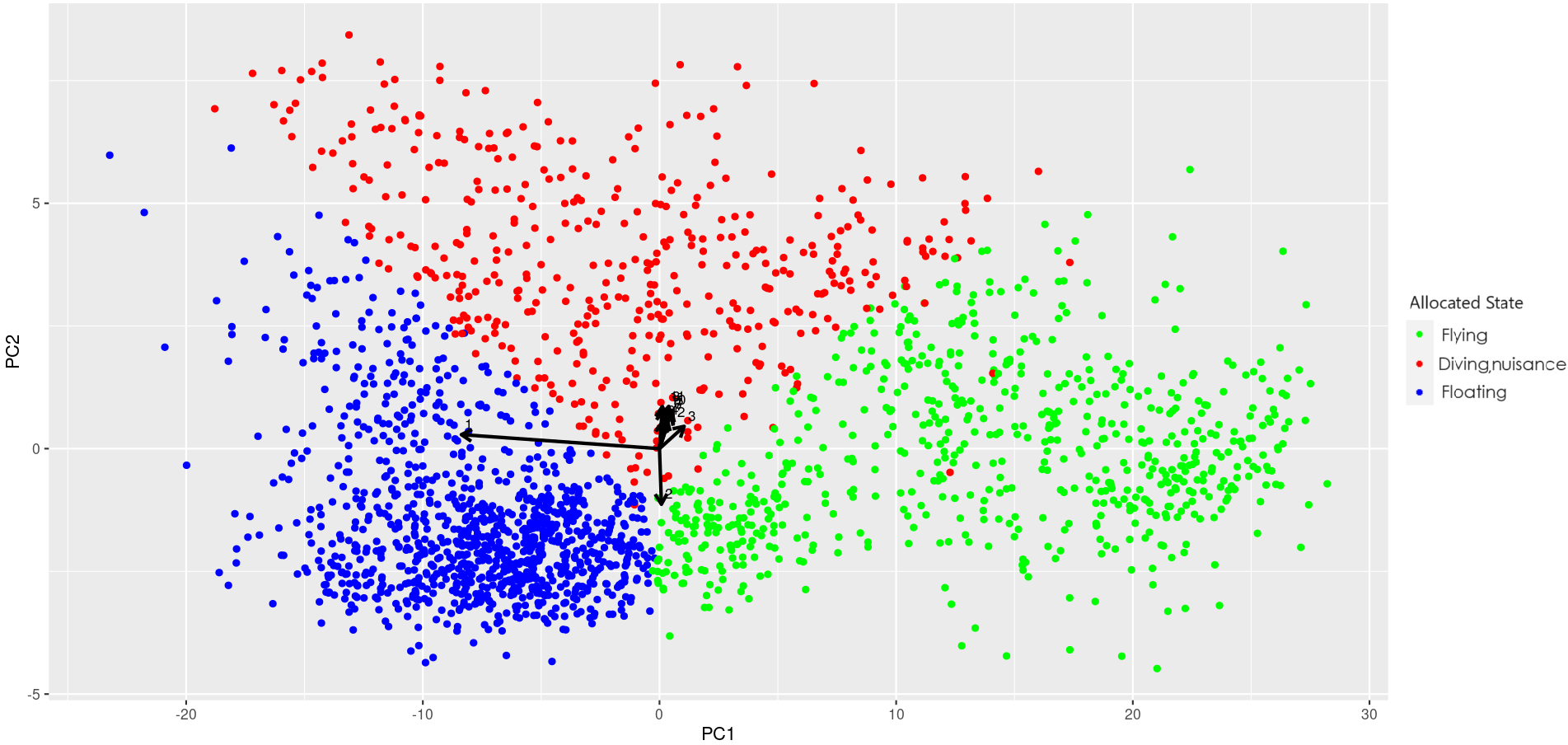}
        \\[\smallskipamount]\includegraphics[height = 6cm, width=1.060\textwidth]{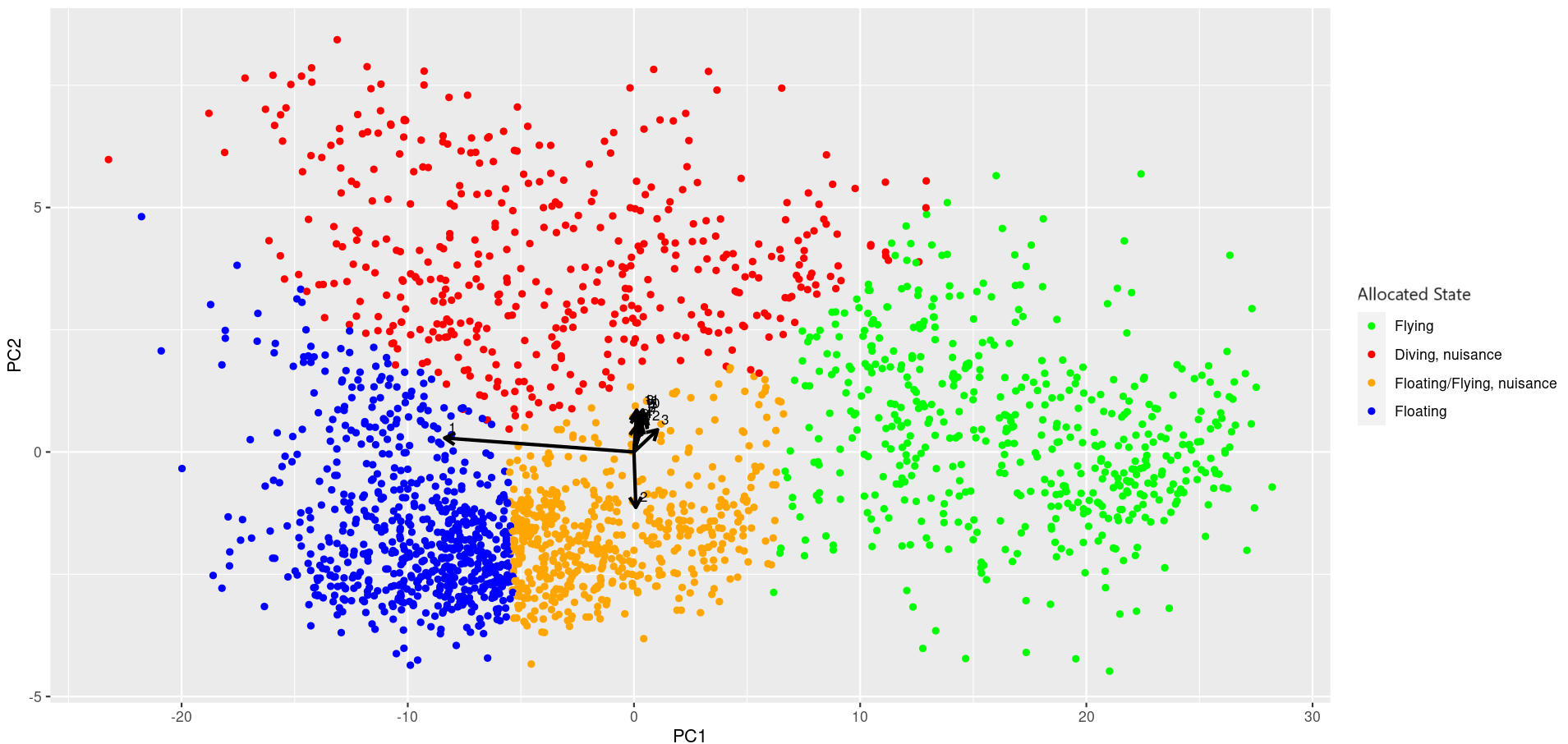}
     \caption{Biplot, with observations coloured according to their modal state allocation, in the case of two  (top row), three (middle row) and four states (bottom row), plotted on the domain of the first two PC.}
 \label{fig:biplot_sup_5}
\end{figure}

For the case were we have $a = \exp(-n^{*}_{5})$ and give rise to a posterior distribution for $N$ for the repulsive prior, $p(2) = 0.11474$, $p(3) = 0.18027$ $p(4) =  0.15573$, $p(5) = 0.12996$, $p(6) = 0.11106$, ... $p(21)=0.00003$, with $\sum_{i=2}^{21}p(i) = 1$.  We observe that the posterior distributions for the parameter $N$ are almost the same either we use percentage $2.5\%$ or $5\%$ with the latter one giving slightly more mass to slightly larger number of states which is expected since it places a smaller penalty.

\bibliographystyle{biom} 
\bibliography{HMMbib.bib}

\end{document}